%% file: stefan_jcp_third.tex
\documentclass[sort&compress,11pt]{elsarticle}
\usepackage[utf8]{inputenc} 
\usepackage[T1]{fontenc}    
\usepackage{hyperref}       
\usepackage{url}            
\usepackage{booktabs}       
\usepackage{amsfonts}       
\usepackage{nicefrac}       
\usepackage{microtype}      
\usepackage{lipsum}		
\usepackage{graphicx}
\usepackage{natbib}
\usepackage{doi}
\usepackage{color}
\usepackage{wrapfig}
\usepackage[small]{caption}
\usepackage{amsmath} 
\usepackage{bm} 

\usepackage{multicol}

\usepackage[linesnumbered,ruled,vlined]{algorithm2e}
\makeatletter
\newcommand{\nosemic}{\renewcommand{\@endalgocfline}{\relax}}
\newcommand{\dosemic}{\renewcommand{\@endalgocfline}{\algocf@endline}}
\let\oldnl\nl
\newcommand{\nonl}{\renewcommand{\nl}{\let\nl\oldnl}}
\SetKwRepeat{Do}{do}{while}
\makeatother

\newcommand{\erf}{\text{erf}}
\newcommand{\x}{\bm{x}}
\newcommand{\normalVect}{\bm{n}}
\newcommand{\jump}[1]{\mbox{$[\![ #1 ]\!]$}}

\newcommand{\dint}{\mbox{\,  \ensuremath{\text{d}}}}

\newcommand{\bfgn}{\mathbf{g}^n}
\newcommand{\bfgnp}{\mathbf{g}^{n+1}}

\newcommand{\tmax}{t_{\text{max}}}
\newcommand{\energy}{e}
\newcommand{\heatSource}{S}
\newcommand{\temp}{T}
\newcommand{\test}{\temp^*}

\newcommand{\speed}{\bm{w}}

\newcommand{\xnodes}{\bm{X}}

\newcommand{\OmegaS}{\Omega_s}
\newcommand{\OmegaL}{\Omega_\ell}
\newcommand{\front}{\Gamma}

\newcommand{\volumicMass}{\rho}
\newcommand{\volumicMassSolid}{\volumicMass_s}
\newcommand{\volumicMassLiquid}{\volumicMass_\ell}
\newcommand{\specificHeat}{c}
\newcommand{\specificHeatSolid}{\specificHeat_s}
\newcommand{\specificHeatLiquid}{\specificHeat_\ell}
\newcommand{\conductivity}{k}
\newcommand{\conductivitySolid}{\conductivity_s}
\newcommand{\conductivityLiquid}{\conductivity_\ell}
\newcommand{\diffusivity}{\alpha}
\newcommand{\diffusivitySolid}{\diffusivity_s}
\newcommand{\diffusivityLiquid}{\diffusivity_\ell}

\newcommand{\latentHeat}{l}

\newcommand{\dirichletBoundary}{\Gamma^d}
\newcommand{\dirichletBoundarySolid}{\dirichletBoundary_s}
\newcommand{\dirichletBoundaryLiquid}{\dirichletBoundary_\ell}
\newcommand{\dirichletTemp}{\temp_d}
\newcommand{\boundaryTempSolid}{\temp_s}
\newcommand{\boundaryTempLiquid}{\temp_{\ell}}
\newcommand{\boundaryTempInterface}{\temp_{0}}

\newcommand{\tempSpace}{\mathcal{U}_{\mathrm{d}}}
\newcommand{\tempSpaceZero}{\mathcal{U}}

\newcommand{\frontPositionEx}{x_f}
\newcommand{\frontPositionExAxisym}{r_f}
\newcommand{\solexParam}{\phi}
\newcommand{\heatSink}{Q}

\newcommand{\meshSet}{\mathcal{M}}
\newcommand{\nodesSet}{\mathcal{N}}

\newcommand{\elemSize}{h_e}

\newcommand{\err}{\mathrm{err}}
\newcommand{\errXf}{\mathrm{\xi}}
\newcommand{\tol}{\epsilon}

\newcommand{\eqspace}[0]{
 \; \;
}

\newcommand{\timeDeriv}[1]{
\ensuremath{ 
#1_{,t}
}
}

\newcommand{\myVectorGrad}[1]{
   \ensuremath{ 
  \bm{\nabla} #1
  }
}

\graphicspath{{./fig/}}

\newcommand{\answ}[1]{{\color{black}{#1}}}

\title{The eXtreme Mesh deformation approach (X-MESH) for the Stefan phase change model}

\hypersetup{
pdftitle={The eXtreme Mesh deformation approach (X-MESH) for the Stefan phase change model},
pdfauthor={Nicolas ~Mo\"{e}s, Jean-François Remacle, Jonathan Lambrechts, Beno\^{i}t L\'{e} },
pdfkeywords={X-MESH, Front relaying, Sharp interface, Phase change, Stefan model, Solidification},
}

\journal{} 

\begin{document}
\begin{frontmatter}

  \author[ECN,IUF]{Nicolas ~Mo\"{e}s}
  \ead{nicolas.moes@ec-nantes.fr}
  \author[UCLouvain]{ Jean-François ~Remacle}
  \ead{jean-francois.remacle@uclouvain.be}
  \author[UCLouvain]{Jonathan Lambrechts}
  \ead{jonathan.lambrechts@uclouvain.be}
  \author[ECN]{Beno\^{i}t L\'{e}}
  \ead{benoit.le@ec-nantes.fr}
  \author[ECN]{Nicolas Chevaugeon}
  \ead{nicolas.chevaugeon@ec-nantes.fr}

\address[ECN]{
	Nantes Université, Ecole Centrale de Nantes, \\
	GeM Institute, UMR CNRS 6183, 
	1 rue de la No\"{e}, 44321 Nantes, France.}
\address[UCLouvain]{
        Université Catholique de Louvain \\
	Institute of Mechanics, Materials and Civil Engineering \\ 
	1348 Louvain-la-Neuve, Belgium}
\address[IUF]{Institut Universitaire de France (IUF)}


%
  
\begin{abstract}
    The eXtreme Mesh deformation approach (X-MESH) is a new paradigm to follow sharp interfaces  without remeshing and without changing the mesh topology. 
    Even though the mesh does not change its topology, it can follow interfaces that do change their topology (nucleation, coalescence, splitting) \answ{and that possibly travel over long distances.}  
    To make this possible, the key X-MESH idea is to allow elements to reach zero measure. This permits interface relaying between nodes
    as well as interface annihilation and seeding in a time continuous manner. 
    The paper targets the Stefan phase change model in which the interface (front) is at a given temperature. Several examples demonstrate the capability of the approach.
\end{abstract}


\end{frontmatter}

\section{Introduction}

The Stefan model is a classical phase change 
model in which motion is neglected: fluid and solid are at rest
(expansion due to freezing is neglected). The phase change takes place
at a given temperature (273.15K for water-ice) and the free energy
is continuous across the front. The internal energy, on the contrary,
is discontinuous across the front and its jump defines the latent heat.
It is the amount of energy released by the fluid as it becomes solid. 
The heat flux is also discontinuous across the front. 
Its jump is equal to the product of the latent heat and the front velocity.
\answ{An extensive description of the Stefan model may be found in the book by Gupta \cite{gupta2003classical} or the more recent 
paper \cite{Koga2020} targeted towards control.
An important seminal work on the mathematical description of the Stefan model can also be found in \cite{friedman1959free}}.

 \begin{figure}
  \centering
  \includegraphics[width=0.9\textwidth]{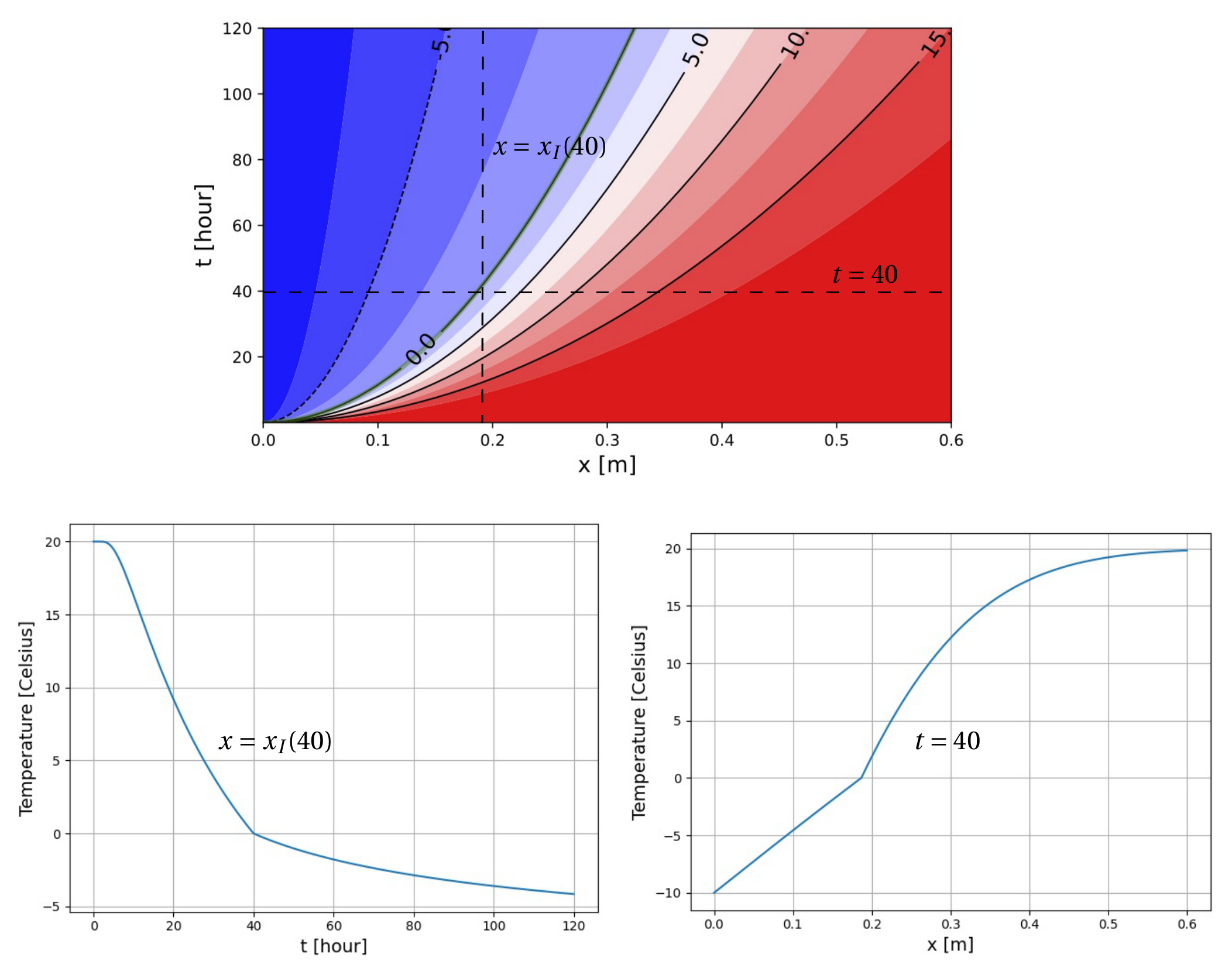}
 \caption{Temperature field for a Stefan problem.}
  \label{fig:stefan}
  \end{figure}
The temperature field for a one-dimensional Stefan model of the freezing 
of water is illustrated on
Figure\ref{fig:stefan}.
The liquid occupies the region $0 \leq x < \infty$. 
We look for a temperature field $T(x,t)$
that is initially at a temperature $T(x,0) = T_{\mathrm{i}} > T_0$ where
$T_0$ is the temperature of phase transition. For all $t>0$, the left side
$x=0$ of the domain is cooled: $T(0,t) = T_c < T_0$.
Solidification begins immediately and the solid phase occupies a region $0
\leq x \leq x_I(t)$ where $x_I(t)$ is the position of the ice front at time $t$: 
$T(x_I(t ),t) = T_0$. Figure \ref{fig:stefan} shows the
solution $T(x,t)$ for physically sound ice and water
parameters ($T_0=0^\circ C$).  Even though the temperature
$\temp(x,t)$ is continuous, both its time and space derivatives are discontinuous across the front.
The derivative jumps are linked with the interface normal velocity, $v$, to ensure temperature continuity. 
More precisely, we have on the front 
\begin{equation}
\nonumber
 \jump{T_{,t}} + v  \jump{T_{,x}} = 0,
\end{equation}
where $T_{,t}$ and $T_{,x}$ indicates the time and space partial derivatives, respectively and $\jump{\cdot}$ is the jump symbol.


The Stefan model addresses the stable solidification of a liquid 
initially at a temperature above the solid-liquid equilibrium 
temperature. In this model, 
surface tension and kinetic mobility are neglected
\cite{Jaafar2017}. 
To model unstable processes as dendritic growth,
the fixed front temperature must be replaced by the so-called Gibbs-Thomson condition. This relation links the front temperature to
its curvature and speed. There also exist anisotropic versions of the Gibbs-Thomson model which take into account the variation of the surface tension and kinetic mobility with respect to the interface orientation.

The Stefan model is known to lead to possible blow-ups in the solution when the initial temperature profile corresponds to an undercooled liquid or a superheated solid \cite{Back2014,King2006}. We will not be considering these type of profiles here.

Numerical schemes to address the Stefan model may be classified 
into tracking and capturing schemes.
In the first category, the mesh moves with the interface.
\answ{This is the case for the Arbitrary \textcolor{black}{Lagrangian Eulerian} (ALE)
approach which was first developed to follow material interfaces in flows \cite{Huerta2004,LOUBERE20104724,boscheri2014high,BARLOW2016603,BURTON2018492}. 
These techniques were then adapted to immaterial interface for phase-change models
\cite{Baines2009,Gros2018,Zhang2019}.}

Unfortunately ALE is not able to handle topological changes of the front and requires remeshing when the mesh is too distorted. Remeshing requires projection of the solution between successive meshes which is highly detrimental to the continuity of the solution in time.
Capturing approaches use a fixed mesh.
This is the case for the level set approach \cite{Chen1997a,Shaikh2016,Vasilev2020}
and the extended finite element method \cite{Merle2002, Ji2002,He2021}.

\answ{
This work starts from an observation: none of the methods proposed in
the state of the art is entirely satisfactory. We therefore decided to
start by listing 4 fundamental properties that would allow us to
design a totally satisfactory scheme:
   \begin{itemize}
\setlength\itemsep{0em}      
    \item[] {\bf P1 - Time continuous deforming mesh with fixed  topology:} the mesh deforms in time in a continuous manner: nodes trajectories form continuous paths and the mesh adjacencies are fixed.
    \item[] {\bf P2 - Conforming mesh to sharp physical interfaces.} 
   \item[] {\bf P3 - Compliant with topological changes:} 
   nucleation, collapse, coalescence and splitting.
   \item[] {\bf P4 - Classical finite element approximation:} no enrichment with additional degrees of freedom.
 \end{itemize}
Properties P1, P2 and P4 ensures that the true solution at
the nodes evolve smoothly over time  
which provides benefits for the accuracy of the
approximation scheme.
 Matching the 4 properties simultaneously is currently not possible
 and very challenging. Up to now, property P3 has always
 been considered as incompatible with the others: large changes
 of front topologies have always been concomitant with mesh adaptation
 or discrete solution enrichment which is incompatible with property
 P4.
 After a lot of thought, we came with a striking conclusion.
The quest for meshes with nice-only element is counter-productive: bad
quality elements or even elements \emph{with a zero measure} opens a new spectrum of
possibilities.
}

We consider in this paper \emph{a new paradigm for mesh movements} allowing  elements to reach zero measure at some instant in their evolution.
This extreme mesh deformation (X-MESH) allows an interface to be
relayed from one node to another in a continuous  fashion.
It also allows interface annihilation or seeding. Moreover, no remeshing is needed and the mesh topology \answ{is} kept fixed. Only node movements are needed. \answ{Over any time-step, the relation between the field change and mesh movement is written in an arbitrary Lagragian Eulerian format.  
The key aspect of the X-MESH is that it keeps the same mesh throughout 
the simulation
whereas traditional ALE simulations do require remeshing for large interface movement and/or topological changes.}

\answ{Note that for different applications than phase-change, 
other works in the literature allow extreme/degenerate mesh deformations to account for large mesh deformations. 
For instance a high-order discontinuous Galerkin method with unstructured space-time meshes for two-dimensional compressible flows on domains with large deformations was considered in \cite{Wang2015} and 
high order direct ALE schemes on moving Voronoi meshes with topology changes were introduced in \cite{GABURRO2020109167,gaburro2021unified}.
Large boundary displacements were also considered with ALE scheme for unsteady inviscid flows in \cite{re2017interpolation}. Finally, it is worth noting the work of Springel in similar directions  \cite{springel2010pur}.
}

The paper is organized as follows. 
Section \ref{twoPhaseStefanModel} describes the equation of the Stefan model both in strong and weak forms. Then it details both time 
and space discretization schemes as well as the nonlinear iterative algorithm.
Section \ref{sec:relaying} describes the X-MESH front relaying process through which the mesh is always matching the front.
In Section \ref{numericalExamples}, a set of 2D simulations are presented.
The straight front semi-infinite 
two-phase problem is considered as well as the axisymmetric case. Both cases are considered with and without latent heat. Finally, the freezing of a fluid subjected to two rotating heat sinks is considered.

\section{Two-phase Stefan model}
\label{twoPhaseStefanModel}
\begin{wrapfigure}{L}{0.4\textwidth}
  \centering
   \includegraphics[width=.37\textwidth]{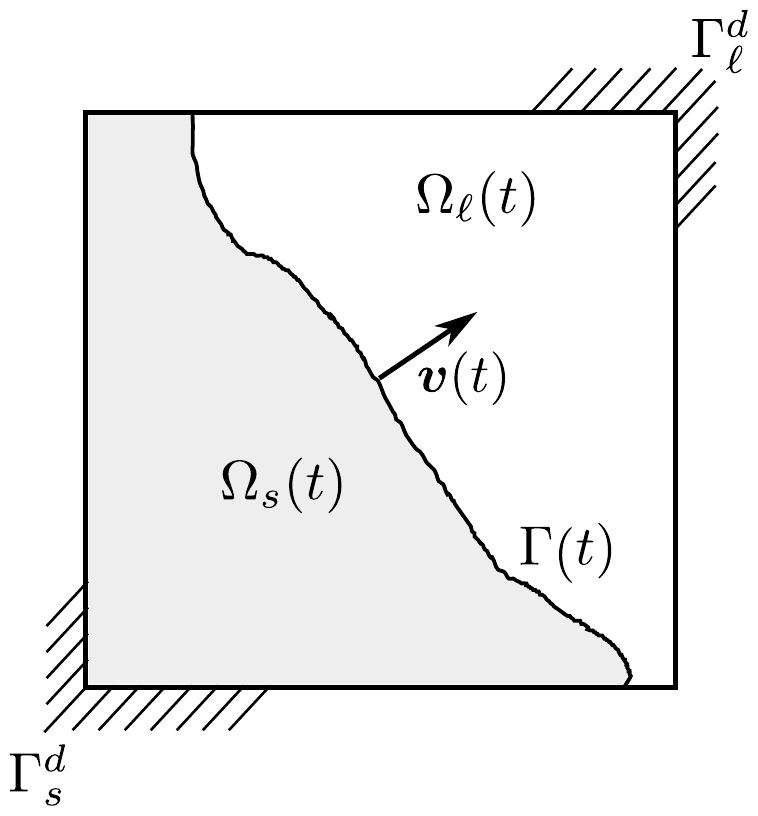}
 \captionof{figure}{Problem definition.}
  \label{fig:problemDefinition}
  \end{wrapfigure}

We consider a domain $\Omega$ depicted in Figure \ref{fig:problemDefinition}. Its boundary, $\partial \Omega$, is assumed fixed in time.
The domain is composed of the solid and liquid  
phases denoted  $\OmegaS$ and $\OmegaL$, respectively.
The front separating the phases, $\front$, 
moves with a velocity denoted $\bm{v}$.

We seek to determine the evolution of the temperature field $\temp(\x,t)$ [K] for $(\x,t) \in \Omega \times [0, \tmax]$, which satisfies the governing equations
 \begin{align}
  \volumicMassSolid \specificHeatSolid \timeDeriv{\temp} = & \myVectorGrad{  (\conductivitySolid \myVectorGrad{ \temp })} + \volumicMassSolid \heatSource \text{ in } \OmegaS, \nonumber \\
   \volumicMassLiquid \specificHeatLiquid \timeDeriv{\temp} = & \myVectorGrad{  (\conductivityLiquid \myVectorGrad{  \temp })} + \volumicMassLiquid \heatSource \text{ in } \OmegaL, \nonumber
 \end{align}
where $\volumicMassSolid$ and $\volumicMassLiquid$ are the densities [kg.m$^{-3}$], $\specificHeatSolid$ and $\specificHeatLiquid$ the specific heat capacities [J.K$^{-1}$.kg$^{-1}$], $\conductivitySolid$ and $\conductivityLiquid$ the isotropic thermal conductivities [W.m$^{-1}$.K$^{-1}$]  of the phases and $\heatSource$ a heat source [W.kg$^{-1}$]. 

Note that to ensure mass conservation within the Stefan model, densities must be considered equal ($\volumicMassSolid = \volumicMassLiquid = \volumicMass$).
The temperature field must also satisfy the following boundary and initial conditions:
\begin{subequations}
\label{dirichletAndInitialConditions}
 \begin{align}
  \temp(\x,t) & = \dirichletTemp, \quad (\x,t) \in \dirichletBoundary \times  [0, \tmax], \\
  \temp(\x,0) & = T_{\mathrm{i}}(\x), \quad  
  \x \in \Omega. 
 \end{align}
\end{subequations}

Here, $\dirichletTemp$ is the temperature imposed on the Dirichlet
boundary $\dirichletBoundary = \dirichletBoundarySolid \cup
\dirichletBoundaryLiquid$. 
To simplify the presentation but without loss of generality, 
the imposed temperature is assumed constant and uniform over $\dirichletBoundary$. The location of $\dirichletBoundary$ is also assumed fixed in time.
The remainder of the
external boundary ($\partial \Omega \setminus 
\dirichletBoundary$) is insulated 
(zero normal temperature gradient).

\answ{At any instant $t$, the location of the phase change front $\Gamma(t)$  is defined by the set of point whose temperature is at the transition temperature 
\begin{equation}
    \Gamma(t) = \{ \x \in \Omega: \temp(\x,t)  = \boundaryTempInterface \}.
\end{equation}}
On the phase change front, the heat flux jump is related to the front speed by
\begin{equation}
 (\conductivitySolid \myVectorGrad{\temp} - \conductivityLiquid \myVectorGrad{\temp}) . \normalVect =  \volumicMass \latentHeat (\bm{v} \cdot \normalVect) \text{\ on\ } \Gamma,
 \label{eq:jump_flux}
\end{equation}
where $\normalVect$ is the normal to $\Gamma$ moving at velocity $\bm{v}$ and $\latentHeat$ is the latent heat [J.kg$^{-1}$].

\begin{figure}[b!]
\centering
 \includegraphics[width=0.8\textwidth]{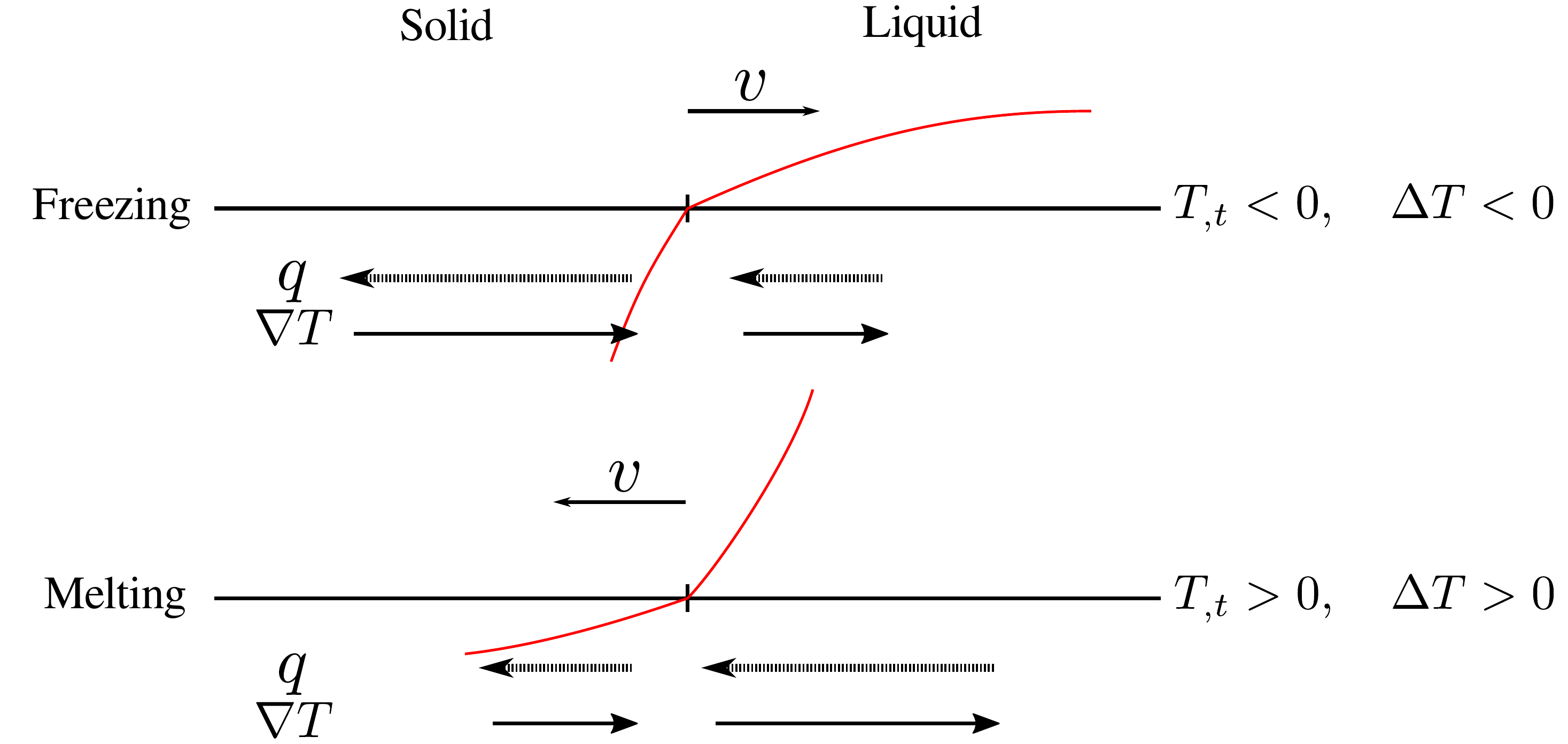}
 \caption{Sketch of the temperature profile across a freezing or melting front in the Stefan model (in the absence of volumetric heat sources). The vector $q$ indicates the heat flux.}
 \label{fig:tempfront}
\end{figure}

A sketch of the temperature profile close to a front in the Stefan model is  depicted in Figure \ref{fig:tempfront}. 
In the absence of volumetric heat sources, 
the profile is concave (convex) in the case of freezing (melting). 
The direction of the heat flux is always from the fluid to the solid.
Its magnitude is lower (higher) in the liquid in the case of freezing (melting). The heat flux jump corresponds to the latent heat released 
by the fluid in the case of freezing or absorbed by the solid as it becomes a  fluid in the case of melting.

\subsection{Variational formulation}
In order to express the variational format of the equations, it is convenient
to introduce the specific internal energy [J.kg$^{-1}$]:
\begin{equation}
 \energy(\temp) =
\begin{cases}
\specificHeatSolid \temp, & \text{\ if \ } T \leq T_0, \\
\specificHeatLiquid \temp +  \overline{l}, & 
\text{\ else. \ }
\end{cases}
\nonumber
\end{equation}
It is discontinuous at the transition temperature and the jump
is the latent heat $l$:
\begin{equation}
 \latentHeat = (\specificHeatLiquid - \specificHeatSolid)  \boundaryTempInterface + \overline{l}. 
 \nonumber
\end{equation}
Finally, it is also convenient to define the conductivity as a function of  the temperature. It is also discontinuous at the transition: 
\begin{equation}
    k(T) = \begin{cases}
    \conductivitySolid, & \text{\ if \ } T \leq T_0, \\
    \conductivityLiquid, & \text{\ else. \ }   
    \end{cases}
\nonumber
\end{equation}

We introduce the \textcolor{black}{Arbitrary} Lagrangian Eulerian frame of reference (see for instance \cite{Boffi2004,Huerta2004}).  
A mapping maps the points $\xnodes^0$ of a reference domain $\Omega^0$ to the points
$\xnodes$ of the domain $\Omega^t$:
\begin{equation}
\xnodes = \xnodes(\xnodes^0, t).
\end{equation}
The mapping is assumed not to 
change the shape of the domain.
The domain velocity  $\bm{w}$, domain deformation gradient $\bm{F}$ and Jacobian 
of the transformation are given by 
\begin{equation}
   \bm{w} = \left.\frac{\partial  \xnodes}{\partial t}
   \right|_{\xnodes^0}, \quad 
   \bm{F} = \left.\frac{\partial \xnodes}{\partial \xnodes^0}\right|_{t}, \quad J = \det \bm{F}.
   \label{eq:kinematicDef}
\end{equation}
The admissible temperature set is defined by 
\begin{equation}
 \tempSpace = \{ \temp: \Omega^0 \rightarrow \mathbb{R}: T \in H^1(\Omega^0), \;
  T  = \dirichletTemp ~~\text{on}~~ \dirichletBoundary
  \nonumber
 \}.
\end{equation}
At any instant $t \in [0,\tmax]$, the temperature field $T$ must belong to the admissible set $\tempSpace$ and satisfy the following variational principle:
 \begin{equation}
\begin{split}
 \frac{\dint}{\dint t} \int_{\Omega^0}  \volumicMass \energy(\temp) \test  J \dint
\xnodes^0  + \int_{\Omega^0} \conductivity(T)  
\bm{g}(\temp) \cdot \bm{g}(\test)   J \dint \xnodes^0
+ \\
 \int_{\Omega^0}
 \volumicMass\energy(\temp) \speed \cdot ~\bm{g}(\test) J \dint \xnodes^0 
= 
\int_{\Omega^0} \volumicMass \heatSource \test J \dint \xnodes^0~, \eqspace \forall \test \in \tempSpaceZero,
\label{eq:variationalFormulationGeneralTwo}
\end{split}
\end{equation}
where the test function space is defined by 
\begin{equation}
 \mathcal{U} = \{ \temp: \Omega^0 \rightarrow \mathbb{R}: T \in H^1(\Omega^0), \;
  T  = 0 ~~\text{on}~~ \dirichletBoundary
  \nonumber
 \}
\end{equation}
and
\begin{equation}
    \bm{g}(T) = \bm{F}^{-1} \nabla_0 T.
\end{equation}

For the sake clarity, we also give 
the above expression integrated on the current domain
 \begin{equation}
\begin{split}
 \frac{\dint}{\dint t} \int_{\Omega^t}  \volumicMass \energy(\temp) \test \dint
\xnodes  + \int_{\Omega^t} \conductivity(T) \myVectorGrad{\temp}
\cdot\myVectorGrad{\test} \dint \xnodes 
 +\\\int_{\Omega^t}
 \volumicMass\energy(\temp) \speed \cdot ~\myVectorGrad{\test} \dint \xnodes 
= 
\int_{\Omega^t} \volumicMass \heatSource \test \dint \xnodes.
\nonumber
\end{split}
\end{equation}

\answ{The variational principle \eqref{eq:variationalFormulationGeneralTwo}
gathers all the equations of the Stefan model.
Both the internal energy $e$ and conductivity functions are discontinuous at the transition temperature $T_0$. 
This temperature defines the location of the interface $\Gamma(t)$ between the solid and liquid phases.}
Finally, contrary to classical tracking approaches, 
we stress the fact that with the X-MESH  the domain velocity $\speed$ on the front may be different from the physical velocity $\bm{v}$, \answ{even in the normal direction.}


\subsection{Time discretization} \label{sec:spadis}
We consider a set of discrete instants. Let $t^n$ be such an instant  and $t^{n+1}$ the next one. We proceed with a $\theta$-scheme to discretize 
\eqref{eq:variationalFormulationGeneralTwo}.
Given the temperature field $T^n$  at  $t^n$, we seek the temperature field $T^{n+1}$  at  $t^{n+1}$ belonging to $\tempSpace$ and satisfying:
\begin{align}
\begin{split}
r(T^*)  \equiv & \int_{\Omega^{0}} \rho e(T^{n+1}) T^{*} J^{n+1} \dint \xnodes^0
   -\int_{\Omega^0} \rho e(T^{n}) T^{*} J^n \dint \xnodes^0  \\
     & +\theta \Delta t^{n+1/2} \int_{\Omega^{0}}
   k(T^{n+1}) \bfgnp(T^{n+1})\cdot \bfgnp(T^{*}) 
  J^{n+1} \dint \xnodes^0 \\
  & +\theta \Delta t^{n+1/2} \int_{\Omega^{0}}
  \rho e(T^{n+1})  \bm{w}^{n+1/2} \cdot  \bfgnp(T^{*})
  J^{n+1} \dint \xnodes^0 \\
   & - \theta\Delta t^{n+1/2}
    \int_{\Omega^0} 
  \rho S^{n+1} T^* J^{n+1} \dint \xnodes^0 \\
  & +(1-\theta)\Delta t^{n+1/2} \int_{\Omega^0} 
  k(T^n) \bfgn(T^n) \cdot \bfgn(T^{*}) 
  J^{n} \dint \xnodes^0   \\
    & +(1-\theta)\Delta t^{n+1/2} \int_{\Omega^0}
 \rho e(T^n)  \bm{w}^{n+1/2} \cdot \bfgn(T^{*})
  J^{n} \dint \xnodes^0  
  \\
 - &  (1-\theta)\Delta t^{n+1/2} \int_{\Omega^0} 
  \rho S^n T^* J^n \dint \xnodes^0 
    \\ = & 0, \quad  \forall T^* \in \tempSpaceZero, \label{eq:resid}
\end{split}
\end{align}
where $\Delta t^{n+1/2} = t^{n+1} - t^n$
and $\bm{w}^{n+1/2}$ is the domain velocity 
discretized as:
\begin{equation}
   \bm{w}^{n+1/2}(\xnodes^0) =
    \frac{\xnodes^{n+1}(\xnodes^0) - 
    \xnodes^n(\xnodes^0)}{t^{n+1} -  t^{n}}.
\nonumber
\end{equation}

\subsection{Space discretization}
The reference domain $\Omega^0$ is partitioned into a triangular mesh $\meshSet^0$. The mesh evolves 
in time. It is denoted $\meshSet^n$ at some later time $t^n$.
In the X-MESH approach, the mesh topology (number of nodes and adjacencies between mesh entities) is kept fixed. 
The mesh evolution is only due to node movements. The set of nodes is denoted  $\nodesSet$. 
The discrete temperature fields at $t^n$ and $t^{n+1}$ are defined as:
\begin{equation}
    T^n(\xnodes^0) = \sum_{i \in \nodesSet} T_i^n \; T_i^{*}(\xnodes^0), \quad T^{n+1}(\xnodes^0) = \sum_{i \in \nodesSet} T_i^{n+1} \; T_i^{*}(\xnodes^0),
\nonumber
\end{equation}
where the scalars $T_i^n$ and $T_i^{n+1}$ designate 
the nodal temperature values at node $i$ while the function $T_i^*$ is
the corresponding finite element approximation function (hat function).
When entered in \eqref{eq:resid}, the approximation function
associated to node $i$ gives a residual denoted $r_i$:
\begin{equation}
r_i  \equiv r(T^*_i).
\nonumber
\end{equation}
The above residual must be zero for all nodes with a free temperature.
These nodes are the ones not belonging to the Dirichlet boundary. The zero residual
condition links the mesh and temperature fields at times
$t_n$ and $t_{n+1}$. We express this condition with the $R$ symbol:
\begin{equation}
    R( (T^{n+1}, \xnodes^{n+1}), (T^n, \xnodes^n)) = 0.
\label{eq:Req}
\end{equation}

Given a mesh at time $t^n$ and a temperature field defined on this mesh, we say that they are compatible if 
the transition temperature value may only be found at nodes 
and not in between nodes.
This implies that the iso-temperature $T^{n}(\x) = T_0$ is exactly represented by the mesh: it explicitly
appears as mesh edges of the triangular mesh. 
We denote this  mesh/temperature compatibility with the $C$ symbol. 
Formally, the X-MESH problem to solve over a time step is 
\begin{align}
\begin{split}
    \text{Given\ } (T^n, \xnodes^n) \in C, \quad \text{find\ } 
     (T^{n+1}, \xnodes^{n+1}) \in C \\ \text{such\ that\ } \quad  R((T^{n+1}, \xnodes^{n+1}), (T^n, \xnodes^n)) = 0.
     \label{eq:step}
\end{split}
\end{align}
 The process is initialized with a compatible couple: $(T^0, \xnodes^0) \in C$. From one time instant to the next, the temperature field evolves and the mesh deforms so that the iso-transition temperature is always part of the mesh. 
It is important to note that this does not force the same nodes to be part of the front from one time step to the next. 
During a time step, a node may stay on the front, leave the front or join the front. 
This is why we stressed the fact in the description of the variational formulation that on the front, the domain (mesh) velocity $\speed$ could be different  from the physical front velocity $\bm{v}$.

Problem \eqref{eq:step} does not have a unique solution (nor shall we try to prove here that the solution exists). The non-uniqueness means that we have some freedom in the mesh evolution. We shall use this freedom to have the mesh return to its initial position progressively after the front has gone away (mesh relaxation).

\subsection{A quasi-Newton scheme}
 \label{semiMonolithicScheme}
 
We detail an iterative process to solve \eqref{eq:step}.
We consider first the issue of finding the temperature 
field $T^{n+1}$ using \eqref{eq:Req} assuming 
given $T^{n}$, $\xnodes^n$ and $\xnodes^{n+1}$.
A direct use of a Newton-Raphson resolution would require to compute the derivative of $e$ and $k$ with respect to the temperature. But those are discontinuous across the interface. Indeed,  a Heaviside function $H$ centered at $T_0$ is involved:
\begin{align}
    e(T) & = ((1-H(T)) \specificHeatSolid + H(T) \specificHeatLiquid) \; T + H(T) \overline{l}, \nonumber \\
    k(T) & = (1-H(T)) \conductivitySolid + H(T) \conductivityLiquid,
    \nonumber
\end{align}
where
\begin{equation}
H(T) = 
\begin{cases}
0, & \text{\ if\ } T \leq T_0,  \\
1, & \text{\ if\ } T > T_0. \\
\end{cases} 
\nonumber
\end{equation}
Even if it would be theoretically possible to consider calculating
the derivative of the Heaviside as Dirac distributions on the interface, using such an
approach poses a problem when an interface  nucleates. Indeed, such
a scheme would not allow to anticipate the creation of a front.

\begin{figure}
\centering
   \includegraphics[width=.5\textwidth]{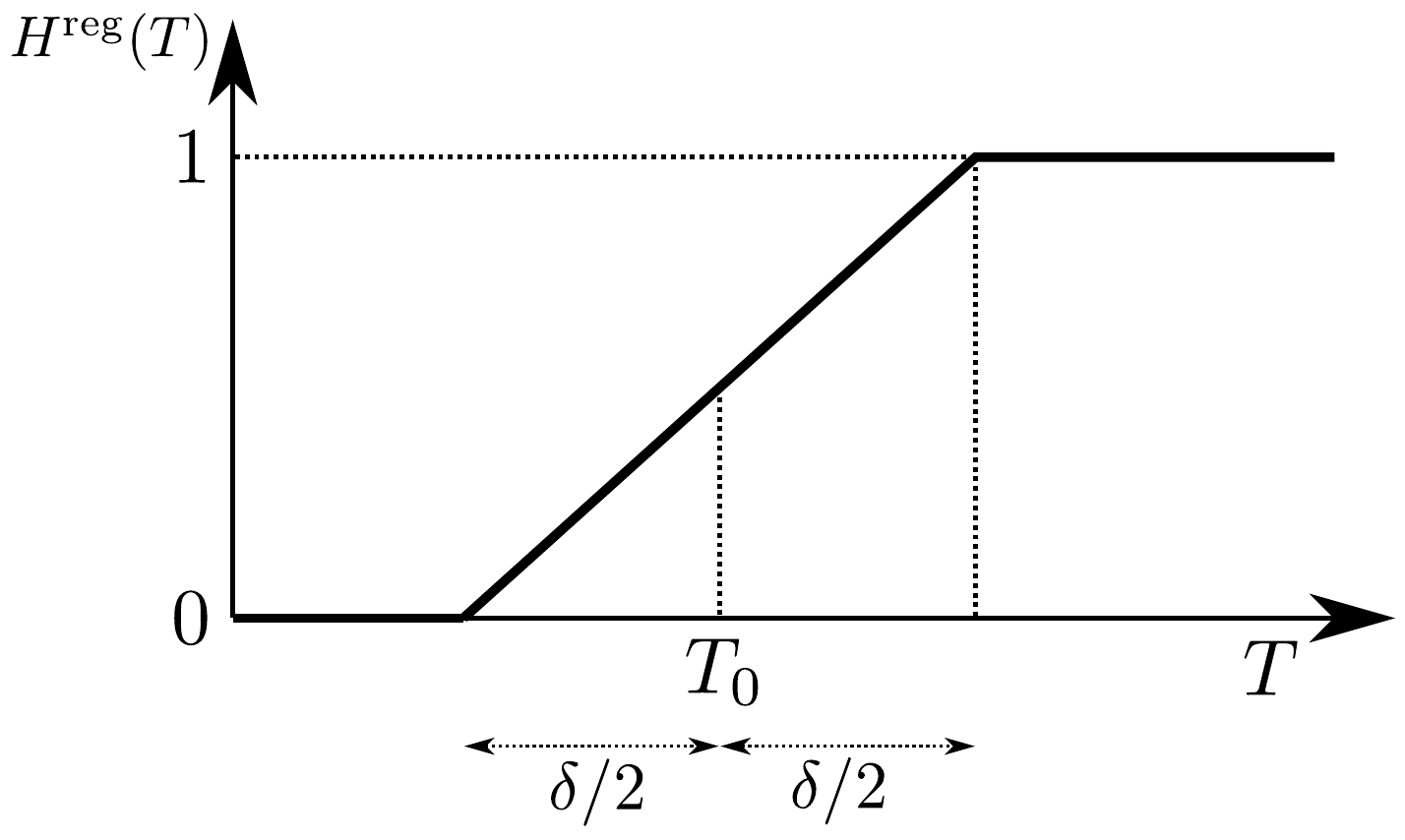}
 \captionof{figure}{Smooth step function $\alpha(T)$ illustration.}
  \label{fig:smoothStepFunction}
\end{figure}

A more robust approach is to smooth the discontinuous quantities
across the interface: such a smoothed approach is widely used in the
literature but its main drawback is accuracy. If
the interface is smoothed over a thickness $\varepsilon$, the convergence to the
exact solution of the Stefan problem cannot be better than
$\sqrt{\varepsilon}$ \cite{azaiez2016two}. Here, we propose a quasi-Newton type scheme
where the tangent matrix is an approximation of the derivative of the
residual while the true sharp residual is kept. The tangent matrix is computed 
as the derivative of a smoothed residual.
For this purpose, we define a smooth
step function $H^{\text{reg}}$ illustrated on Figure \ref{fig:smoothStepFunction}:
\begin{equation}
H^{\text{reg}}(T) = 
\begin{cases}
0, \quad \quad \quad \quad \quad \quad & \text{\ if\ } T < T_0 - \delta/2, \\
 \displaystyle \frac{ T - (T_0 - \delta/2)}{\delta}, \quad \,
& \text{\ if\ }  T_0 - \delta/2 \leq T \leq 
T_0 + \delta/2, \\
1, \quad \quad \quad \quad \quad \quad  & \text{\ if\ } T > T_0 + \delta/2, \\
\end{cases} 
\nonumber
\end{equation}
where $\delta$ [K] is a numerical parameter.  
\answ{We checked that parameter $\delta$ has no influence on the converged results
but has some slight influence on the convergence rate of the non linear
process.}
From this function, we can define continuous energy and conductivity functions:
\begin{align}
    e^{\text{reg}}(T) & = ((1-H^{\text{reg}}(T)) \specificHeatSolid + H^{\text{reg}}(T) \specificHeatLiquid) \; T + H^{\text{reg}}(T) \overline{l}, \nonumber \\
    k^{\text{reg}}(T) & = (1-H^{\text{reg}}(T)) \conductivitySolid + H^{\text{reg}}(T)\conductivityLiquid. \nonumber
\end{align}

 We denote by $r^{\text{reg}}_i$ the residual expression in which the sharp functions $e(T)$ and $k(T)$ are replaced by the regularized ones. Quasi-Newton iterations are denoted by the $k$ index. Note that other smoothing functions could be used, see for instance \cite{Vasilev2020}. 

The staggered solution procedure for a step is detailed in the Algorithm \ref{alg:scheme}. 
After initialization, the temperature 
update (quasi-Newton step) alternates with the mesh update (projection step) until convergence. To be precise the projection step also 
modifies the temperature field but only for the nodes located on the front after the projection.

The initialization step copies the previously converged temperature 
field and,  for nodes not located on the front,
relaxes their location with the following formula:
\begin{equation}
    \xnodes^{n+1,0} = 0.9 \xnodes^{n} + 0.1 \xnodes^{0}.
\label{eq:relax}
\end{equation}
The initial couple $(T^{n+1,0}, \xnodes^{n+1,0})$  belongs to $C$.
As explained earlier, the Newton step uses a smoothed gradient operator but with the sharp residual.  
The projection step on $C$ will be detailed in the next section. 
The idea is to move nodes close to the iso-$T_0$ on the iso-$T_0$ and then 
set the temperature at these nodes to $T_0$. 
\begin{algorithm}

\caption{The X-MESH solution procedure for a time step.
The Einstein summation convention rule holds 
with indices $i$ and $j$ spanning all nodes except the ones on the Dirichlet boundary.
The symbol $\epsilon$  [K] denotes the user given 
tolerance  for convergence. The field $\Delta$ is a linear interpolation 
of the $\Delta_i$ nodal values  
(except for the Dirichlet nodes on which a zero value is set).}
\label{alg:scheme}
 Step Initialization: \\

   { \nonl \begin{center}
           $ k = 0 \text{ and } (T^{n+1,0}, \xnodes^{n+1,0} )  =  \text{relax\_in\_C}(T^n, \xnodes^{n}) $
        \end{center} }
\Do{$\err$ > $\tol$}
{
 Quasi-Newton step: \label{alg:newtonstep} \\
   { \nonl
   \answ{
  \begin{align*}  
  A_{ij} & = \frac{\mathrm{d} r^{\text{reg}}_i}{\mathrm{d}T^{n+1}_j} \Bigr|_{(T^{n+1,k},\xnodes^{n+1,k})}
  \\
  A_{ij} \Delta_j & = r_j \mid_{(T^{n+1,k},\xnodes^{n+1,k})}
  \\
  T^{'}_i & = T^{n+1,k}_i + \Delta_i
  \end{align*} 
  }
  }\\
  Projection step: \\
   { \nonl \begin{equation*}
   (T^{n+1,k+1}, \xnodes^{n+1,k+1})  =  \text{project\_on\_C}(\xnodes^{n+1,k}, T^{'})  
\end{equation*} } \\ 
 Stopping criteria:  \\
    { \nonl  \begin{equation*}
  \mathrm{err} = \left( 
  \frac{\int_{\Omega^0} \Delta^2 J^{n+1,k+1} \dint  \xnodes^0}{\int_{\Omega^0}   \dint \xnodes^0}
    \right)^{1/2} 
  \end{equation*} } \\
  
\If{ $\err$ < $\tol$}
 {
 $(T^{n+1}, \xnodes^{n+1}) = (T^{n+1,k+1}, \xnodes^{n+1,k+1})$
 }
 \Else
 {
 $(T^{n+1,k}, \xnodes^{n+1,k}) \leftarrow (T^{n+1,k+1}, \xnodes^{n+1,k+1})$ \\
 $k \leftarrow k+1 $
 }
 }

\end{algorithm}

One could think of a simpler scheme than the one described in Algorithm \ref{alg:scheme} that we shall call fixed point.
In this scheme, 
the temperature at the centroid of each element obtained from $(T^{n+1,k},\xnodes^{n+1,k})$, and denoted $T_{\mathrm{elt}}$,
is used to decide the phase for the whole element. The expression 
of the internal energy and conductivity are then for a given element
\begin{align}
    e^{\text{fix}}(T) & = ((1-H(T_\mathrm{elt})) \specificHeatSolid + H(T_{\mathrm{elt}}) \specificHeatLiquid) \; T + H(T_{\mathrm{elt}}) \overline{l}, \nonumber \\
    k^{\text{fix}} & = (1-H(T_{\mathrm{elt}})) \conductivitySolid + H(T_{\mathrm{elt}}) \conductivityLiquid. \nonumber
\end{align}
With these expressions, the residual is now a linear expression of $T$ and setting it to zero is equivalent to solving a linear system.
With the fixed point algorithm, the step iterates between a linear system solve and a projection on $C$. 
The strength of the fixed point algorithm is its simplicity. Unfortunately, except for zero latent heat, this algorithm is not robust.

\answ{\input{relaying}}

\section{Numerical examples}
\label{numericalExamples}

This section begins with a dimensional study of
the equations to define useful non-dimensional parameters.

\subsection{Non-dimensional form}
We will be considering examples of solidification of a liquid domain initially at a temperature $T_\ell$.  
Non-dimensional  temperature, coordinates and time are denoted by tildas:
\begin{equation}
    \tilde{T} = \frac{T - T_0}{T_\ell - T_0}, \quad \tilde{x} = {x \over
      L_r}, \quad \tilde{t} = {t \over t_r},
\nonumber
\end{equation}
where $L_r$ and $t_r$ are reference length and time, respectively.
We define also the following non-dimensional quantities 
\begin{equation}
  \label{eq:adimparam}
    \kappa = {k_s \over k_\ell}, \quad \alpha = {\alpha_s \over \alpha_\ell}, \quad \gamma = 
    \frac{l}{c_\ell (T_\ell-T_0)}, \quad
    \tilde{\bm{v}} = \frac{t_r}{L_r} \bm{v}, \quad 
    \tilde{S} = \frac{\rho S L_r^2}{k_\ell (T_\ell - T_0)},
\end{equation}
where we have used the thermal diffusivity of the solid and liquid phase  [m$^2$/s]:
\begin{equation}
       \diffusivitySolid = \frac{k_s}{\rho c_s},
    \quad \diffusivityLiquid = \frac{k_\ell}{\rho c_\ell}.
    \nonumber
\end{equation}
Note that $\kappa <= 1$  and $\alpha <= 1$. The coefficient $\gamma$, usually called the Stefan number,  is the ratio between the latent heat and sensible heat.
Choosing the following relation between the reference time and length, 
\begin{equation}
  \nonumber
    t_r = {L_r^2 \over \diffusivityLiquid},
\end{equation}
the non-dimensional equations are 
\begin{align}
   \frac{\tilde{T}_{,\tilde{t}}}{\alpha}   & =  \tilde{\Delta} \tilde{T} + \frac{\tilde{S}}{\kappa}   \quad \text{(solid phase)}, \nonumber\\
    \tilde{T}_{,\tilde{t}} & = \tilde{\Delta}
    \tilde{T} + \tilde{S} \quad \text{(liquid phase)} \nonumber
\end{align}
and the interface condition is 
\begin{equation}
    \gamma \tilde{\bm{v}} \cdot \normalVect = \left( \kappa \tilde{\bm{\nabla}} \tilde{T} \mid_{s} - \tilde{\bm{\nabla}}\tilde{T}  \mid_{\ell} \right)  \cdot \normalVect. \nonumber
\end{equation}

The X-MESH approach will now be tested on several examples. 
Unless otherwise noted, 
the values of the physical properties
are taken from 
table \ref{table:physicalProperties} and the 
numerical parameters from table 
\ref{table:numProperties}.

\begin{table}
\centering
\begin{tabular}{llcccc}
  \hline
  Property  & Units     &  Symbol    &  Solid  & Liquid  &  Interface \\ \hline
  Density & kg.m$^{-3}$ &  $\volumicMass$    &  1000   & 1000    &  \\
  Specific heat capacity & J.K$^{-1}$.kg$^{-1}$ & $\specificHeatSolid \mid \specificHeatLiquid$ & 2090 & 4185 &    \\
  Conductivity  & W.m$^{-1}$.K$^{-1}$  & $\conductivitySolid \mid \conductivityLiquid$ & 2.1  & 0.6  &    \\
  Latent heat   & J.kg$^{-1}$         & $\latentHeat$ &   &    &  $3.3 \times 10^{5}$  \\
  Phase change temperature & K     & $\boundaryTempInterface$ & &  & 273.15 \\  
  \hline
 \end{tabular}
 \captionof{table}{Physical properties for test cases.}
 \label{table:physicalProperties}
\end{table}

\begin{table}
\centering
\begin{tabular}{llcc}
  \hline
  Parameter  & Units     &  Symbol    &  Value  \\ \hline
  $\theta$-scheme & - & $\theta$ & 0.5 \\
  Regularization temperature & K & $\delta$  & 8 \\
  Minimum allowed area & m$^{2}$  &   $A_\mathrm{tol}$ & $5\times10^{-9}$ \\
  Convergence tolerance & K & $\tol$ &  $10^{-5}$ \\
  \hline
   \end{tabular} 
\captionof{table}{Numerical parameters for the test cases.}
\label{table:numProperties}
\end{table}

\subsection{The straight front semi-infinite two-phase problem}
\label{straightFrontStaggeredNoLatentHeat}

\begin{figure}[t!]
\begin{center}
  \includegraphics[width=0.45\textwidth]{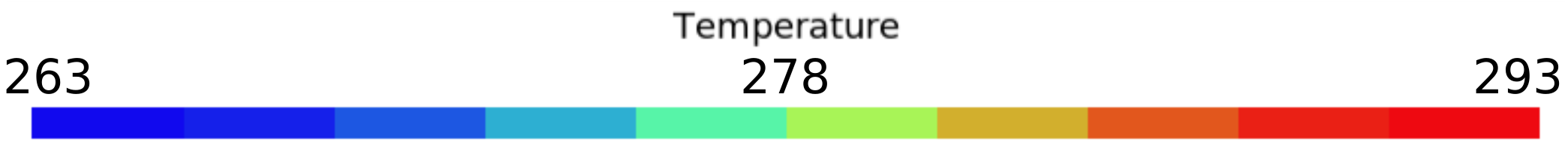}
  \begin{tabular}{ccc}
    \includegraphics[width=0.3\textwidth]{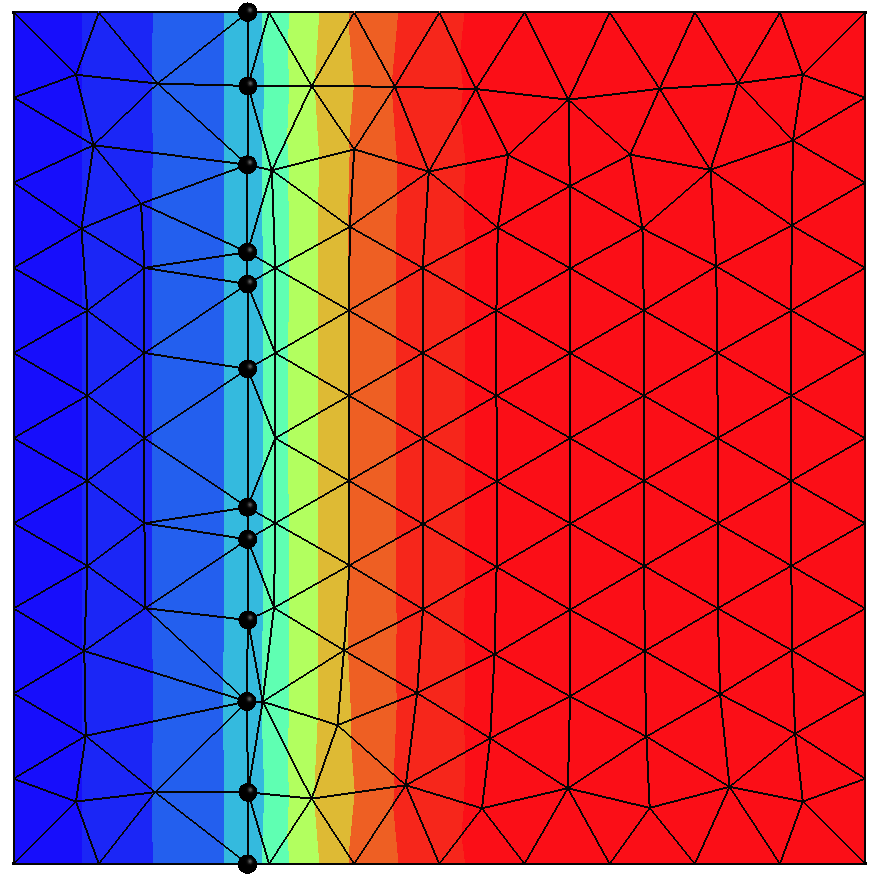}&
    \includegraphics[width=0.3\textwidth]{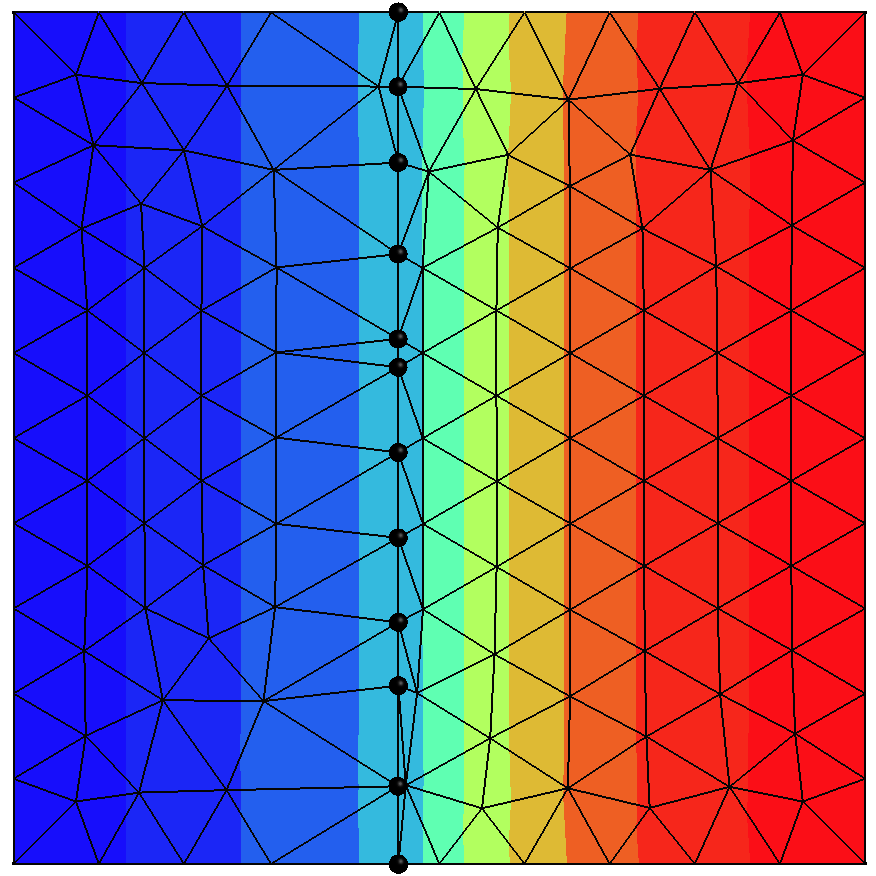}&
    \includegraphics[width=0.3\textwidth]{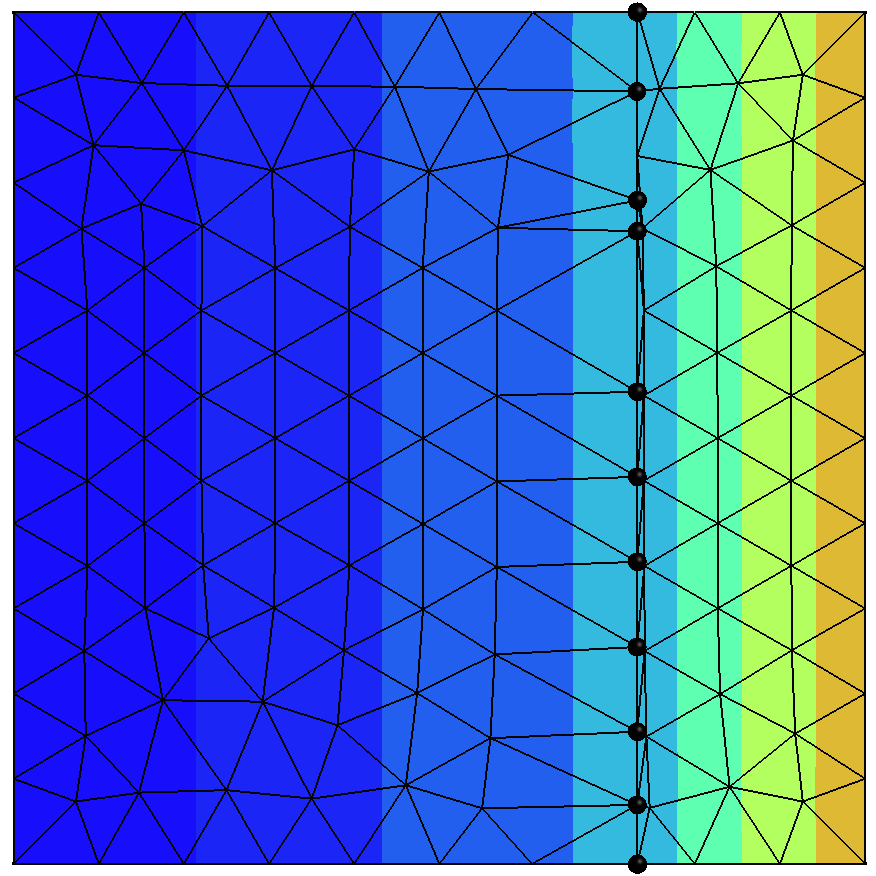}\\
    (A) & (B) & (C)
  \end{tabular}  
  \begin{tabular}{cc}
 \includegraphics[width=0.45\textwidth]{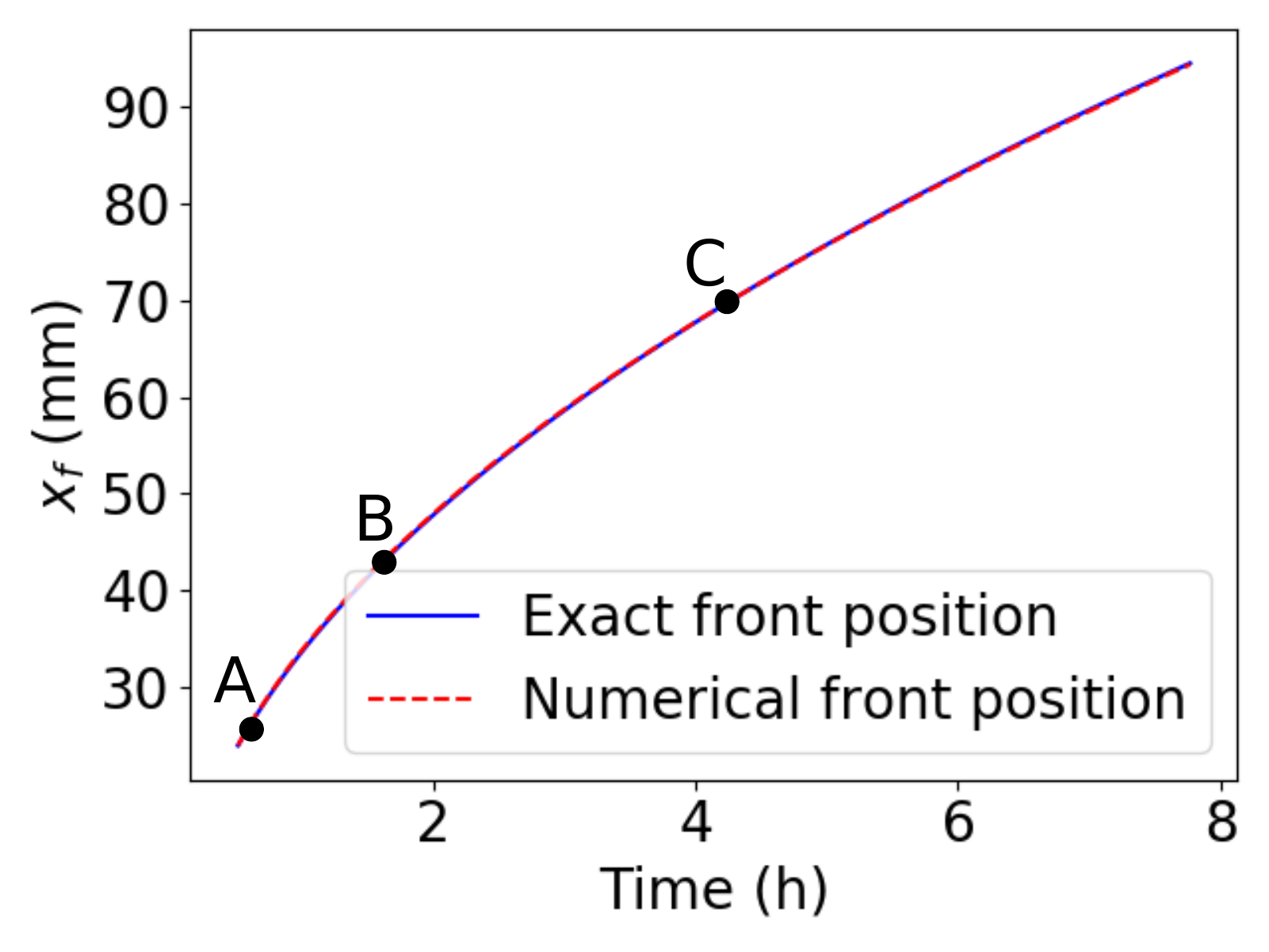}&
 \includegraphics[width=0.45\textwidth]{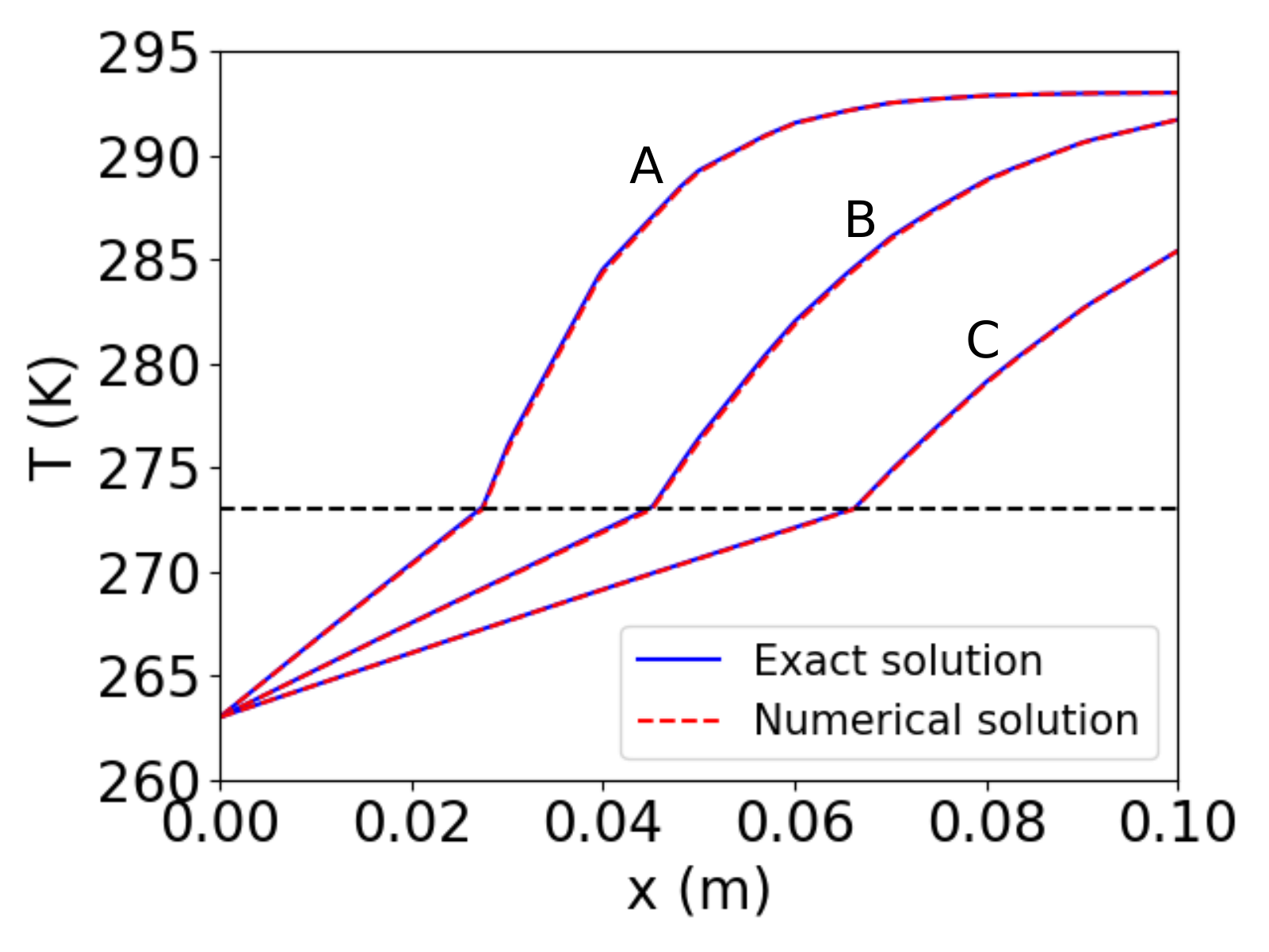} \\
  (a) & (b)    
  \end{tabular}  
\end{center}
\caption{ \answ{Straight front semi-infinite problem}: numerical results for $\latentHeat = 0$ [J.kg$^{-1}$] and $\elemSize = 0.01$ [m].
Temperature fields and fronts (\answ{front vertices are represented by big, black dots}) at $t = 0.66$ [h] (A), $1.78$ [h] (B) and 
$4.68$ [h] (C).
Comparison of exact and numerical front positions (a) and temperature profiles along the horizontal direction (obtained by averaging the temperature vertically) (b). }
\label{fig:stephan1DResultsCoarseNoLatentHeat}
\end{figure}

The straight front two-phase Stefan problem can be expressed as a heat
conduction problem in a semi-infinite domain represented by $\Omega
=[0,+\infty[$. There is no volumetric heat source. The domain is
initially liquid at temperature $\boundaryTempLiquid$ = 293 [K] and the
temperature is abruptly lowered at the temperature
$\boundaryTempSolid$ = 263 [K] at the left side of the domain. A
solidification front located at $\frontPositionEx(t)$ 
moves continuously in the right direction. It is given (see
\cite{Carslaw1959}) by a classical square root expression
\begin{equation}
  \frontPositionEx(t) = 2 \solexParam \sqrt{\diffusivitySolid t}.
\label{eq:xexa}
\end{equation}
The dimensionless coefficient $\solexParam$ is a  obtained by
solving the following transcendental equation:
\begin{equation}
 \displaystyle \frac{\specificHeatSolid (\boundaryTempInterface - \boundaryTempSolid)}{\latentHeat}  \frac{e^{-\solexParam^2}}{\erf({\solexParam})} - \frac{1}{\gamma \sqrt{\alpha}}\frac{e^{- \alpha \solexParam^2} }{ (1-\erf(\solexParam \sqrt{\alpha})) } - \solexParam \sqrt{\pi}  = 0,
\nonumber
\end{equation}
where $\erf(x) = {\frac {2}{\sqrt {\pi }}} \displaystyle \int
_{0}^{x}e^{-t^{2}} \dint t$ is the error function. 
The temperature field in the solid phase is given by
\begin{equation}
 \temp(x,t) = \boundaryTempSolid + \frac{(\boundaryTempInterface - \boundaryTempSolid  )}{\erf(\solexParam)} \erf\left( \frac{x}{2 \sqrt{\diffusivitySolid t} }  \right), \text{ for } x \leq \frontPositionEx
\nonumber
\end{equation}
and in the liquid phase by
\begin{equation}
 \temp(x,t) = \boundaryTempLiquid - \frac{( \boundaryTempLiquid - \boundaryTempInterface  )}{(1-\erf(\solexParam\sqrt{\alpha)}  )} \left(1 - \erf\left( \frac{x}{2 \sqrt{\diffusivityLiquid t} }  \right) \right), \text{ for } x >
 \frontPositionEx.
 \nonumber
\end{equation}

\begin{figure}[t!]
\begin{center}
\includegraphics[width=0.45\textwidth]{fig/stephan1DTemperatureNoLatentHeatTemperatureScale.pdf}
  \begin{tabular}{ccc}
    \includegraphics[width=0.3\textwidth]{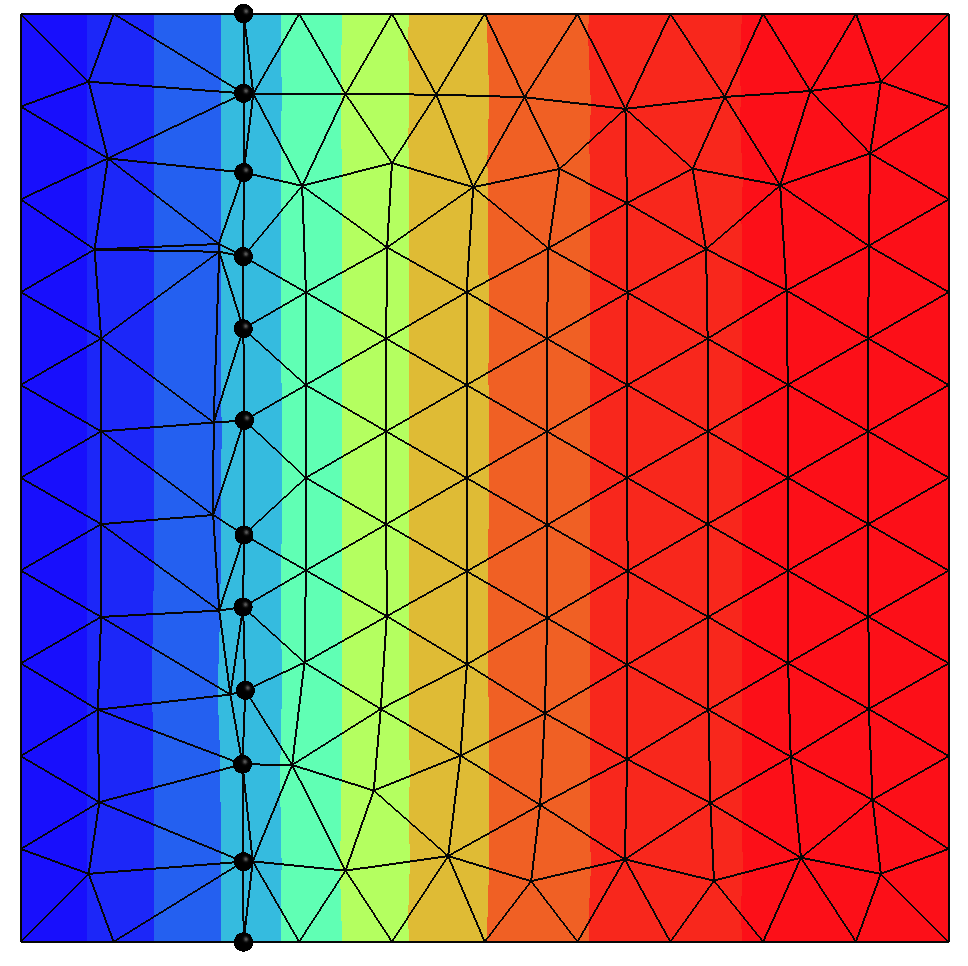}&
    \includegraphics[width=0.3\textwidth]{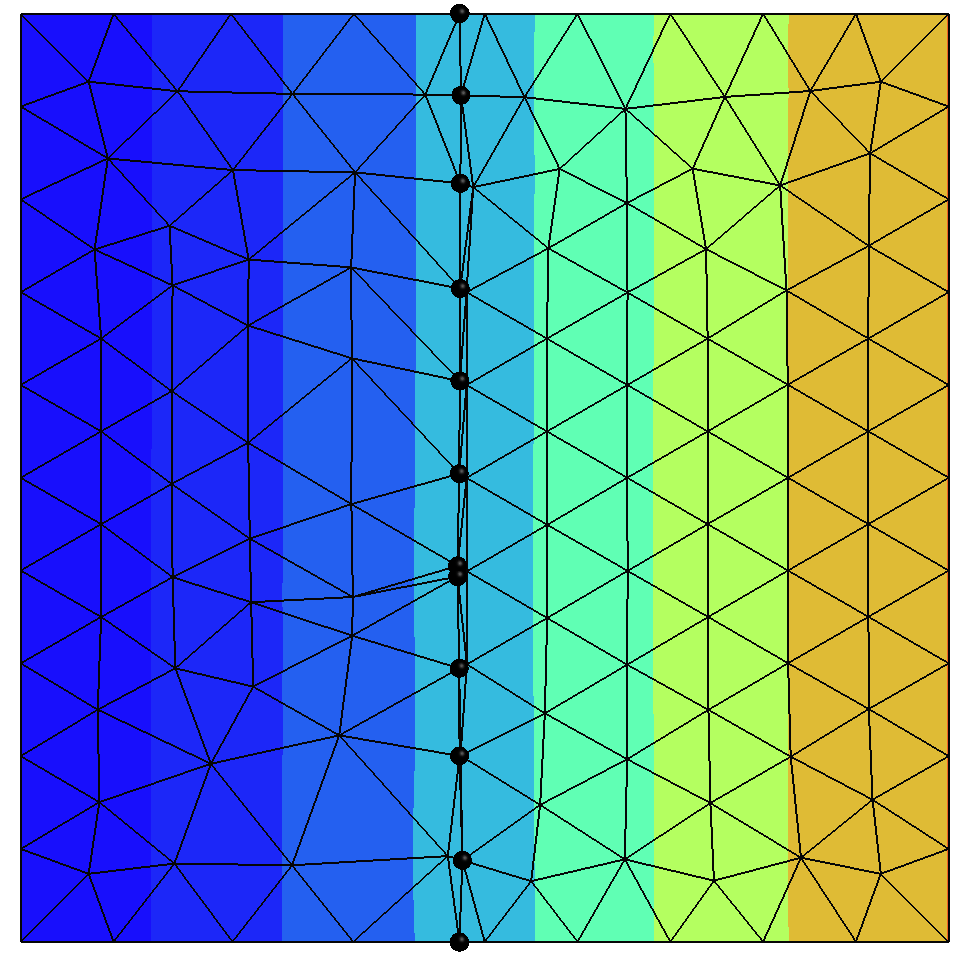}&
    \includegraphics[width=0.3\textwidth]{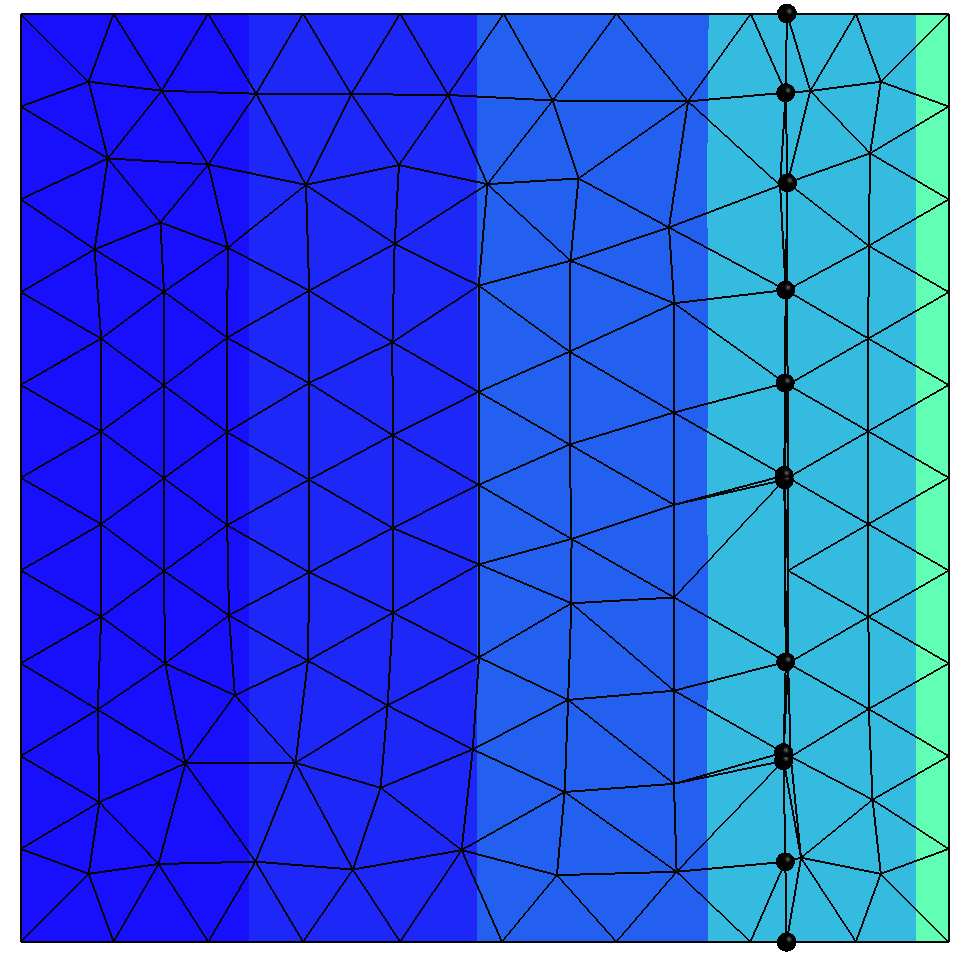}\\
    (A) & (B) & (C)
  \end{tabular}  
  \begin{tabular}{cc}
 \includegraphics[width=0.45\textwidth]{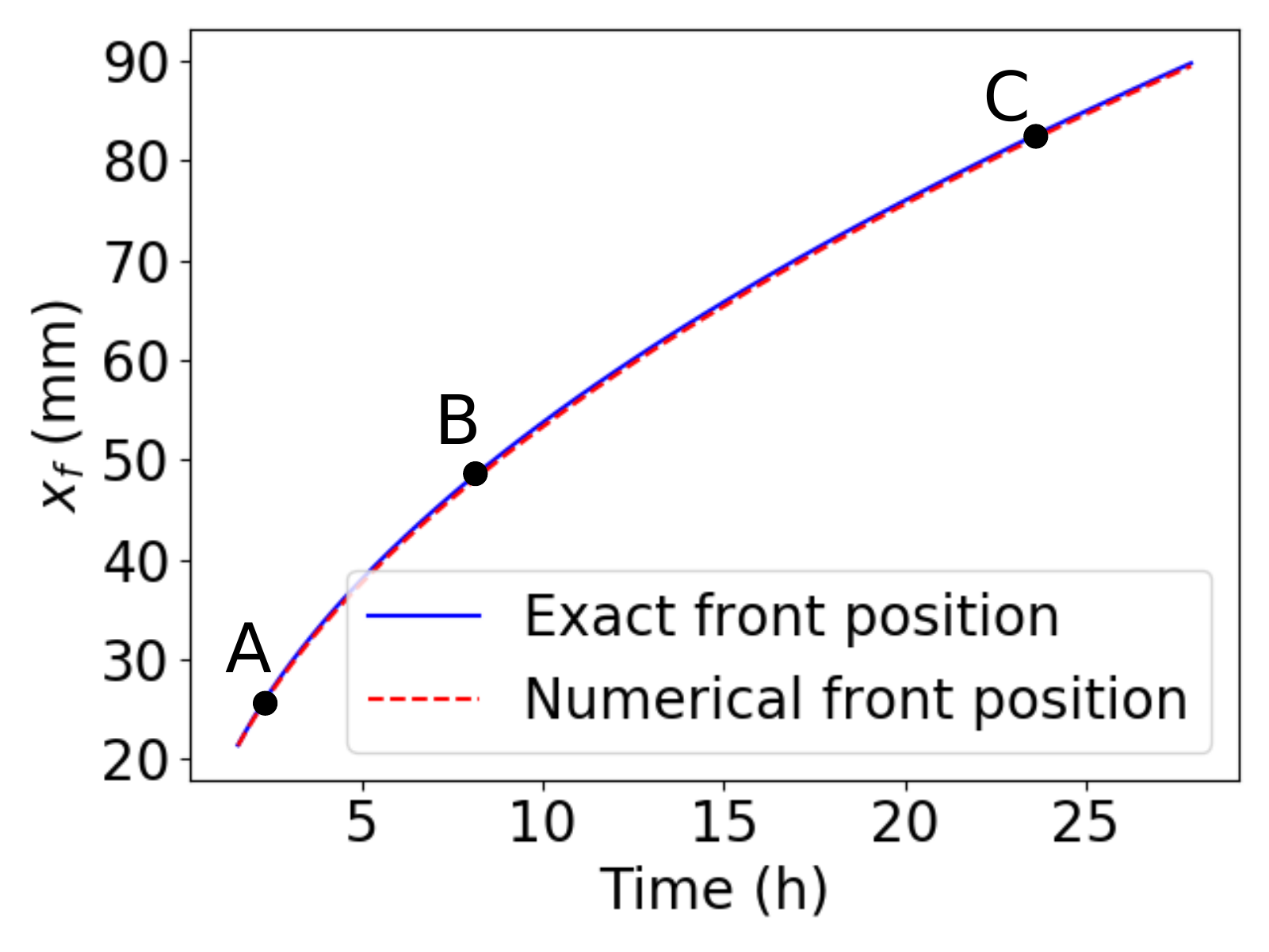}&
 \includegraphics[width=0.45\textwidth]{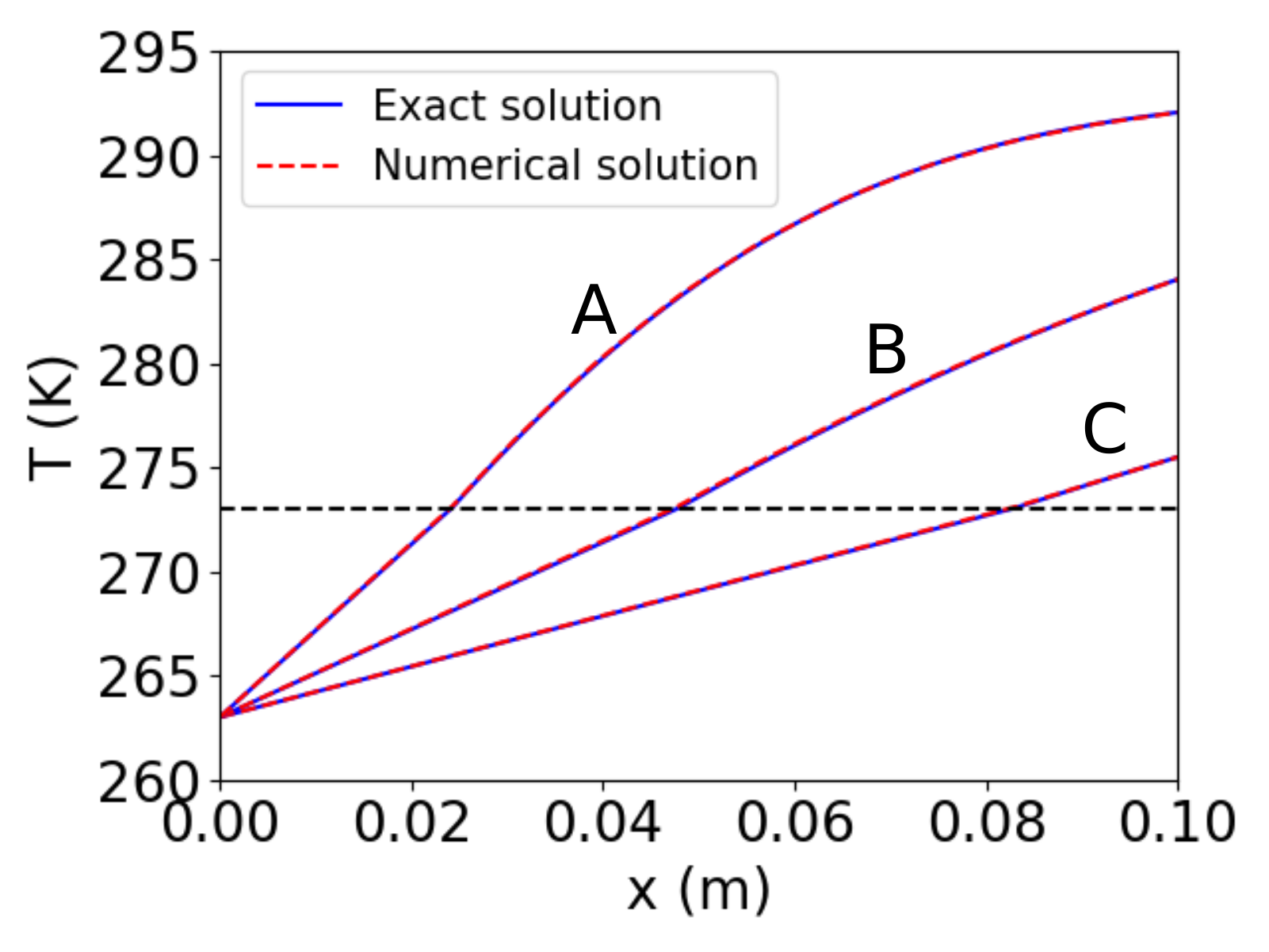} \\
  (a) & (b)    
  \end{tabular}  
\end{center}
\caption{ \answ{Straight front semi-infinite problem}: numerical results for $\latentHeat = 3.3 \times 10^{5}$ [J.kg$^{-1}$] and $\elemSize = 0.01$ [m]. 
Temperature fields and fronts (\answ{front vertices are represented by big, black dots}) at $t = 3.01$ [h] (A), $7.89$ [h] (B) and $23.71$ [h] (C).
Comparison of exact and numerical front position (a).  Analytical and numerical temperature solutions along the horizontal direction (b). }
\label{fig:stephan1DResultsCoarse}
\end{figure}

Although the geometry of the problem is 1D and semi-infinite, we use a
square domain of dimensions 0.1 $\times$ 0.1 [m$^2$]\answ{. To simulate the infinite boundary
condition, the temperature is imposed on the left (constant equal to $\boundaryTempSolid$) and on the right boundary (varying with time according to the above formula evaluated for $x = 0.1$ [m]).} Top and bottom sides are insulated.
The domain is meshed with linear triangular elements with a characteristic size of $\elemSize = 0.01$ [m].
Computations are performed with 
$\latentHeat = 0$ [J.kg$^{-1}$] (Figure
\ref{fig:stephan1DResultsCoarseNoLatentHeat})
and with $\latentHeat = 3.3 \times 10^{5}$ [J.kg$^{-1}$] (Figure
\ref{fig:stephan1DResultsCoarse}). 
We detail now how the time step for these simulations has been chosen.
From the analytical solution \eqref{eq:xexa}, we can compute the front velocity:
\begin{equation}
{d x_f \over dt} = \phi {\sqrt{\diffusivitySolid \over t}}.
\label{eq:velox}
\end{equation}
It is  proportional 
to the $\phi$ coefficient.
With the ice-water parameters of Table \ref{table:physicalProperties}, 
$\solexParam = 0.282$ for $\latentHeat = 0$ [J.kg$^{-1}$] 
and $\solexParam = 0.141$ for $\latentHeat = 3.3 \times 10^{5}$ [J.kg$^{-1}$].
The latent heat tends to slow down the freezing front compared to the no-latent heat case but without changing the order of magnitude of the velocity. We observe in \eqref{eq:velox} that the front velocity is infinite at $t=0$ [s].
To avoid singularity issues in the numerical simulation we start from a 
time $t^0>0$ and use the analytical solution at that time as initial condition. We also 
make sure that the mesh represents exactly the initial front (by applying the project\_on\_C operator).
Relation \eqref{eq:velox} may then be used to estimate the numerical front advance $\Delta x$ over the time step  $\Delta t$:
\begin{equation}
    \Delta x = \phi \sqrt{\diffusivitySolid \over t} \Delta t.
    \nonumber
\end{equation}
A variable time step can be chosen to ensure a constant front advance:
\begin{equation}
    \Delta t^{n+1/2} = \sqrt{ \beta \; t^{n}} \quad  \Rightarrow \quad \Delta x \simeq  \phi \sqrt{\beta \diffusivitySolid},
    \label{eq:dt}
\end{equation}
where $\beta$ [s] is a parameter.

\begin{figure}
\centering
\begin{tabular}{cc}
 \includegraphics[height=4cm]{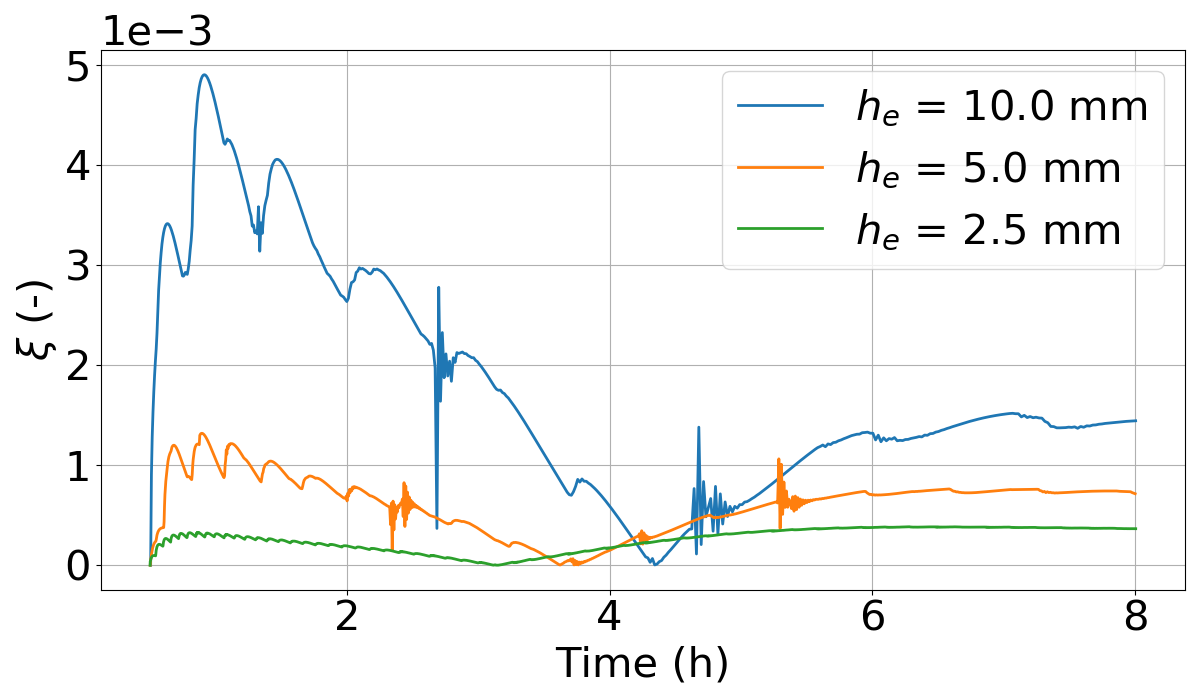}
& 
\includegraphics[height=4cm]{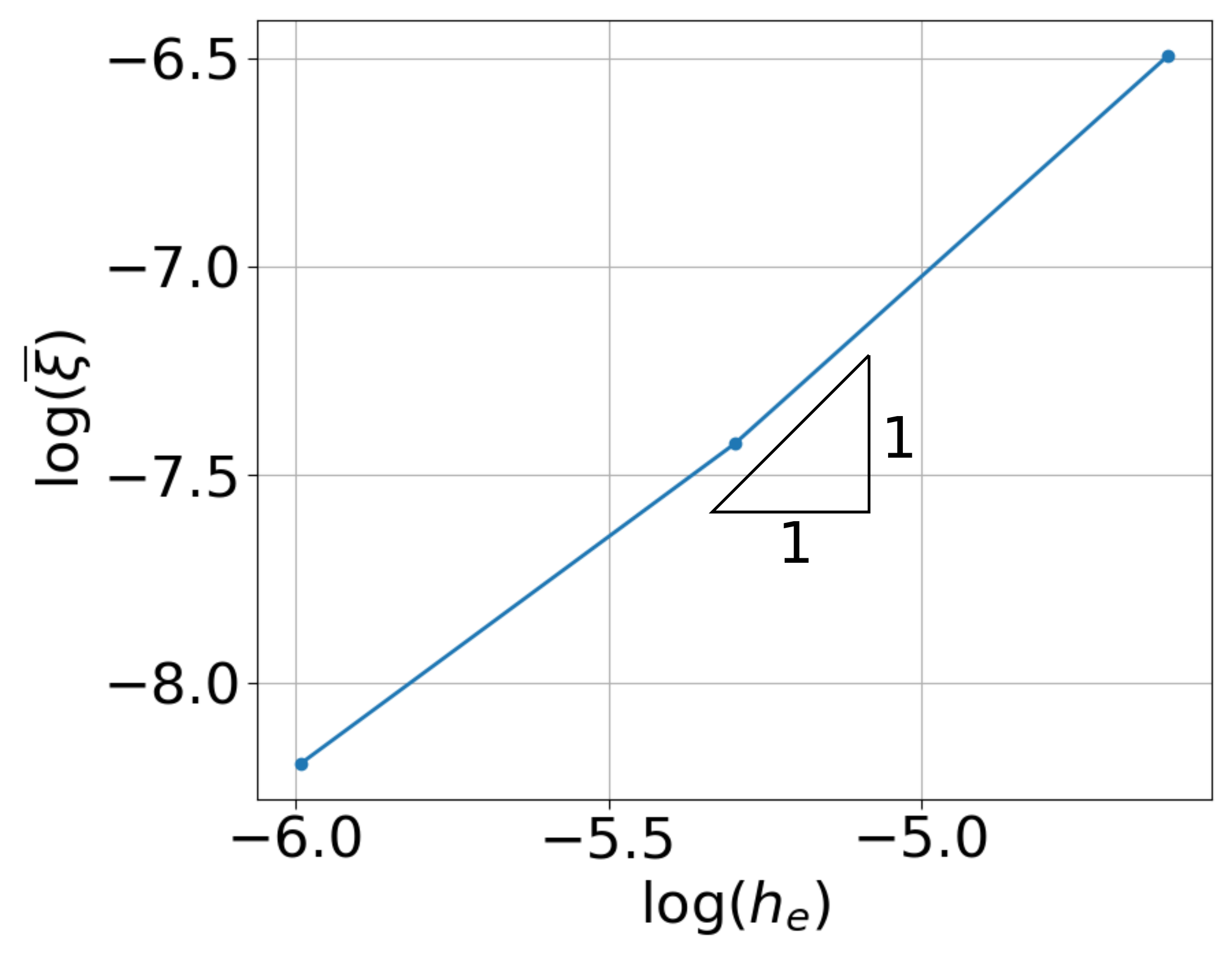}
 \\
(a) & (b)
\end{tabular}
\caption{ \answ{Straight front semi-infinite problem: error comparisons for $\latentHeat = 0$ [J.kg$^{-1}$]. Relative front location error evolution with respect to time for different mesh refinements (a). Integrated error convergence with respect to the mesh element size (b).}}
\label{fig:stephan1DErrorComparisonNoLatentHeat}
\end{figure}

In the case $\latentHeat = 0$ [J.kg$^{-1}$],
Figure \ref{fig:stephan1DResultsCoarseNoLatentHeat}, 
we choose $t^0 = 500$ [s] and $\beta = 100$ [s] which corresponds
to a \emph{small time step} i.e. a time step that only allows the front
to advance of a length $\Delta x \simeq \elemSize/3 = 0.01/3$ [m]. 
In the simulations, we observe that 
the front
relaying algorithm acts as we had imagined it: elements of very
small sizes are created and nodes of the front pass the relay (or the
baton) to other nodes as they move along the edges downstream of the front.
One can observe that front nodes are
quasi-aligned on a vertical straight line.  The \answ{numerical} position $x_f^{\text{num}}$ of the front (Figure
\ref{fig:stephan1DResultsCoarseNoLatentHeat}, bottom left)
is computed by taking the average $x$ coordinates of the
phase change front nodes. The temperature values plotted along the $x$ direction (Figure
\ref{fig:stephan1DResultsCoarseNoLatentHeat}, bottom right) are
obtained by interpolating the temperature fields on a 500
($x$-direction) $\times$ 20 ($y$-direction) uniform grid, and
averaging along the $y$ direction. Figure
\ref{fig:stephan1DErrorComparisonNoLatentHeat} (a) shows for different
values of $\elemSize$, the \answ{relative} error on the front
position with respect to time \answ{, noted $\errXf(t)$ [no unit], and  computed as}
\begin{equation}
 \errXf(t) = \displaystyle \frac{| x_f^{\text{num}}(t) - x_f(t) |}{x_f(t)},
 \nonumber
\end{equation}
\answ{while Figure \ref{fig:stephan1DErrorComparisonNoLatentHeat} (b) shows a relative error integrated over the total duration of the simulation: }
\begin{equation}
 \overline{\errXf} = \displaystyle \frac{ \displaystyle \int_0^{t_{\text{max}}} | x_f^{\text{num}}(t) - x_f(t) | \dint t }{ \displaystyle \int_0^{t_{\text{max}}} x_f(t) \dint t },
 \nonumber
\end{equation}
\answ{where $t_{\text{max}}$ is the total duration of the simulation.}
Linear mesh convergence is observed,
which makes sense because the front velocity depends on the jump of
temperature gradients which converge at best linearly.
\answ{In their paper \cite{Ji2002}, Ji, Chopp and Dolbow find the same
  convergence rate using XFEM. Yet, the error values are significantly
lower with X-MESH.}

\begin{figure}
\centering
\begin{tabular}{cc}
 \includegraphics[height=4cm]{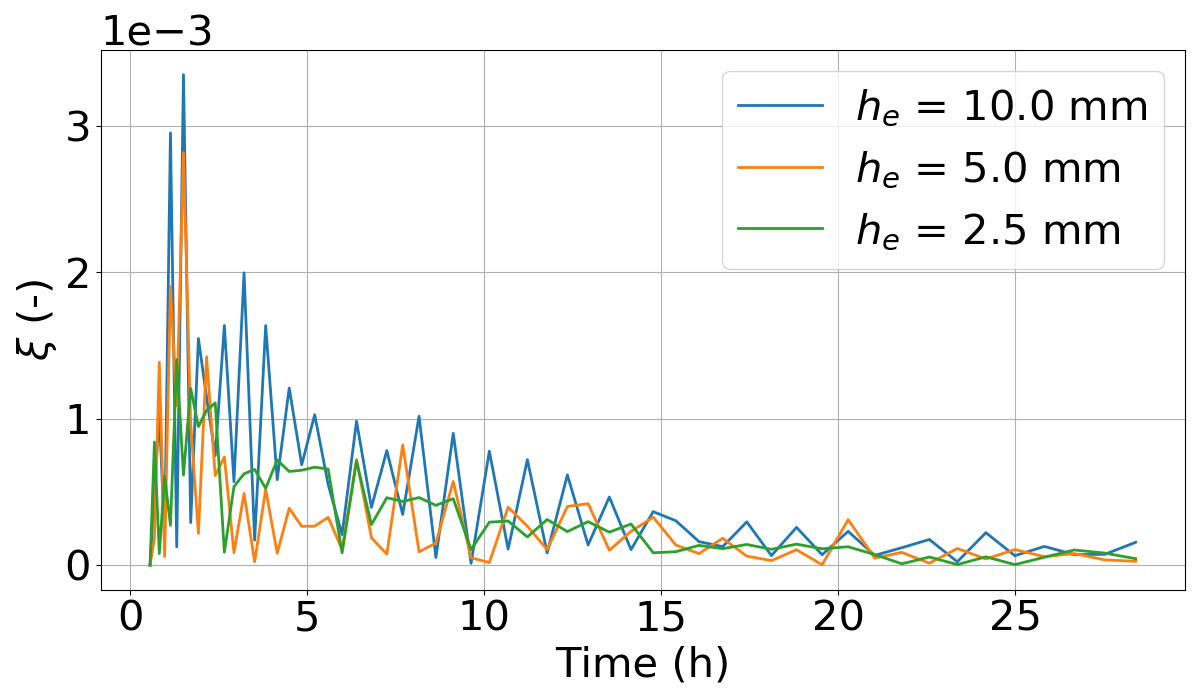}
& 
\includegraphics[height=4cm]{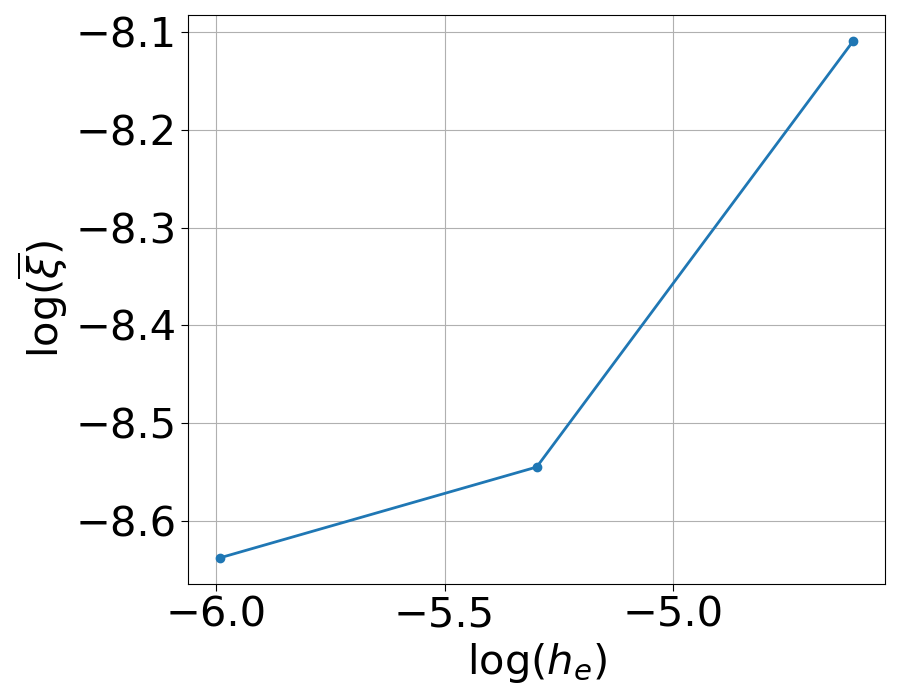}
 \\
(a) & (b)
\end{tabular}
\caption{ \answ{Straight front semi-infinite problem: error comparisons for $\latentHeat = 3.3 \times 10^{5}$ [J.kg$^{-1}$].  Relative front location error evolution with respect to time for different mesh refinements (a). Integrated error convergence with respect to the mesh element size  (b).}}
\label{fig:stephan1DErrorComparisonLatentHeat}
\end{figure}

Regarding the case $\latentHeat =3.3 \times 10^{5}$ [J.kg$^{-1}$],
Figure \ref{fig:stephan1DResultsCoarse}, 
we choose $t^0 = 1000$ [s] and $\beta = 100$ [s]. It 
leads again to a \emph{small time step} i.e. a time step that only allow the front
to advance of a length $\Delta x \simeq \elemSize/6 = 0.01/6$ [m].
Convergence results are \answ{shown in Figure \ref{fig:stephan1DErrorComparisonLatentHeat}. Contrary to the case $\latentHeat = 0$ [J.kg$^{-1}$], more oscillations of $\errXf(t)$ are observed, and the order of convergence of $\overline{\errXf}$ is not exactly 1. The differences may be explained by the solving method used for each problem}. 
While a simple fixed point scheme as described at the end of
\S\ref{semiMonolithicScheme}
was sufficient to converge
the case without latent heat, a more sophisticated quasi-Newton
approach is
required to converge the case with latent heat as described in Algorithm 1.
Indeed, a 
fixed point scheme where the initial temperature update does not take
into account the variation of energy due to latent heat  has the effect of predicting a front location 
that is way
too far from its converged position. Then, the front oscillates without
converging. The quasi-Newton scheme acts like a damper and allows to
converge in less that 10 iterations, without oscillations. 
\answ{More effort is necessary to solve the case where $\latentHeat > 0$, which can be due to the fact that the heat flux is not continuous (see equation \ref{eq:jump_flux}). It may also explain why the error ont the front position has more oscillations (see Figure \ref{fig:stephan1DErrorComparisonLatentHeat}), although the error levels are absolutely acceptable.}

\begin{figure}[t!]
\begin{center}
\includegraphics[width=0.45\textwidth]{fig/stephan1DTemperatureNoLatentHeatTemperatureScale.pdf}
  \begin{tabular}{ccc}
    \includegraphics[width=0.3\textwidth]{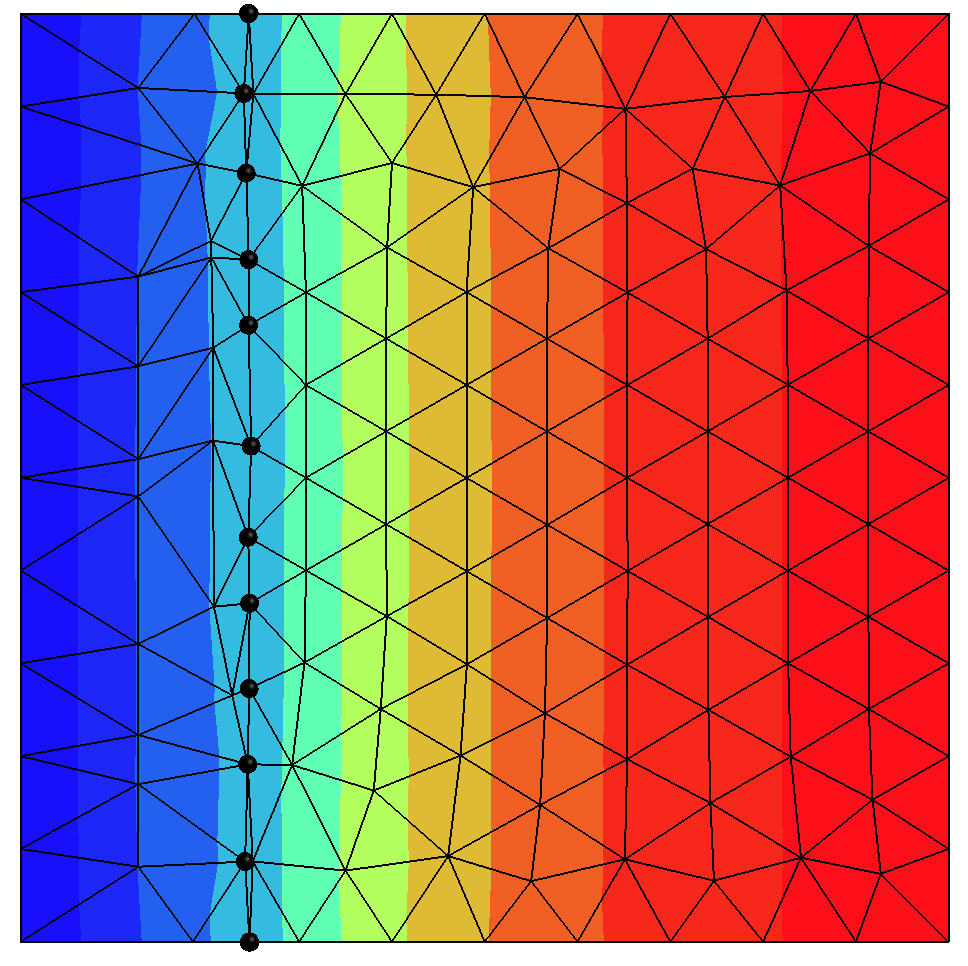}&
    \includegraphics[width=0.3\textwidth]{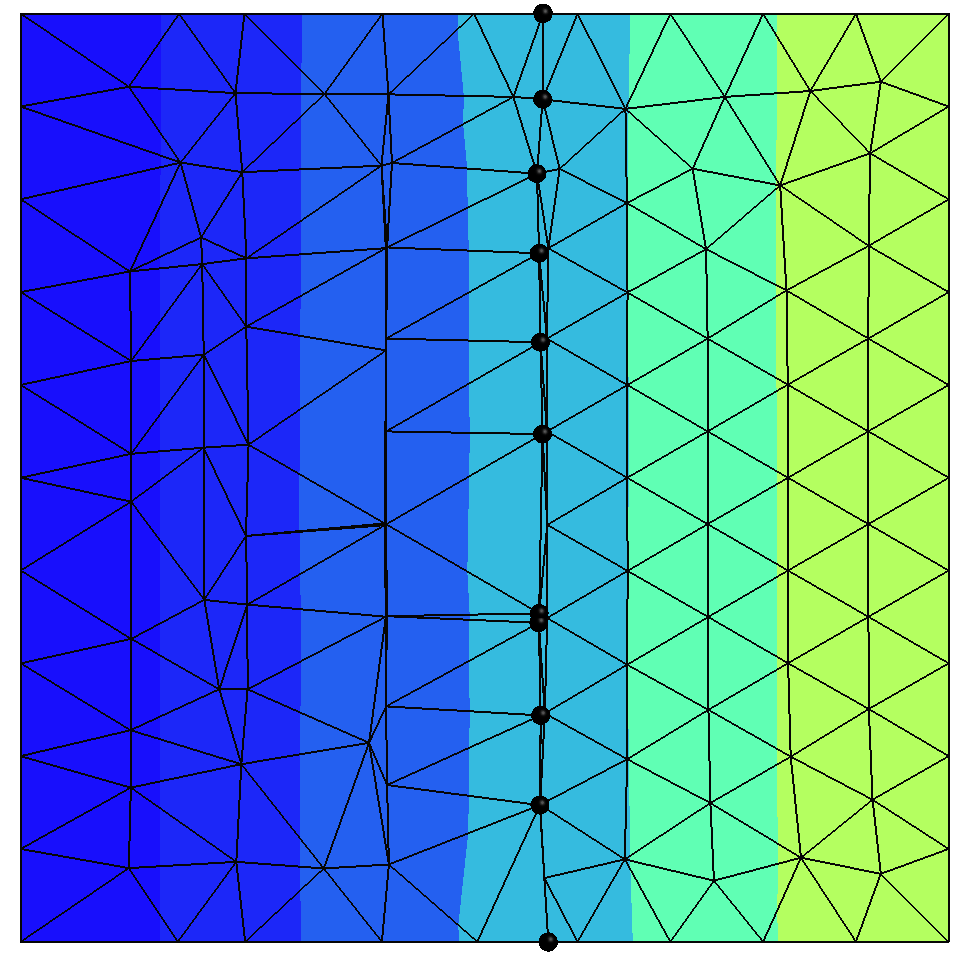}&
    \includegraphics[width=0.3\textwidth]{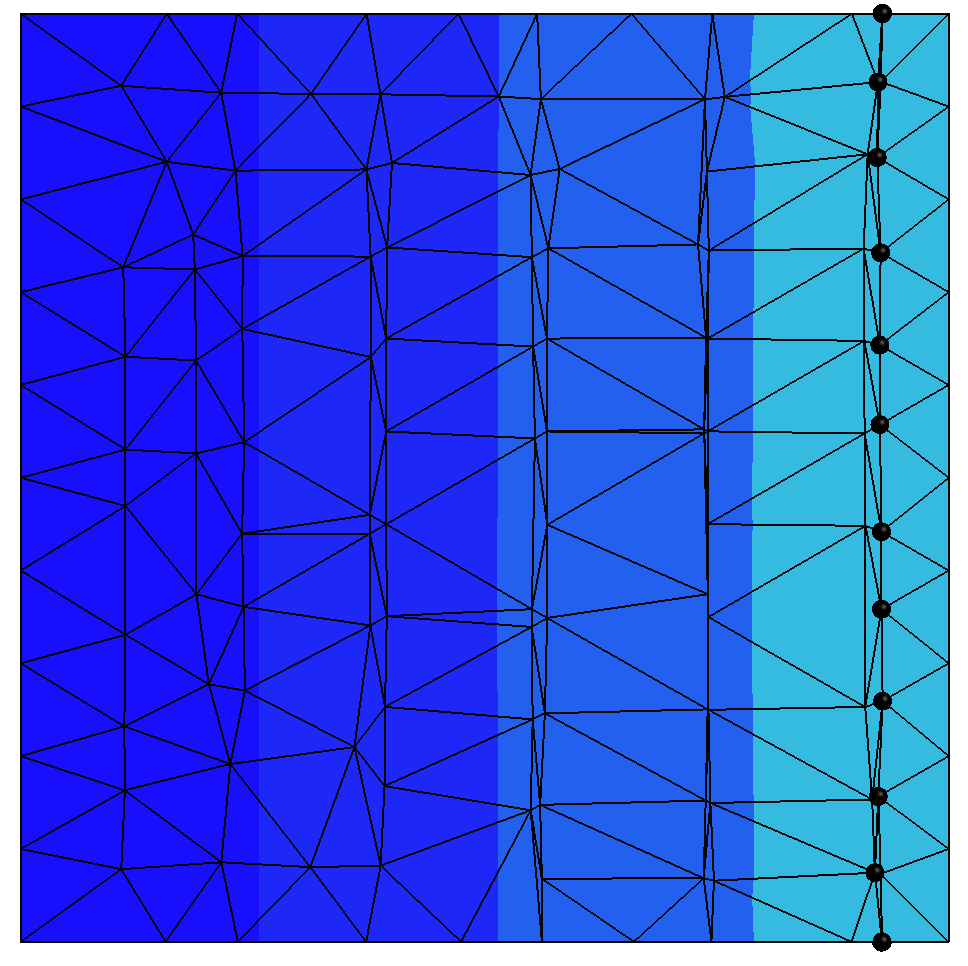}\\
    (A)  & (B)  & (C) 
  \end{tabular}  
  \begin{tabular}{cc}
 \includegraphics[width=0.45\textwidth]{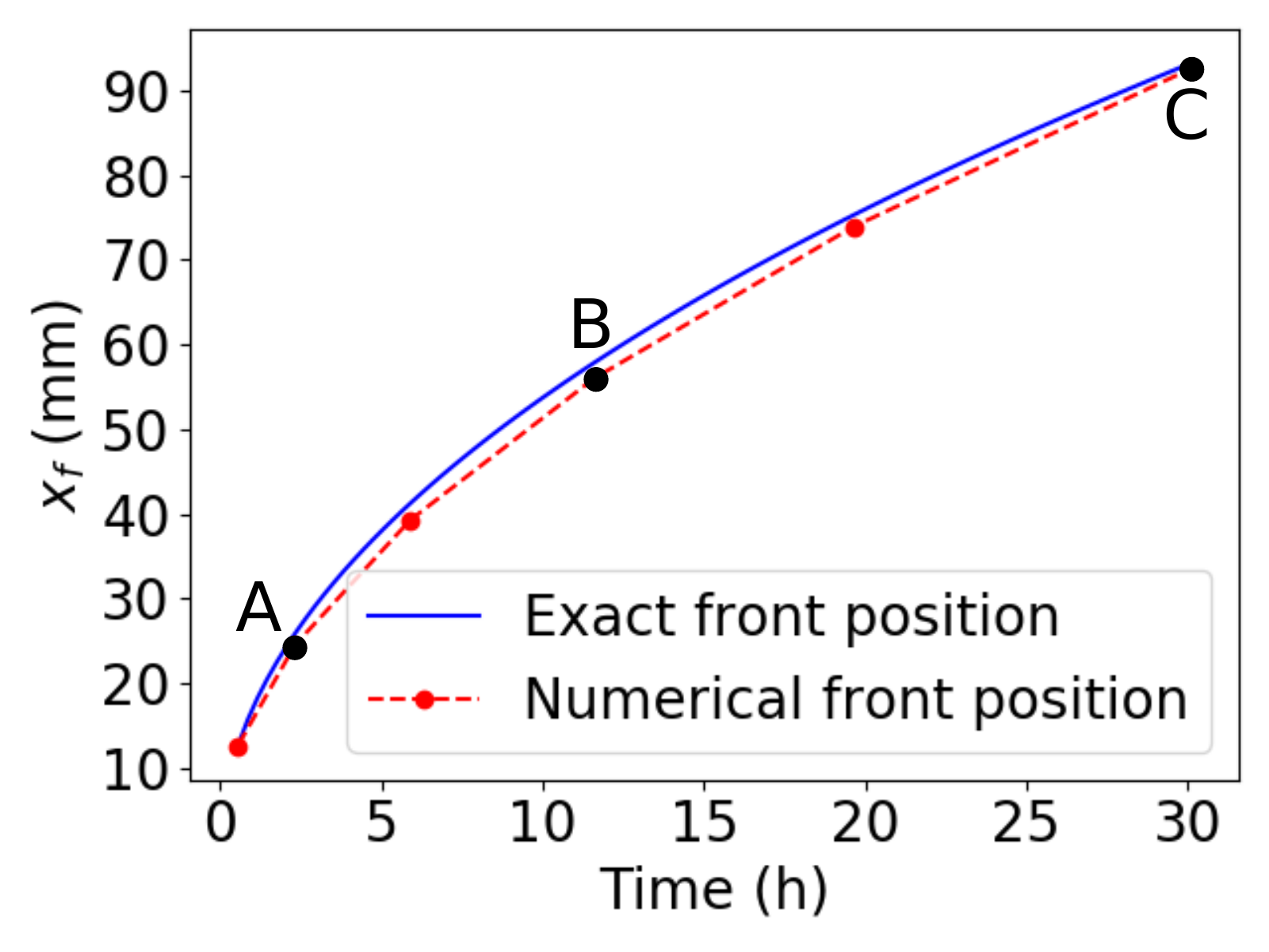}&
 \includegraphics[width=0.45\textwidth]{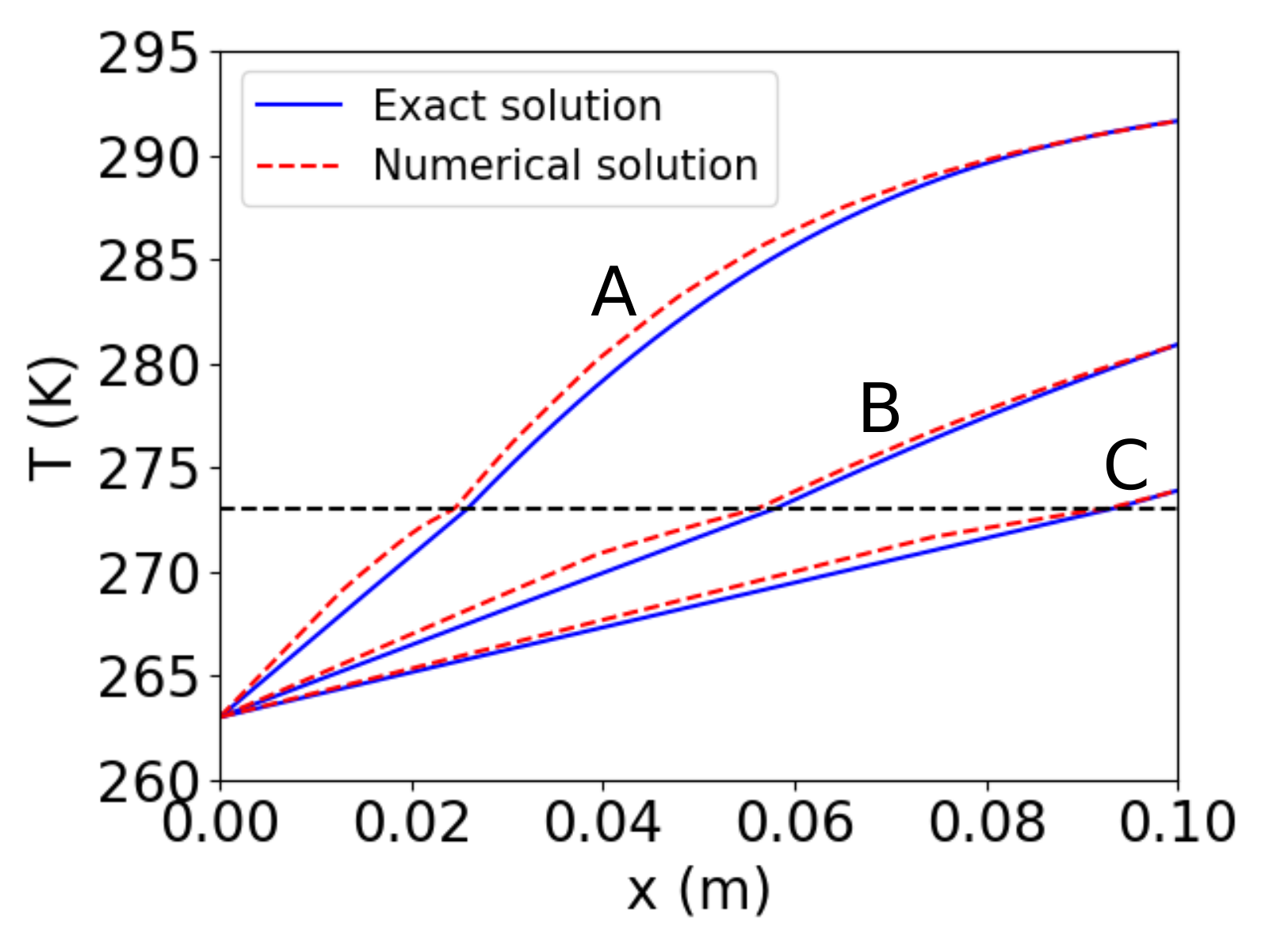} \\
  (a) & (b)    
  \end{tabular}  
\end{center}
\caption{ \answ{Straight front semi-infinite problem} solved with five time steps and $\theta = 1$: numerical results for $\latentHeat = 3.3 \times 10^{5}$ [J.kg$^{-1}$] and $\elemSize = 0.01$ [m].
Temperature fields and fronts (\answ{front vertices are represented by big, black dots}) after the first, third and fifth time-step ($t = 2.31$ [h] (A), $11.62$ [h] (B) and $30.1$ [h] (C).
Comparison of exact and numerical front positions (a). Analytical and numerical solutions along the horizontal direction (b). }
\label{fig:stephan1DResultsBiggerDtTheta1}
\end{figure}

\begin{figure}[t!]
\begin{center}
\includegraphics[width=0.45\textwidth]{fig/stephan1DTemperatureNoLatentHeatTemperatureScale.pdf}
  \begin{tabular}{ccc}
    \includegraphics[width=0.3\textwidth]{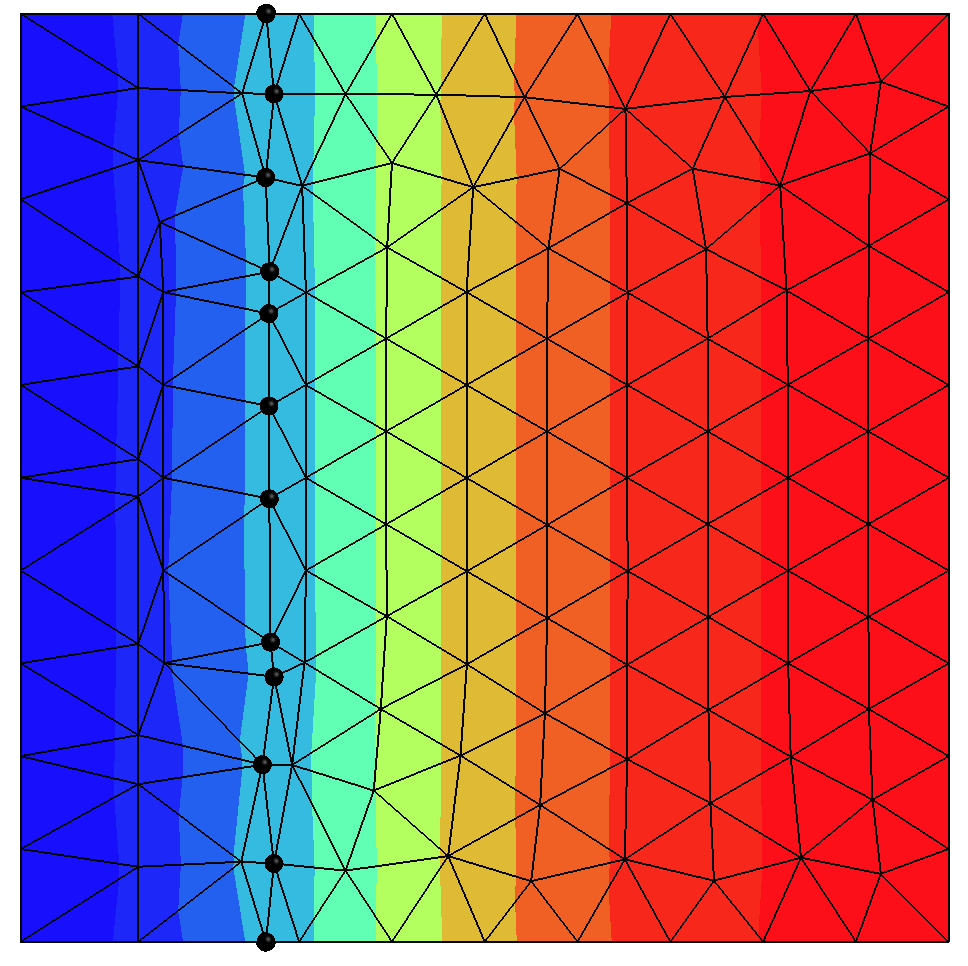}&
    \includegraphics[width=0.3\textwidth]{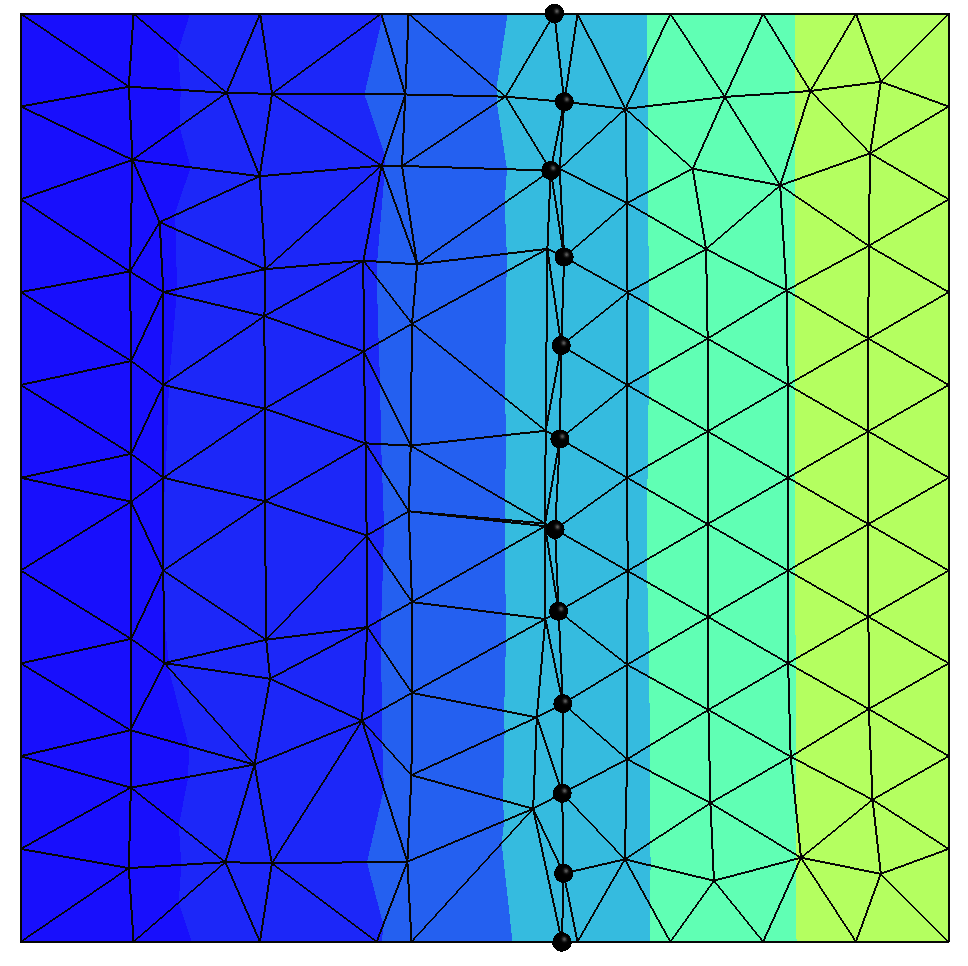}&
    \includegraphics[width=0.3\textwidth]{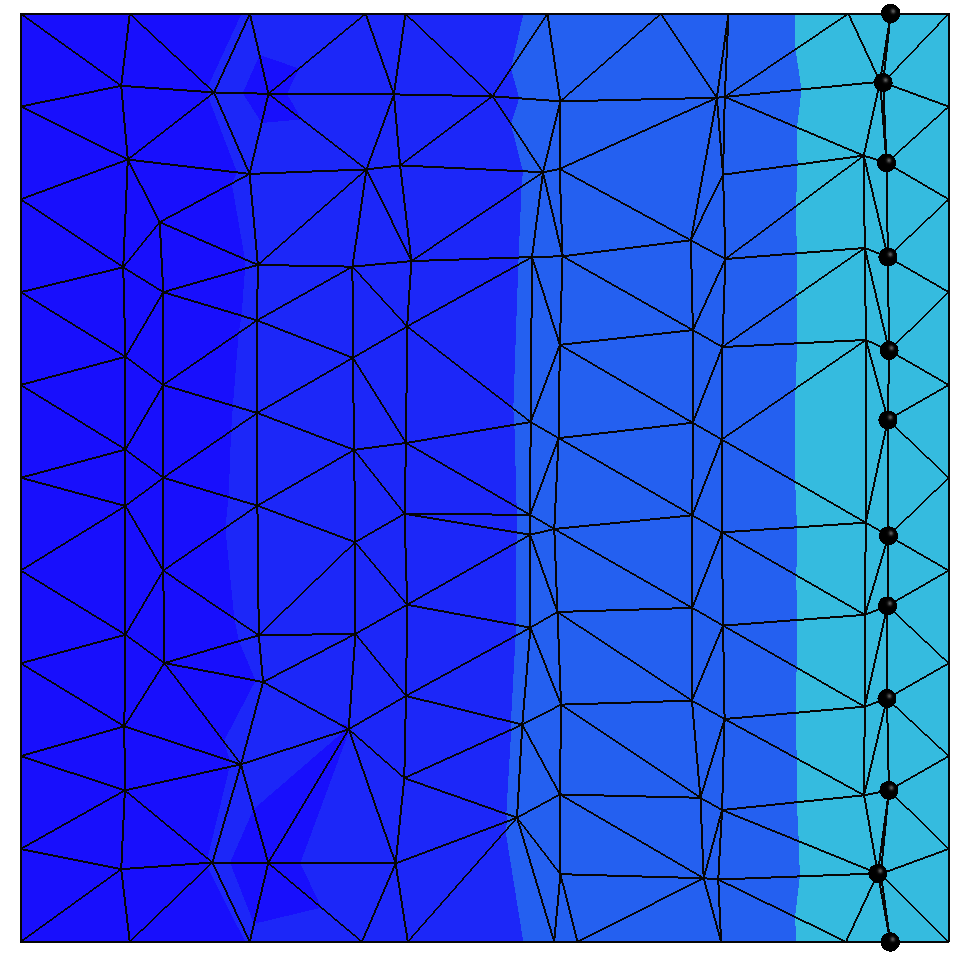}\\
    (A)  & (B)  & (C) 
  \end{tabular}  
  \begin{tabular}{cc}
 \includegraphics[width=0.45\textwidth]{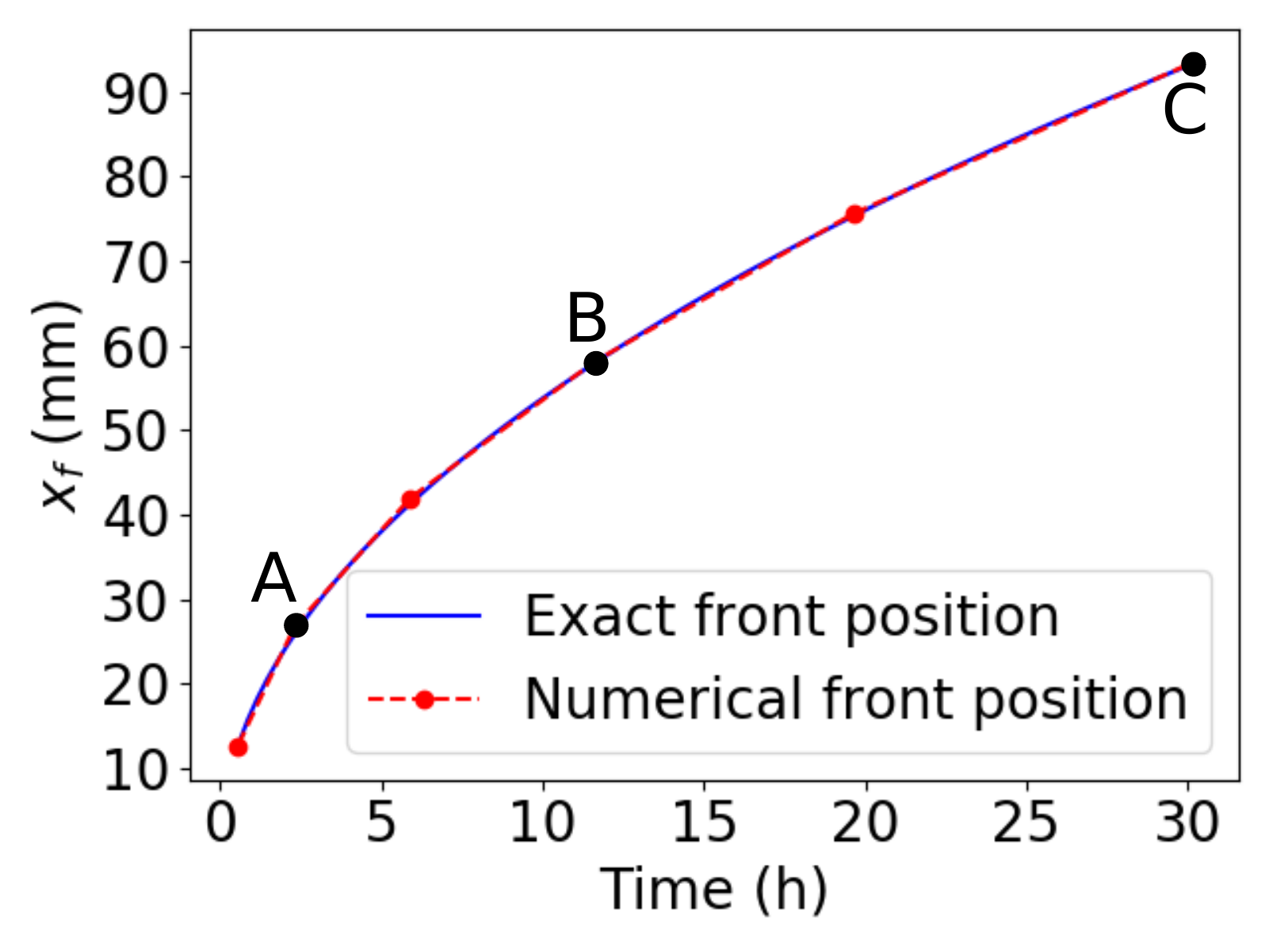}&
 \includegraphics[width=0.45\textwidth]{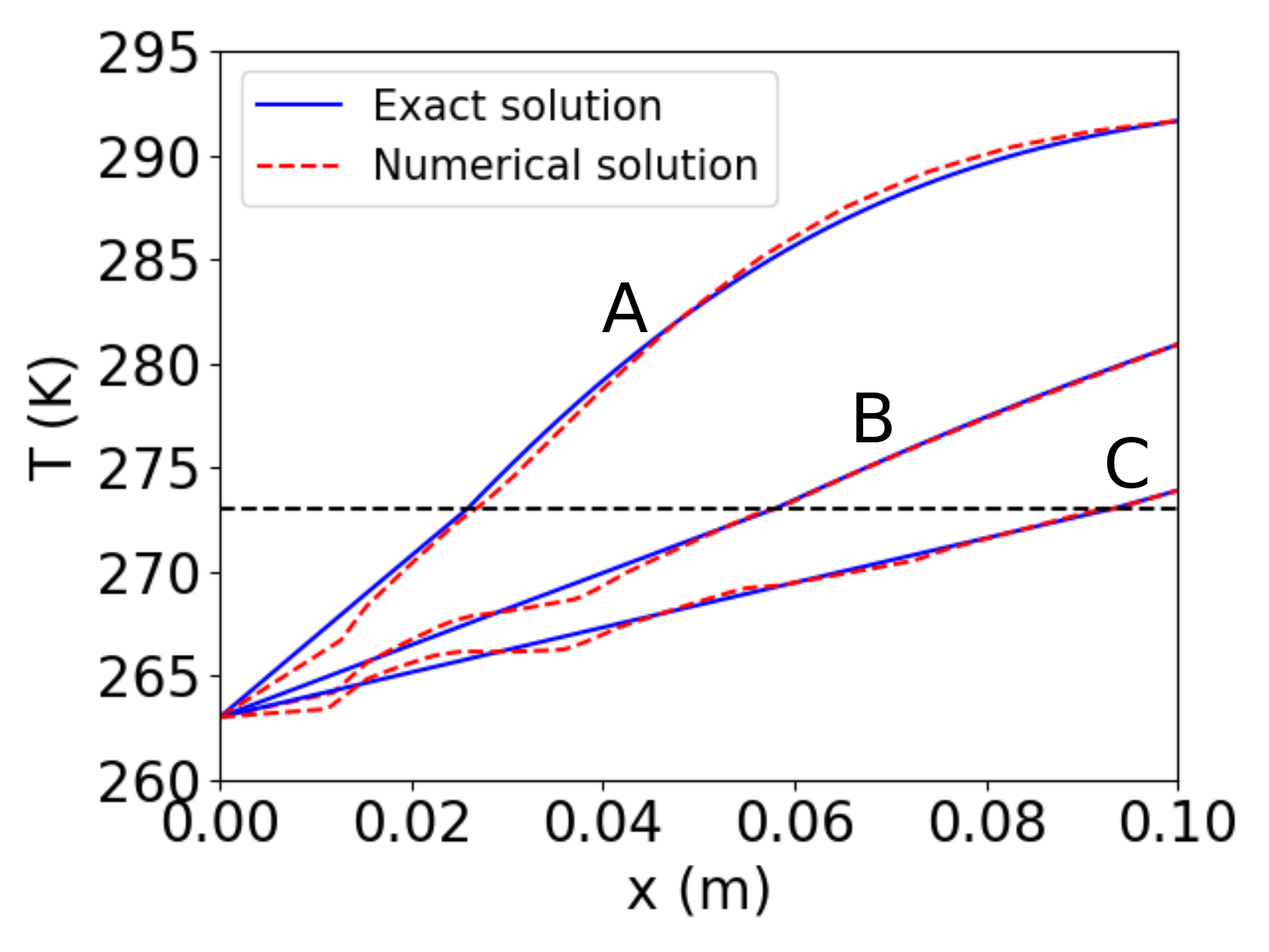} \\
  (a) & (b)    
  \end{tabular}  
\end{center}
\caption{\answ{Straight front semi-infinite problem} solved with five time steps and $\theta = 0.5$: numerical results for $\latentHeat = 3.3 \times 10^{5}$ [J.kg$^{-1}$] and $\elemSize = 0.01$ [m]. 
Temperature fields and fronts (\answ{front vertices are represented by big, black dots}) after the first, third and fifth time-step ($t = 2.31$ [h] (A), $11.62$ [h] (B) and $30.1$ [h] (C)). Comparison of exact and numerical front position (a).  Analytical and numerical solutions along the horizontal direction (b). }
\label{fig:stephan1DResultsBiggerDt}
\end{figure}

A tempting question is now: 
can we do large time steps i.e. time steps that allows the front to
advance of more than one element: $\Delta x > h_e$? The answer is surprisingly \emph{yes}.
Figure \ref{fig:stephan1DResultsBiggerDtTheta1} shows three meshes corresponding to three different time steps where $t^0 = 2000$ [s] and $\beta = 20000$ [s] were chosen so that the phase change front crosses the domain in 5 time steps and more than three elements are crossed by the
front in one time step. In this case, the nodes of the new front at
$t^{n+1}$ are not necessarily connected by edges to the nodes of the front at time
$t^n$. Nevertheless, the front relaying algorithm can be applied as is and
it is not necessary to compute the set of relays that allowed the
transition from the configuration at time $t^n$ to the one at time $t^{n+1}$. 
Moreover, the number of iterations required to converge the
quasi-Newton algorithm remains reasonable (under 25 iterations per time step) even in the case of very large time steps. 
Note that to obtain the above results a value $\theta = 1$ was used (giving a fully implicit resolution). 
If one chooses  $\theta = 0.5$, the front location agreement is even better but some oscillations are observed in the wake of the front as shown in Figure \ref{fig:stephan1DResultsBiggerDt}.
Finally, note that as the front advances quickly, the relaxation has not
enough time to take effect and the mesh has not recovered yet in the wake of the front 
at the final step of propagation. 
In a general approach, the relaxation coefficients 0.9 and 0.1  in \eqref{eq:relax} should ideally involve the front speed and element size.

\subsection{The axisymmetric two-phase problem}
\label{axisymmetric}

The next example is the axisymmetric problem of the solidification of a liquid domain due to a Dirac heat sink $\heatSink$ (100 [W.m$^{-1}$])
located at $r = 0$ [m]. At
$t = 0$ [s], the temperature is uniform: $T_\ell= 293$ [K]. Then,
for later times $t \in [0, \tmax]$, the position of the solidification
front (see \cite{Carslaw1959}) is given by
  \begin{equation}
  \frontPositionExAxisym(t) = 2 \solexParam \sqrt{\diffusivitySolid t}.
  \nonumber
 \end{equation}
The coefficient $\solexParam$ is obtained by solving
\begin{equation}
 \frac{\heatSink}{4 \pi \volumicMass \diffusivitySolid \latentHeat} e^{-\solexParam^2}  + \frac{1}{\alpha \gamma} \frac{\exp(-\solexParam^2 \alpha)}{\operatorname{Ei}(-\solexParam^2 \alpha)}  - \solexParam^2 = 0,
 \nonumber
\end{equation}
where $\text{Ei}$ is the exponential integral function ${\displaystyle
  \operatorname {Ei} (x) = \int_{-\infty }^{x}{\frac {e^{t}}{t}}\dint t.}$
We have
$\solexParam = 0.162$ for $\latentHeat = 0$ [J.kg$^{-1}$] and
$\solexParam = 0.094$ for $\latentHeat = 3.3 \times 10^{5}$ [J.kg$^{-1}$].
The temperature field for the solid 
($r \leq \frontPositionExAxisym$) and liquid phase 
($r \geq \frontPositionExAxisym$) is given by
\begin{equation}
\temp(r,t) = 
    \begin{cases}
    \boundaryTempInterface + \frac{\heatSink}{4 \pi \conductivitySolid} \left[ \operatorname{Ei} \left( - \frac{r^2}{4 \diffusivitySolid t} \right) - \operatorname{Ei} (-\solexParam^2) \right] , \text{ for } r \leq \frontPositionExAxisym, \\
    T_\ell - \frac{T_\ell - \boundaryTempInterface}{\text{Ei}(-\solexParam^2 \alpha )} \operatorname{Ei} \left( - \frac{r^2}{4 \diffusivityLiquid t} \right) , \text{ for } r \geq \frontPositionExAxisym.
    \end{cases}
    \label{eq:liquid}
\end{equation}

Two versions of this problem are now presented, one that is easy and a harder one. 

\subsubsection{The easy version}
\begin{figure}
\centering
  \begin{tabular}{cc}
 \includegraphics[width=0.5\textwidth]{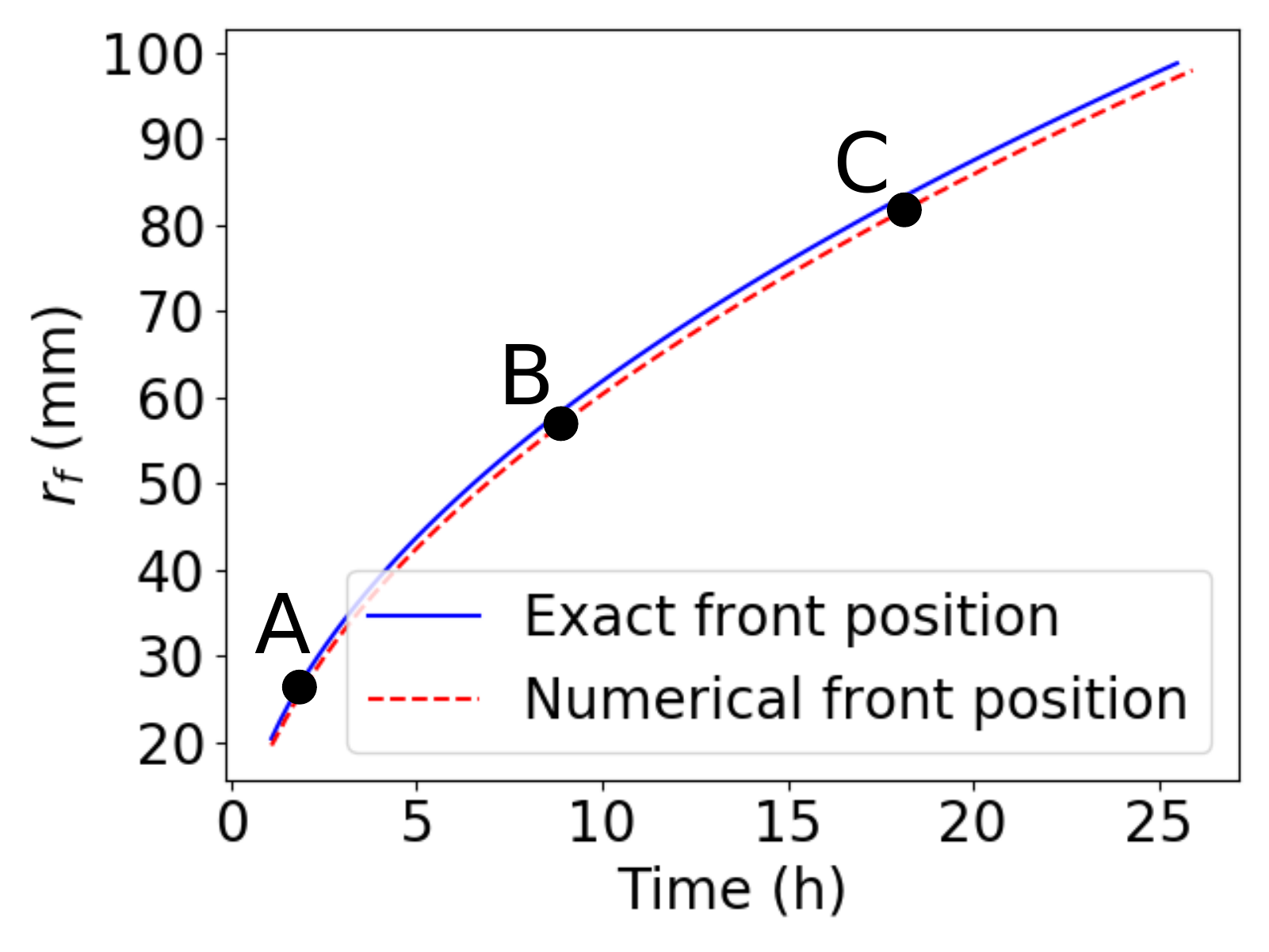}&
 \includegraphics[width=0.35\textwidth]{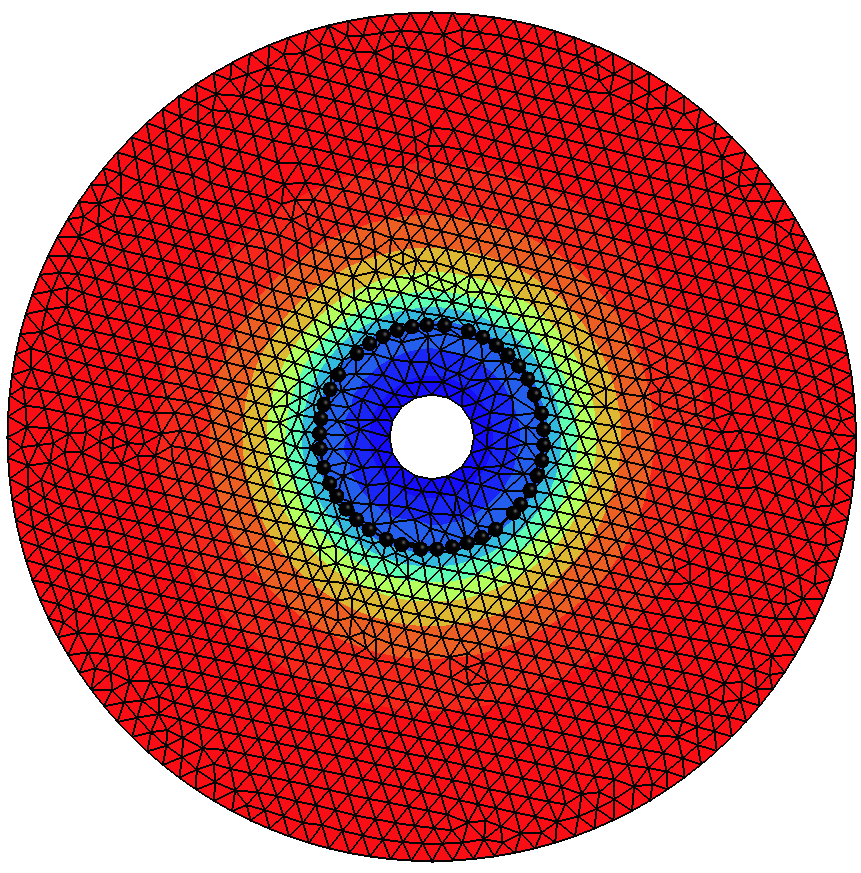} \\
  (a) & (A) \\
 \includegraphics[width=0.35\textwidth]{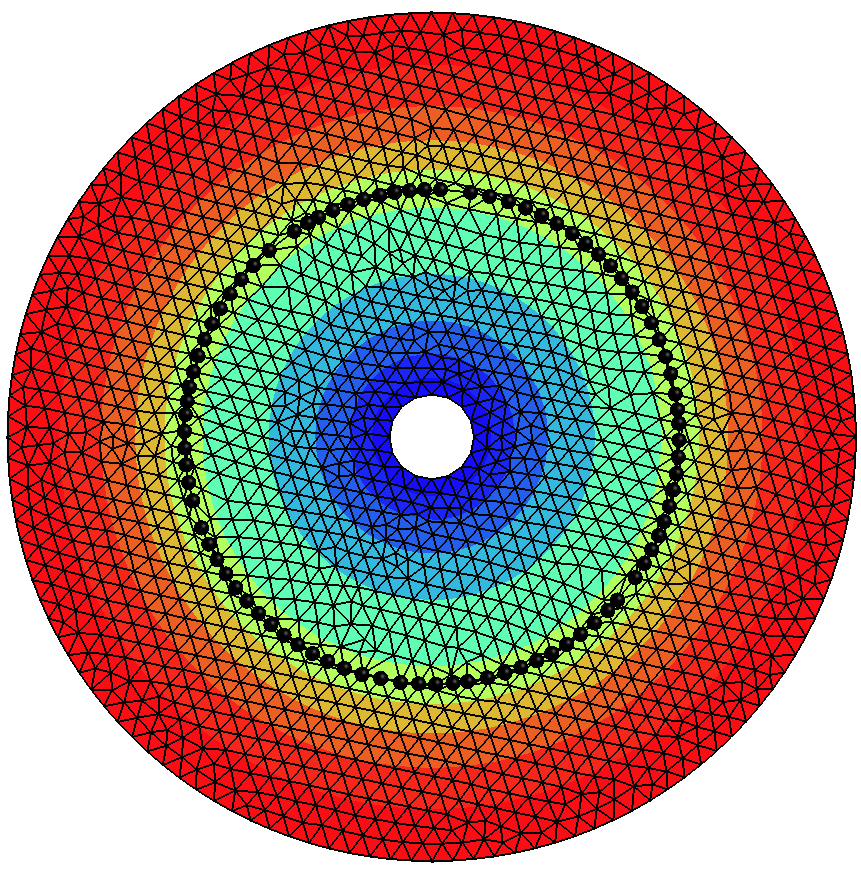}&
 \includegraphics[width=0.35\textwidth]{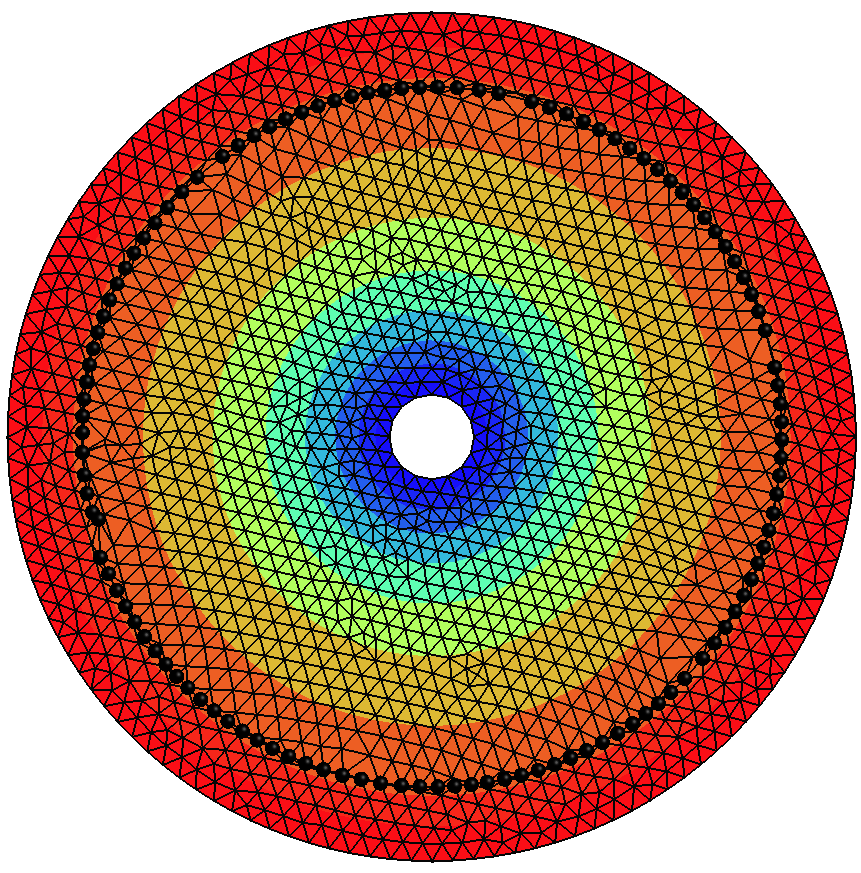} \\
  (B) & (C) \\
\end{tabular}
\includegraphics[width=0.7\textwidth]{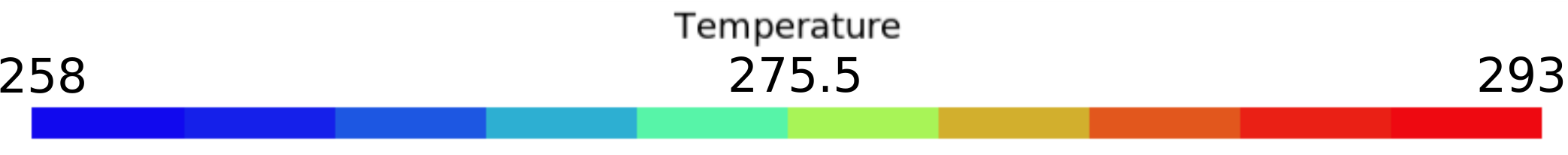}
\caption{ Axisymmetric problem: numerical results for $\latentHeat = 0$ [J.kg$^{-1}$] for $\elemSize = 0.005$ [m].  Comparison of exact and numerical front positions (a). Temperature fields and fronts at $t = 2.1$ [h] (A), $9.6$ [h] (B) and $18.9$ [h] (C). \answ{The front vertices are represented by big, black dots.}}
\label{fig:stephan1DAxisymResultsCoarseNoLatentHeat}
\end{figure}

\begin{figure}
\centering
\begin{tabular}{cc}
 \includegraphics[height=4cm]{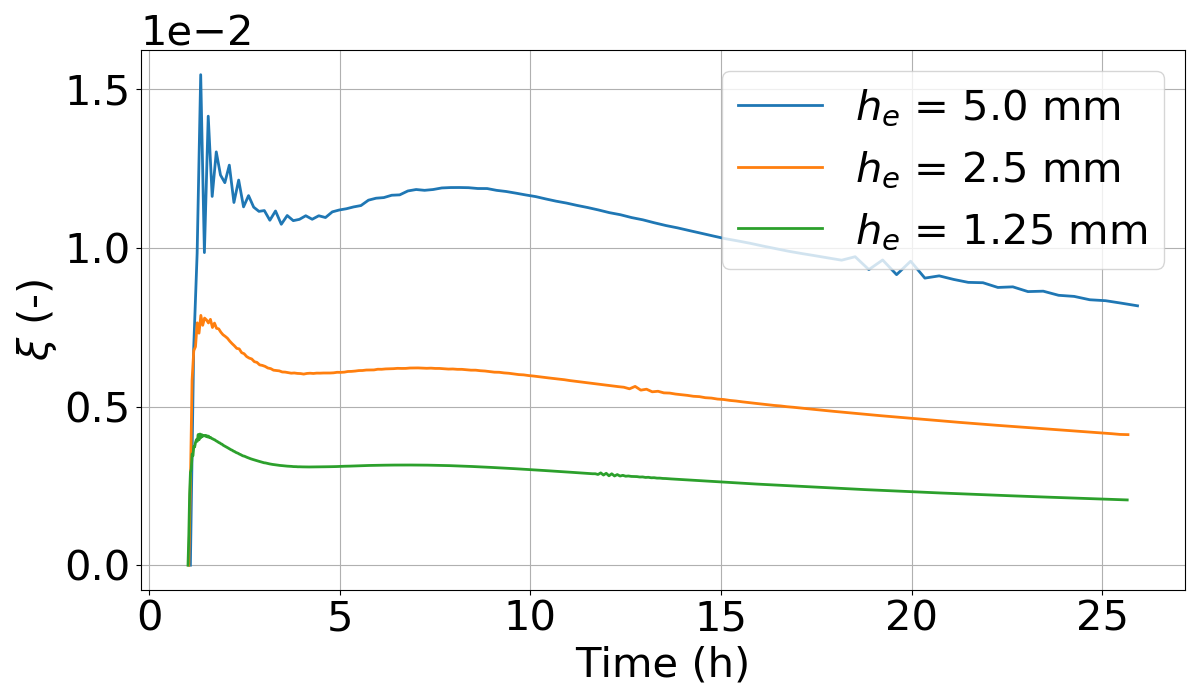}
& 
\includegraphics[height=4cm]{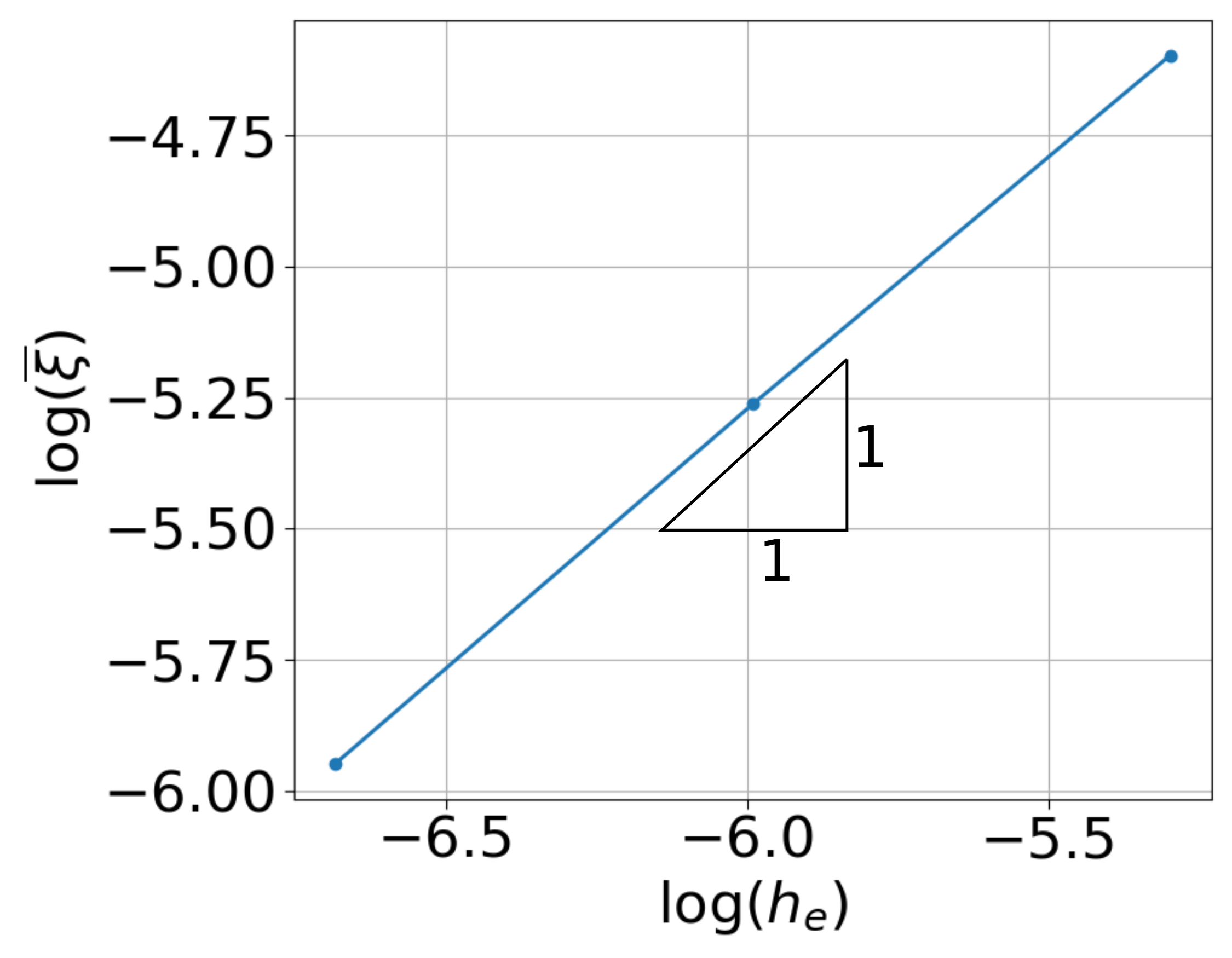}
 \\
(a) & (b)
\end{tabular}
\caption{Axisymmetric problem: error comparisons for $\latentHeat = 0$ [J.kg$^{-1}$]. Relative front location error evolution with respect to time for different mesh refinements (a). Integrated error convergence with respect to the mesh element size  (b).}
\label{fig:stephan1DAxisymAxisymErrorComparisonNoLatentHeat}
\end{figure}

To avoid the infinite initial front speed, we start the simulation at 
$t^0=3600$ [s] with the exact solution  as initial temperature. 
We also  consider a ring geometry to avoid the direct 
effect of the Dirac sink. The exact temperature evolution is imposed to an
inner radius $R_{\text{int}} = $ 0.01 [m] and an external radius $R_{\text{int}} = $ 0.1 [m].
For this easy case a zero latent heat is considered. 
Finally, a variable time step $\Delta t^{n+1/2} = \sqrt{25 t^n}$ [s] is chosen.
The results obtained for a coarse mesh with a finite element size
$\elemSize = 0.005$ [m] are given in Figure
\ref{fig:stephan1DAxisymResultsCoarseNoLatentHeat} while the errors for
different mesh sizes are plotted in Figure
\ref{fig:stephan1DAxisymAxisymErrorComparisonNoLatentHeat}. The 
numerical value of the front position which is plotted
and used to compute the error is obtained by taking the average of the
$r$ coordinate of the nodes of the phase change front. Here again, the
front is correctly captured by the X-MESH approach. 

\subsubsection{The harder version}
\begin{figure}
\begin{center}
\begin{tabular}{cc}
\includegraphics[width=0.4\linewidth]{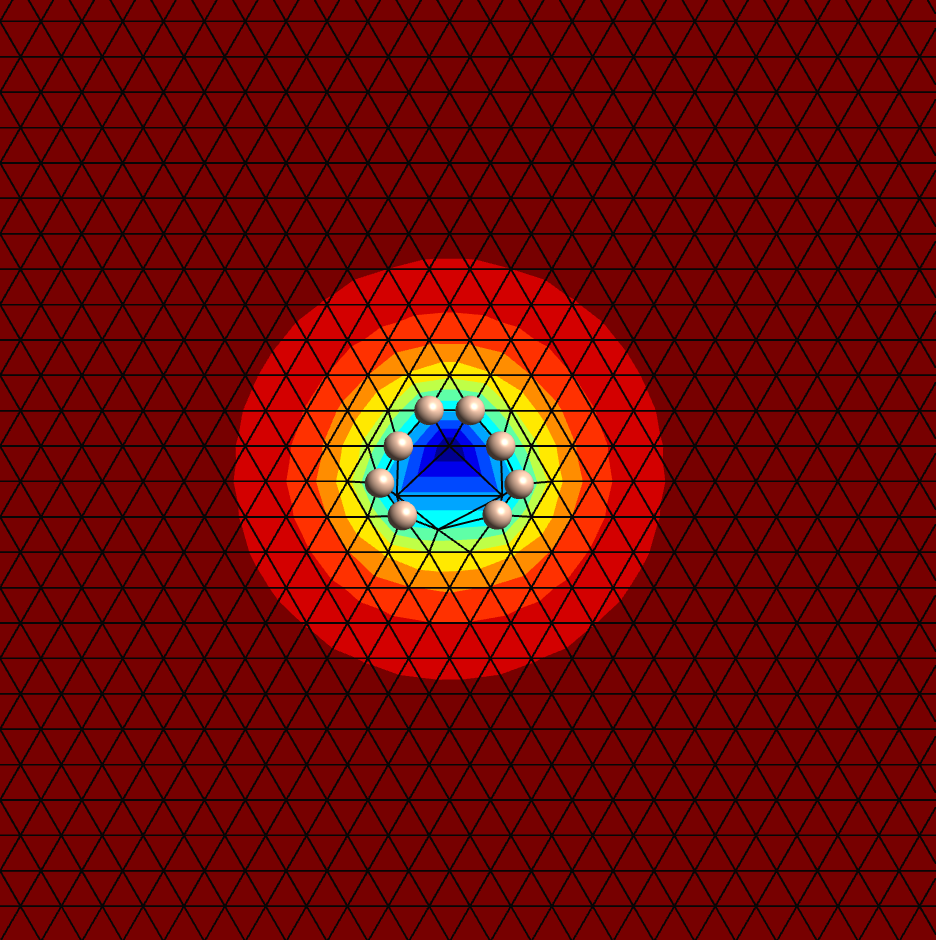}&
\includegraphics[width=0.4\linewidth]{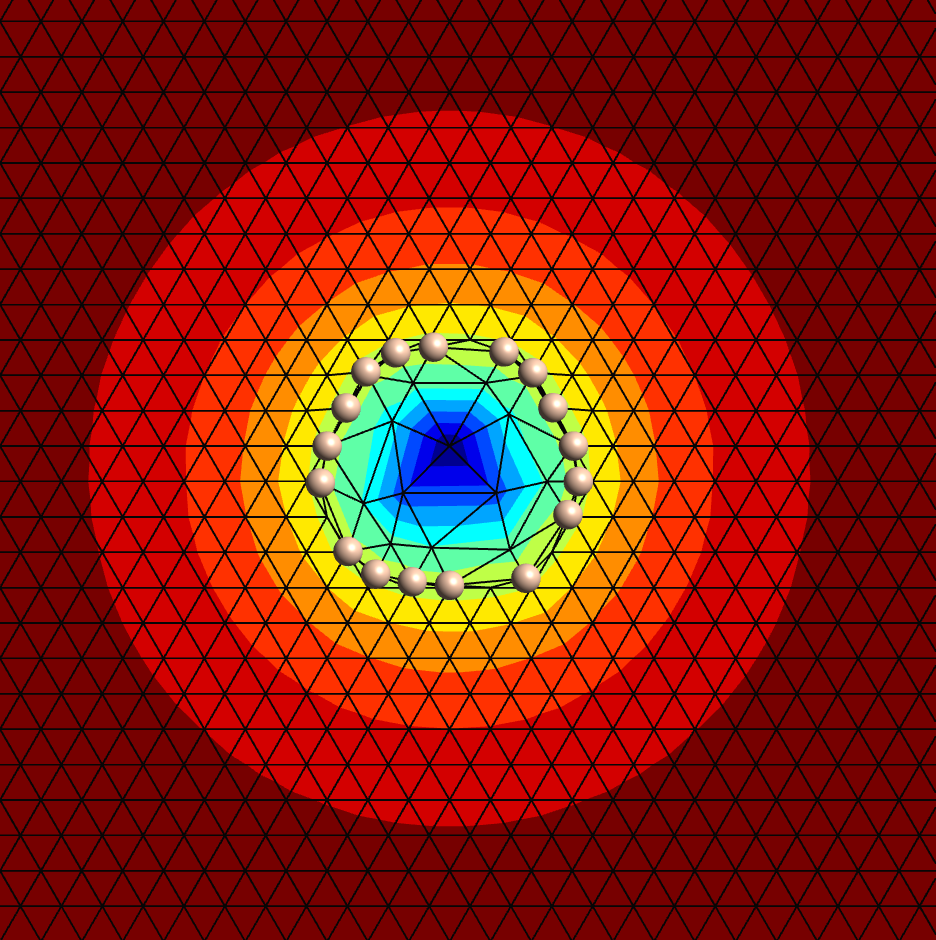}\\
\includegraphics[width=0.4\linewidth]{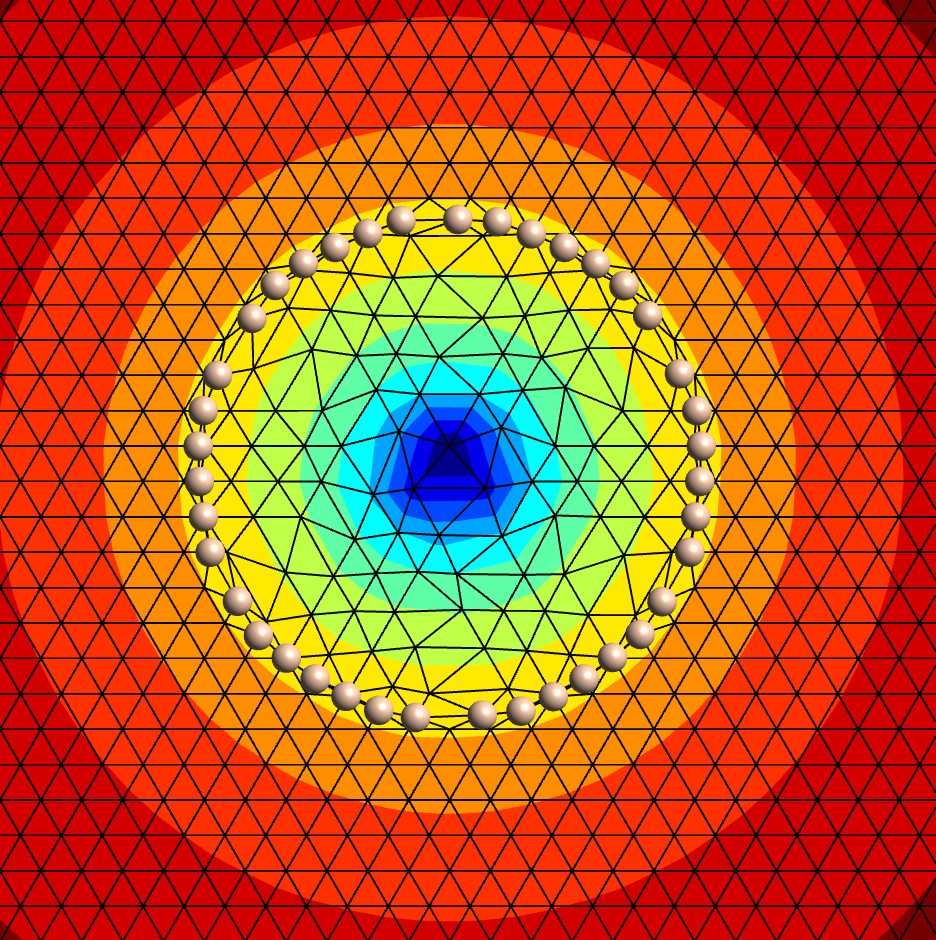}&
\includegraphics[width=0.4\linewidth]{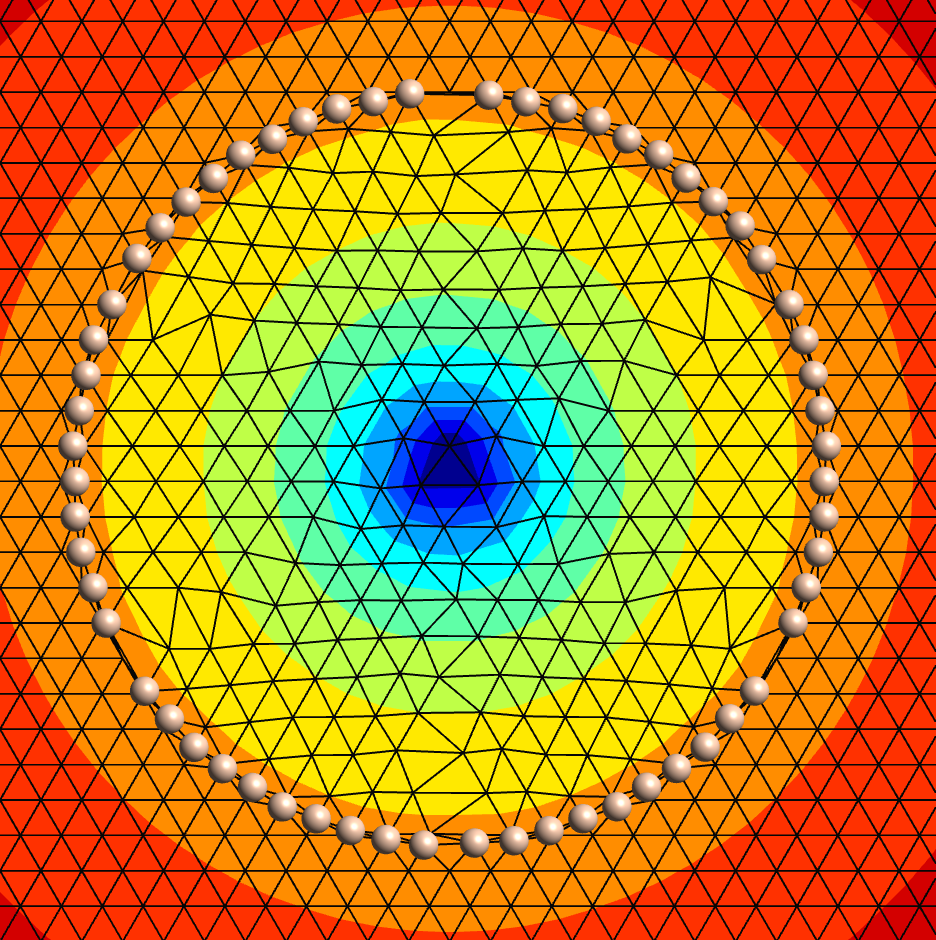}
\end{tabular}
\end{center}
\begin{center}
\includegraphics[width=0.75\linewidth]{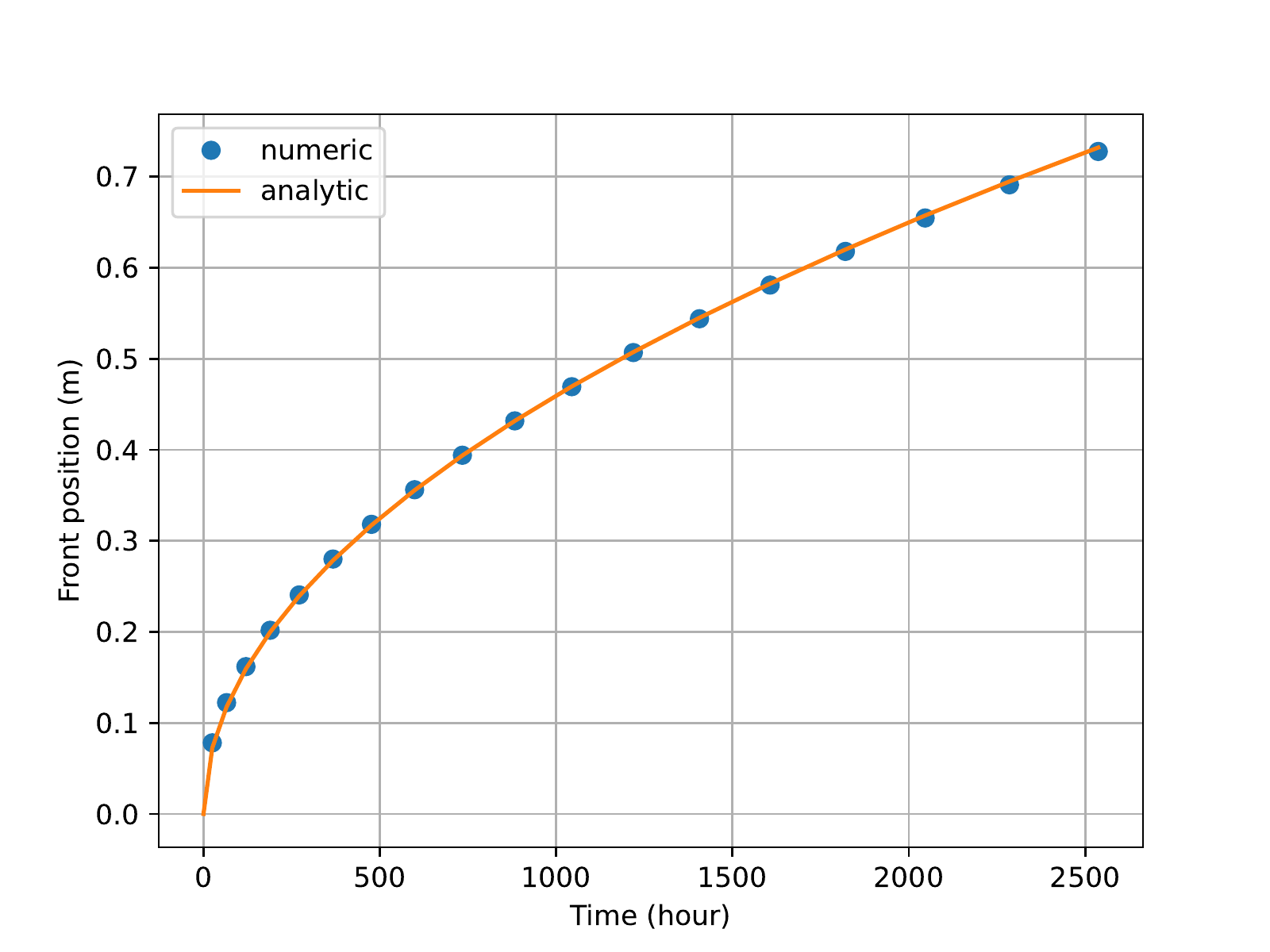}
\end{center}
\caption{Axisymmetric problem
with $\latentHeat = 3.3 \times 10^{5}$ [J.kg$^{-1}$]: the freezing front is nucleated and then
  propagates (top). \answ{The front vertices are represented by big, yellow dots.}
  Comparison of exact and numerical front positions (bottom).
\label{fig:axi2}}
\end{figure}

This test case is there to push further the capabilities of our
method. We use again the axisymmetric test case, but with a latent
heat of $3.3 \times 10^{5}$ [J.kg$^{-1}$] and by explicitly modeling the Dirac 
heat sink in the center of the circle. The Dirac sink ($100$ [W.m$^{-1}$])
contributes to the three nodes 
of the element in which it is located proportionally to its barycentric coordinates.
Instead of considering an infinite domain, we consider a square with dimensions 3 $\times$ 3 [m$^2$]
and apply the analytical solution
to the outer contour.
We start the simulation at $t^0 = 0$ [s] to see if the method can handle the nucleation and face the infinite initial velocity of the analytical solution.
Figure \ref{fig:axi2} shows solutions at different times as well as a
comparison with the analytical solution. 
An average of  $9$ quasi-Newton iterations
were necessary to reduce the residual by a factor $10^6$. Figures show
that our method is able to nucleate a front. At early stages, the
shape of the front is not circular due to the insufficient resolution
of the mesh. The Stefan model is very stable: when the front grows, it
becomes quasi-circular and our numerical solution is very close to the analytical
solution. Meshes of Figure \ref{fig:axi2} show the effect of mesh
relaxation: the mesh at the center of the circle that was initially affected
by the front relaying algorithm  progressively returns to its initial state.

Regarding the time step, we could not rely on formula \eqref{eq:dt}
since $t^0=0$. We used instead the implicit relation 
\begin{equation}
    \Delta t^{n+1/2} = \sqrt{\Delta t^{1/2}  \; t^{n+1}} 
\end{equation}
in which the $\beta$ parameter has been replaced by the choice of the first time step $\Delta t^{1/2}$.
Noting that $t^{n+1} = t^n + \Delta t^{n+1/2}$, we get the time step value for any instant $t^n$:
\begin{equation}
    \Delta t^{n+1/2} = \Delta t^{1/2} \left( \frac{1 + \sqrt{1 + 4 t^n/\Delta t^{1/2}}}{2} \right).  
    \label{eq:dtimp}
\end{equation}
In the simulation, the value $\Delta t^{1/2} = 25$ [h] has been used.

\subsection{Two rotating heat sinks}
\label{multipleHeatSources}

\begin{figure}
 \begin{center}
 \begin{tabular}{|ccc|}
\hline 
   (1) & (2) & (3)   \\
  \includegraphics[width=3.8cm]{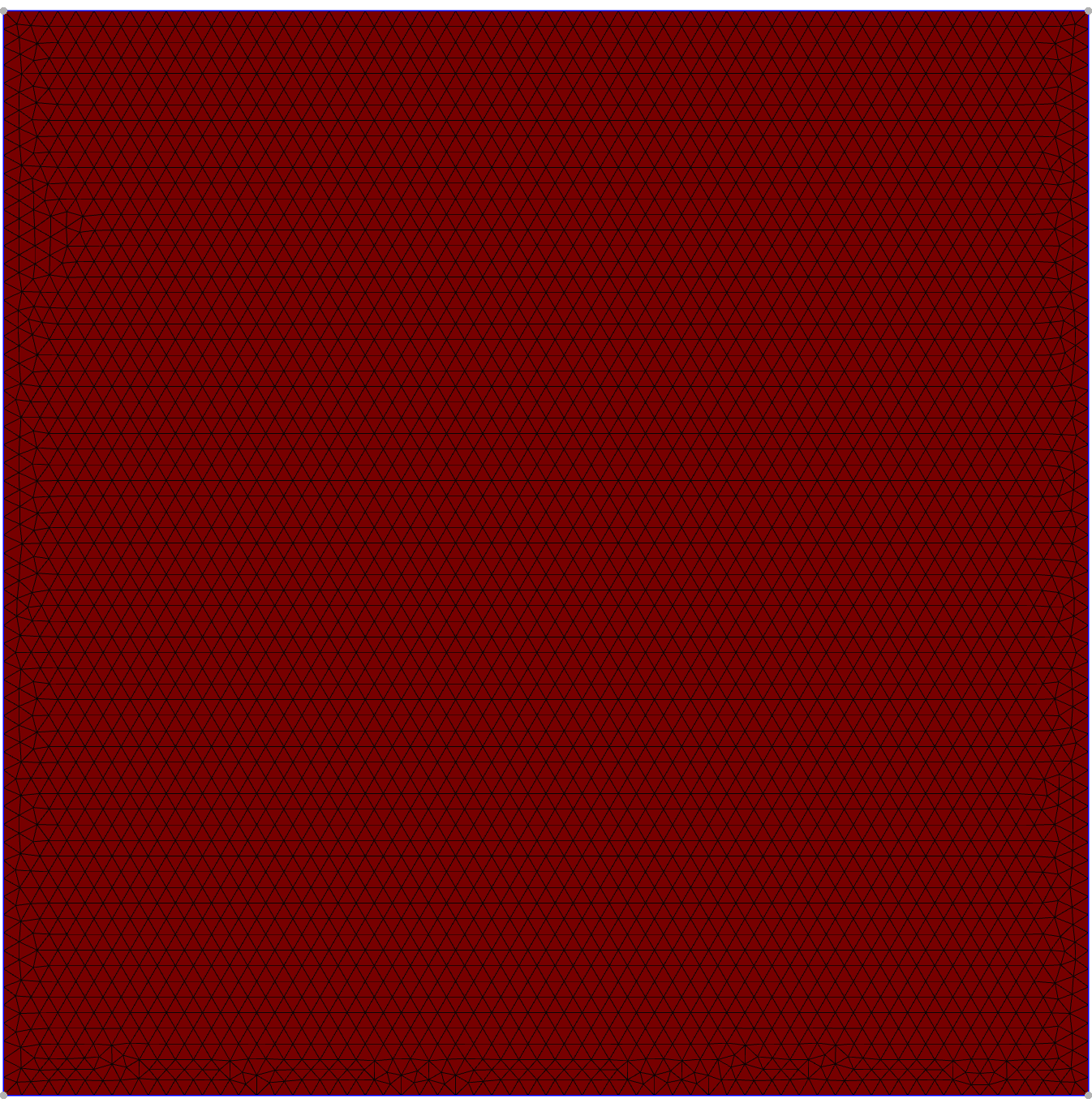} & 
  \includegraphics[width=3.8cm]{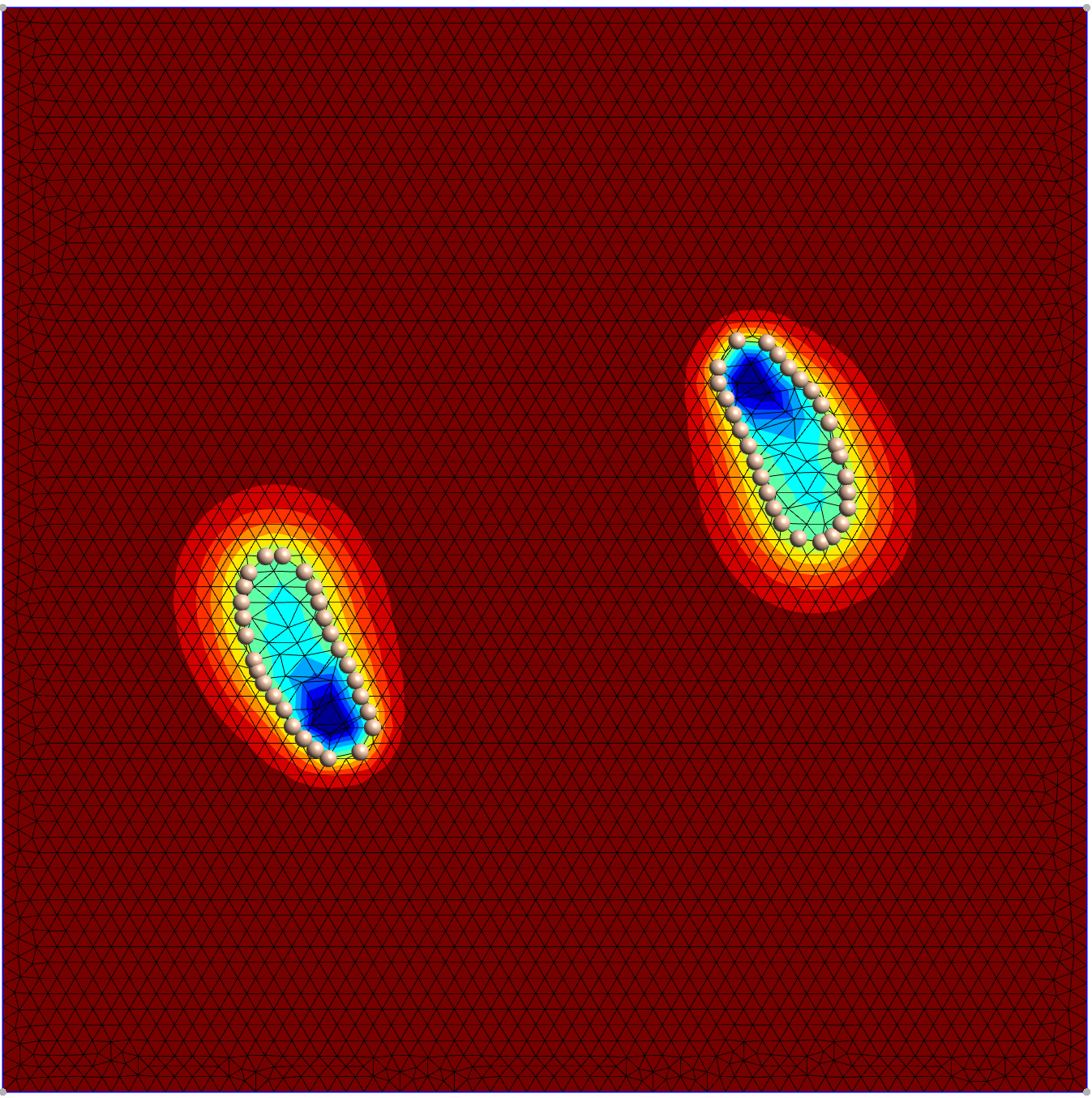} & 
  \includegraphics[width=3.8cm]{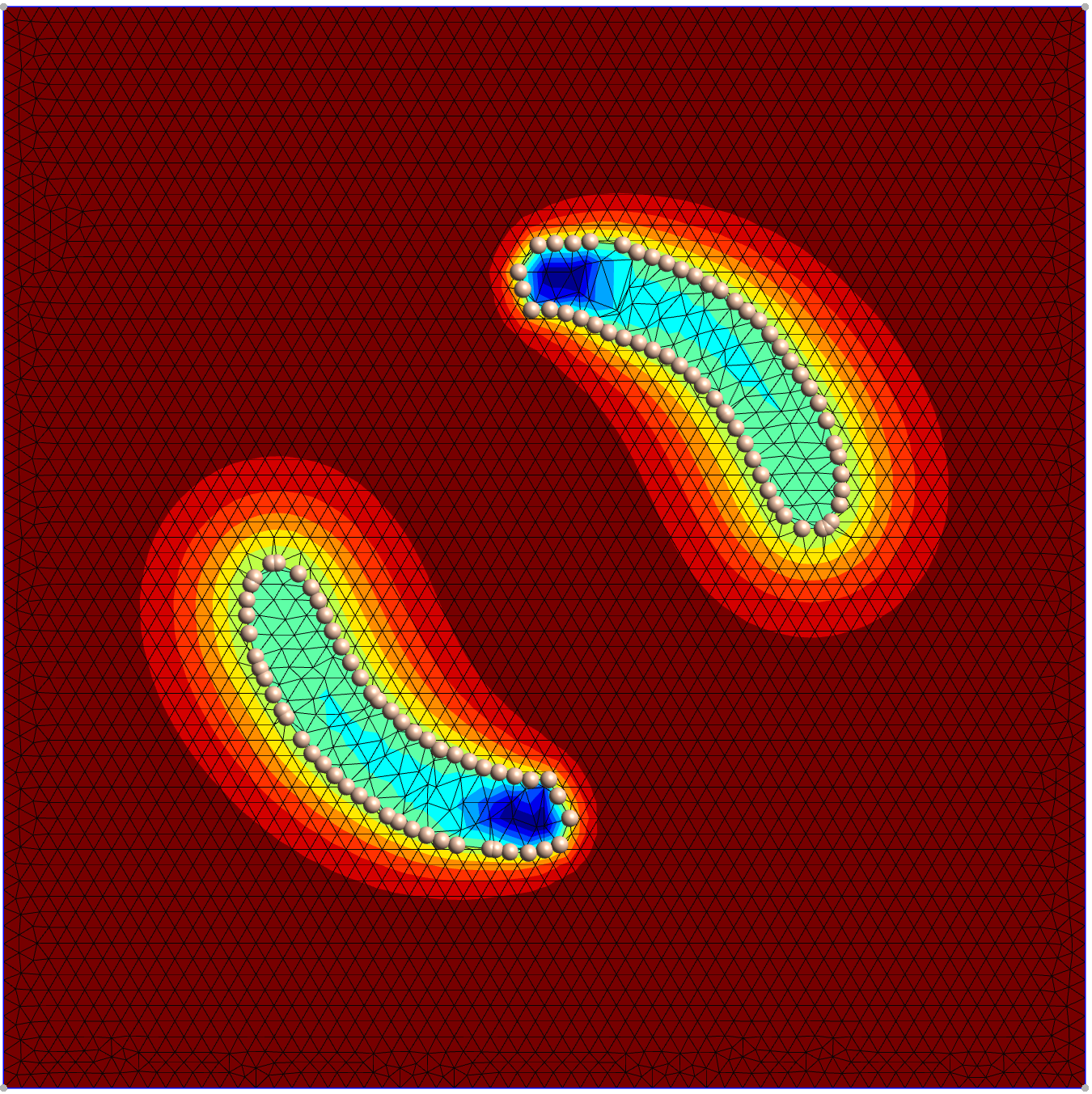} \\
\hline 
   (4) & (5) & (6)  \\
  \includegraphics[width=3.8cm]{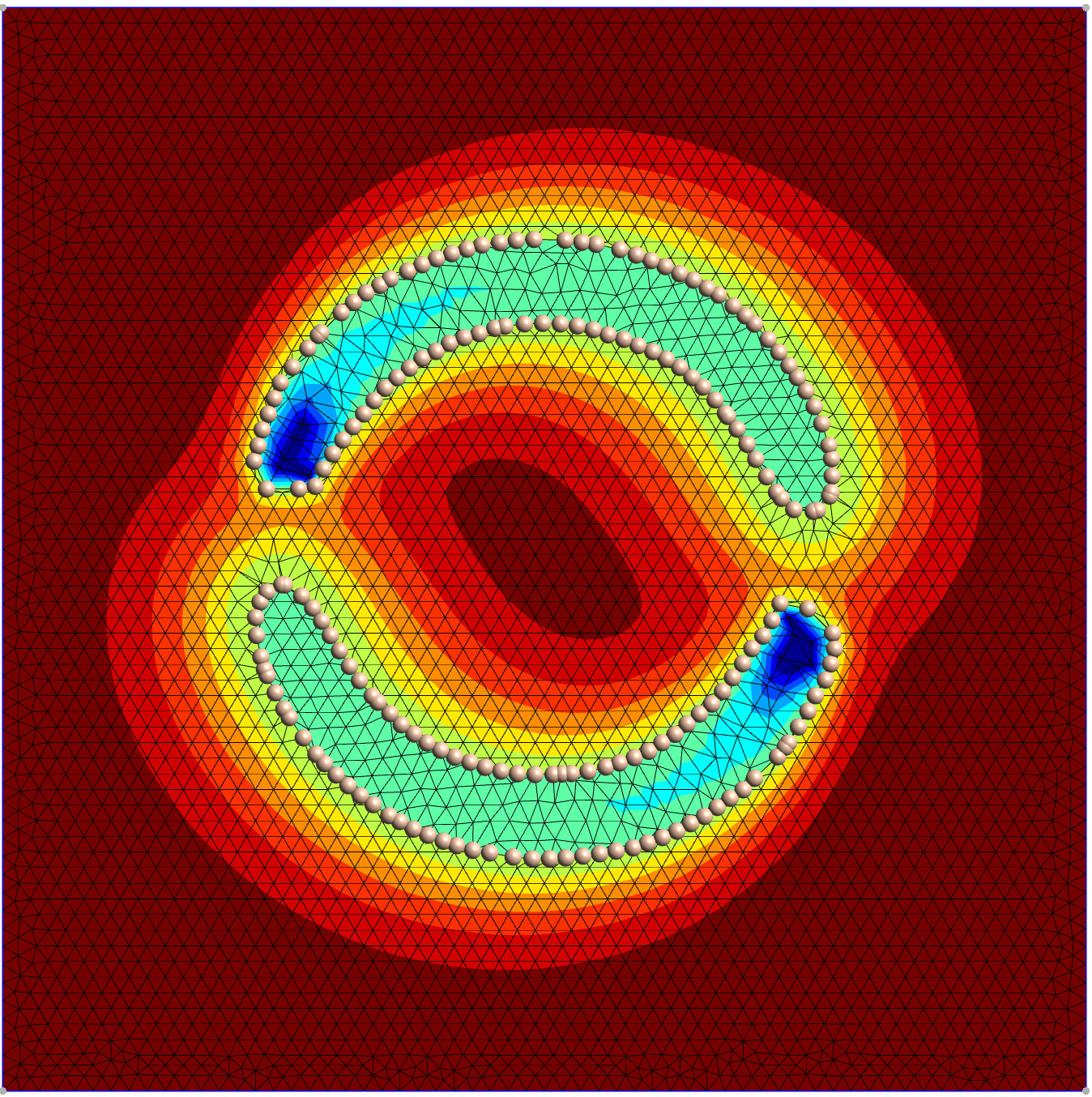} &
  \includegraphics[width=3.8cm]{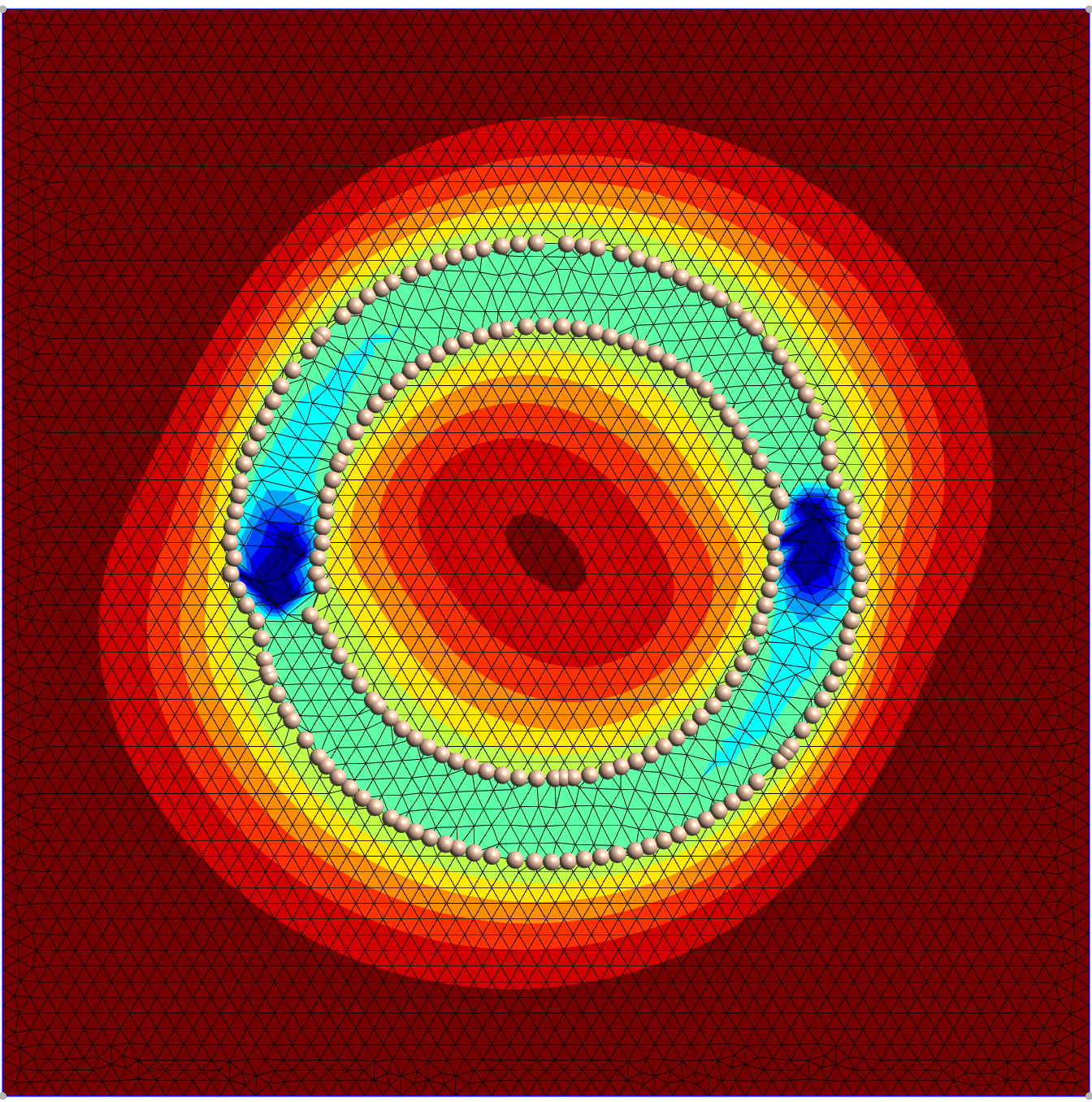} & 
  \includegraphics[width=3.8cm]{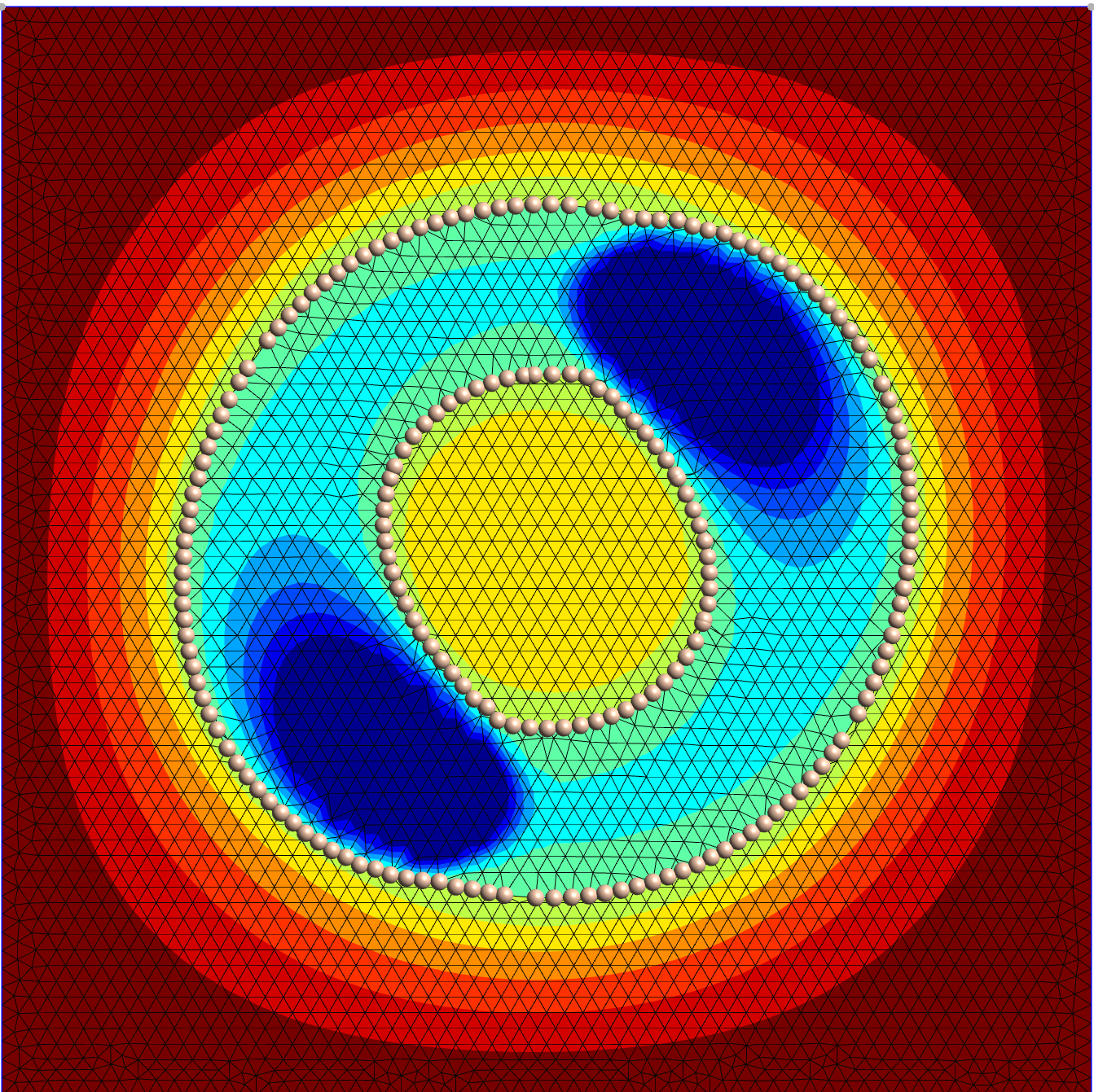} \\ 
\hline 
   (7) & (8) & (9)  \\
  \includegraphics[width=3.8cm]{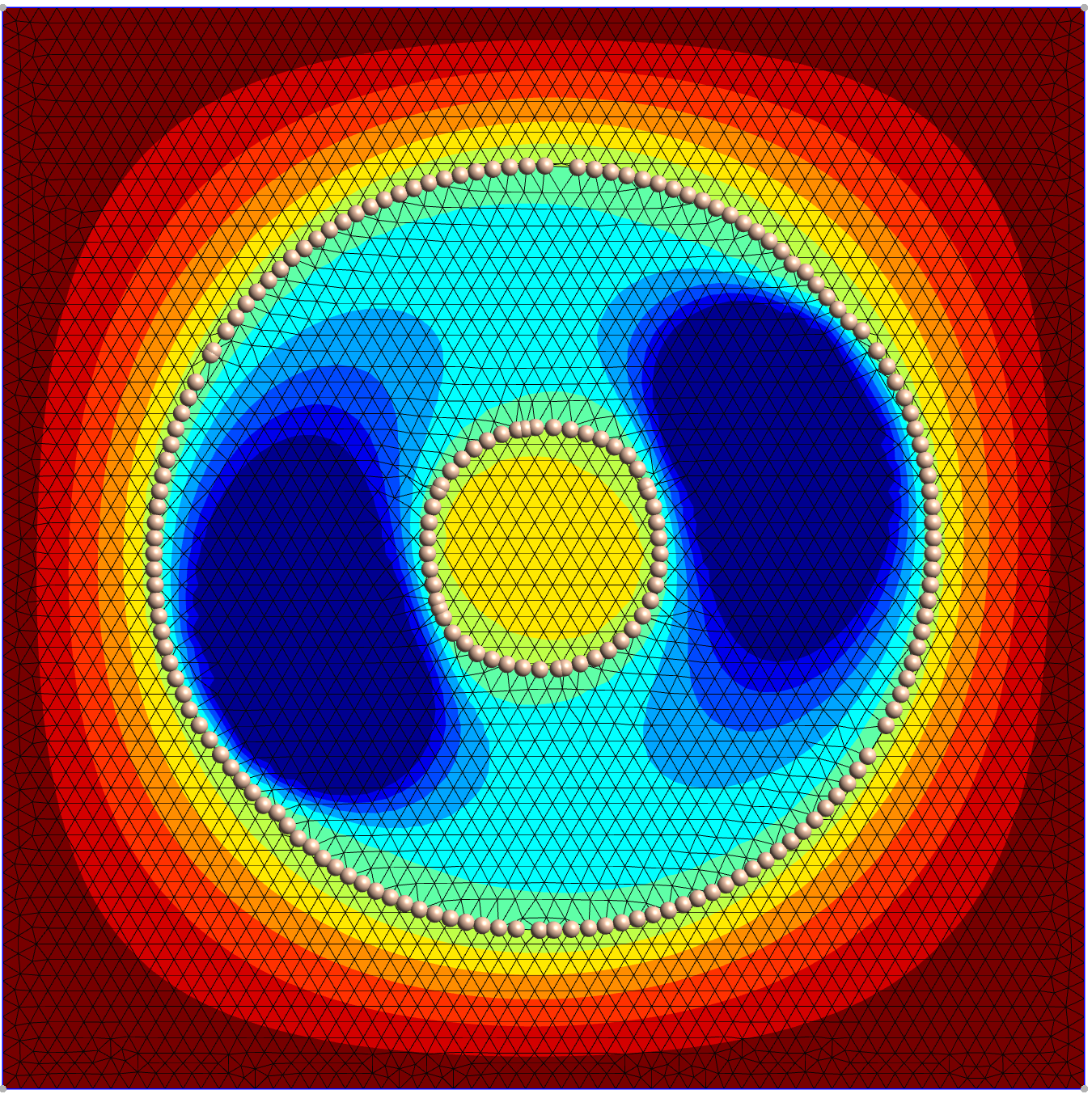} & 
  \includegraphics[width=3.8cm]{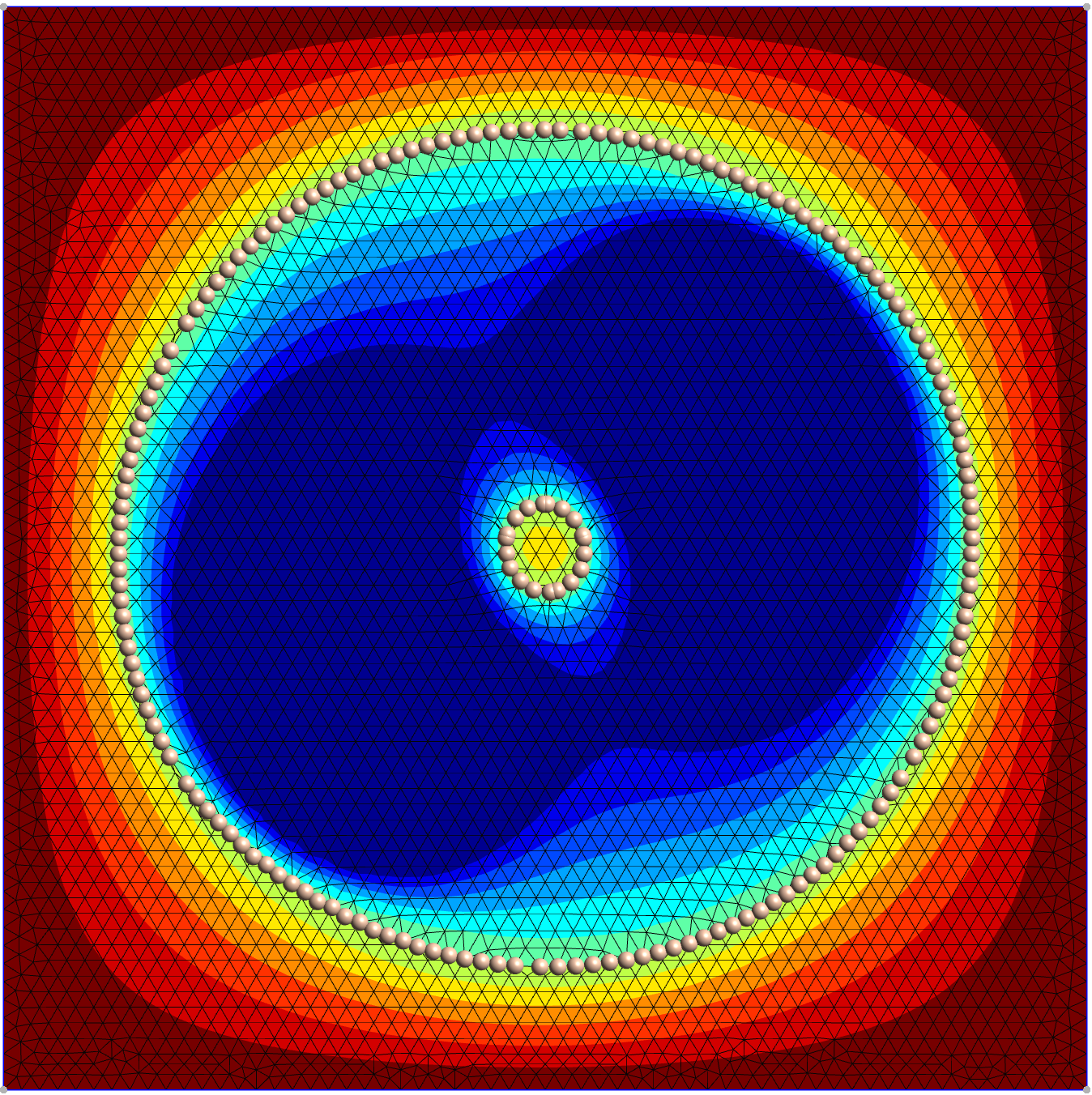} & 
  \includegraphics[width=3.8cm]{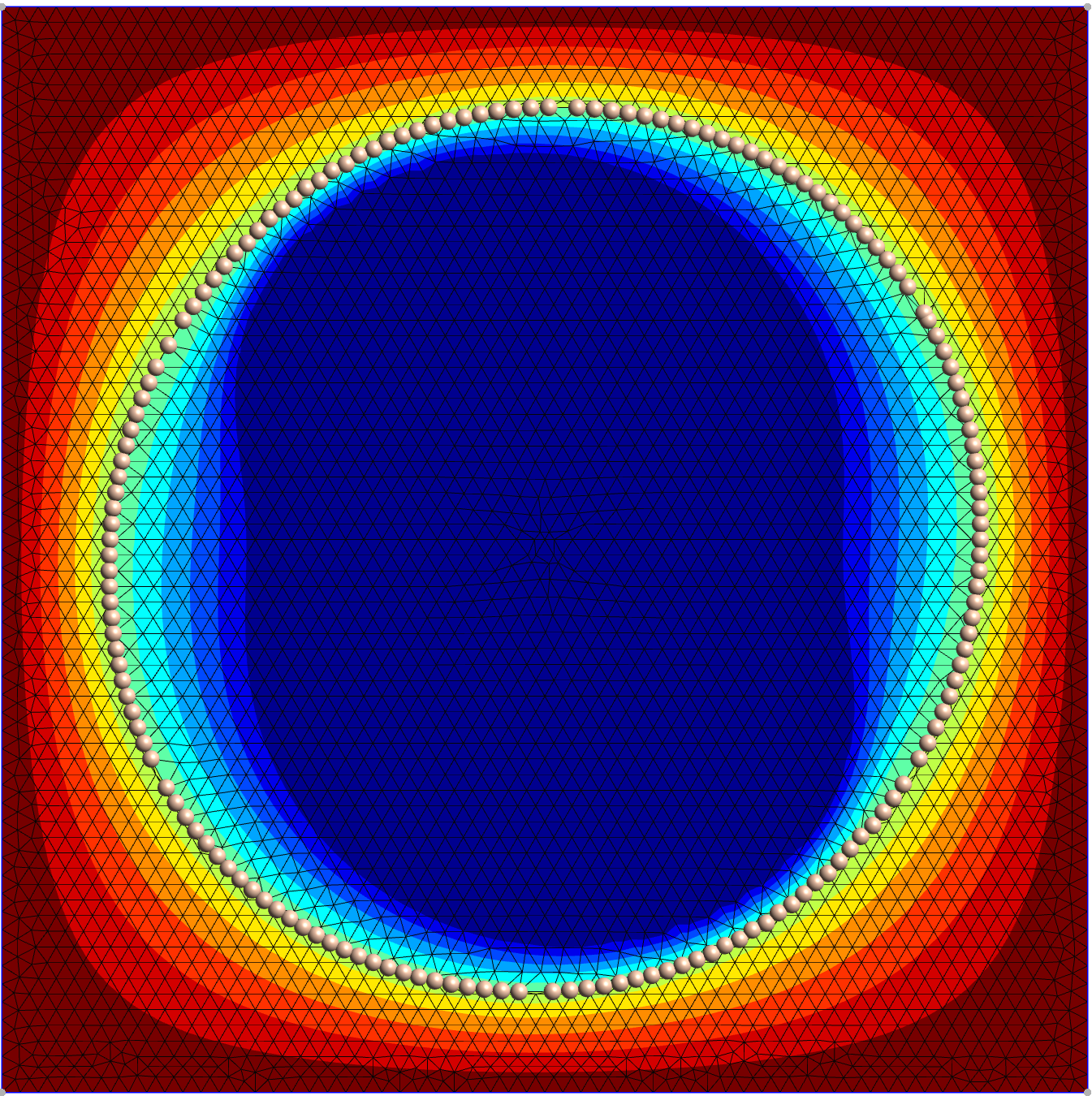} \\
\hline 
\end{tabular}                                            
 \end{center}
 \caption{Stefan problem with two rotating heat sinks.
   Front vertices are represented as yellowy dots. The color scale is
   saturated from below: any temperature below $-10$ degrees Celsius
   is represented in dark blue.
   At time 0 [s], the temperature is uniform and no
   front is yet present (Figure (1)).
   Two solidification  fronts rapidly nucleate (Figure (2)), grow
   (Figures (3) $\rightarrow$ (4)) and
 eventually merge (Figure (5)), forming two solidification fronts. The internal
 solidification front progresses towards the center of the domain
 (Figures (5) $\rightarrow$ (8)) and is finally annihilated, leaving the center of the
 domain entirely solidified (Figure (9)).  \answ{(An animated version of this simulation is provided as a supplementary material to this paper).}}
 \label{fig:multipleHeatSourcesTemperature}
 \end{figure}

In this last example, we start from an initial uniform temperature field $T_\ell = 283$ [K] on a square domain
of dimensions 3 $\times$ 3 [m$^2$]. This corresponds to image (1) on Figure \ref{fig:multipleHeatSourcesTemperature}. The same $T_\ell$ temperature is used as boundary condition on the square contour.
Two Dirac heat sinks of
intensity $500$ [W.m$^{-1}$], initially situated at positions
$x = \pm 0.75$ [m] and $y=0$ [m],  rotate on a circle of radius $0.75$ [m] with a period of 240 [h]. A constant time step of 4 [h] is used in the simulation.

The problem is solved with $\latentHeat =
3.3 \times 10^{5}$ [J.kg$^{-1}$].
Initially, there is no front. 
Then, due to the heat sinks, the
temperature decreases below $\boundaryTempInterface$, and two
fronts start to nucleate. Afterwards, two icy zones grow until merging with each other. At a later time, the middle of the domain is
totally frozen. This example illustrates the 
X-MESH capability to nucleate, merge  and annihilate fronts. 
The nonlinear solver converged in
less than 15 iterations at every time step.
It took less than two minutes to complete the simulation with a code that
is far from being optimized. 

\section{Conclusions}
In this paper, a new method to address the Stefan problem has been introduced. 
The extreme mesh deformation approach (X-MESH) tracks the solidification/melting 
front without remeshing and with a classical finite element formulation.
Moreover,  the mesh topology is kept fixed.
The X-MESH introduces the notion of 
front relaying allowing to keep an optimal mesh quality, 
at least for elements which are not directly connected to the front. 
The elements in contact with the front may have zero size during the relay.
To deal with these zero- or close to zero-measure element, we used an added 
volume approach but we plan to replace it with dualization which will not require any volume approximation.
The iterative process used in the solver is based on a quasi-Newton scheme which is quite easy to implement. The tangent operator uses smoothed energy and conductivity functions while
the residual uses the true sharp functions.

The numerical experiments on two problems with analytic solutions showed the quality of the obtained X-MESH solution. The final experiments with rotating heat sinks did demonstrate the X-MESH capability to have front nucleation, merge and annihilation in the same simulation without remeshing.

The front relaying methodology proposed here, although quite basic,
allows to solve problems that are far from being basic. Even if, at
convergence of the quasi-Newton, we found that the obtained meshes are
all valid, there is no guarantee that the proposed approach does
not produce inverted triangles. The front-relaying scheme could be
improved in this sense.

\answ{Finally, note that for the Stefan model considered, no history variable are needed. If history variable were present (as cumulative plasticity in plasticity models), these variables would need to be transferred as the mesh moves even though the mesh topology is kept fixed throughout the simulation. 
This, also, needs to be analyzed in further studies.}


\appendix

\bibliography{references}  

\end{document}

%% file: relaying.tex
\section{Front relaying}
\label{sec:relaying}
\begin{figure}[h!]
  \begin{tabular}{cc}
     (a) Mesh and function at $t=0$ &  
     (b) Active nodes at $t=\pi/5$  \\  
    \includegraphics[width = 0.38 \textwidth]{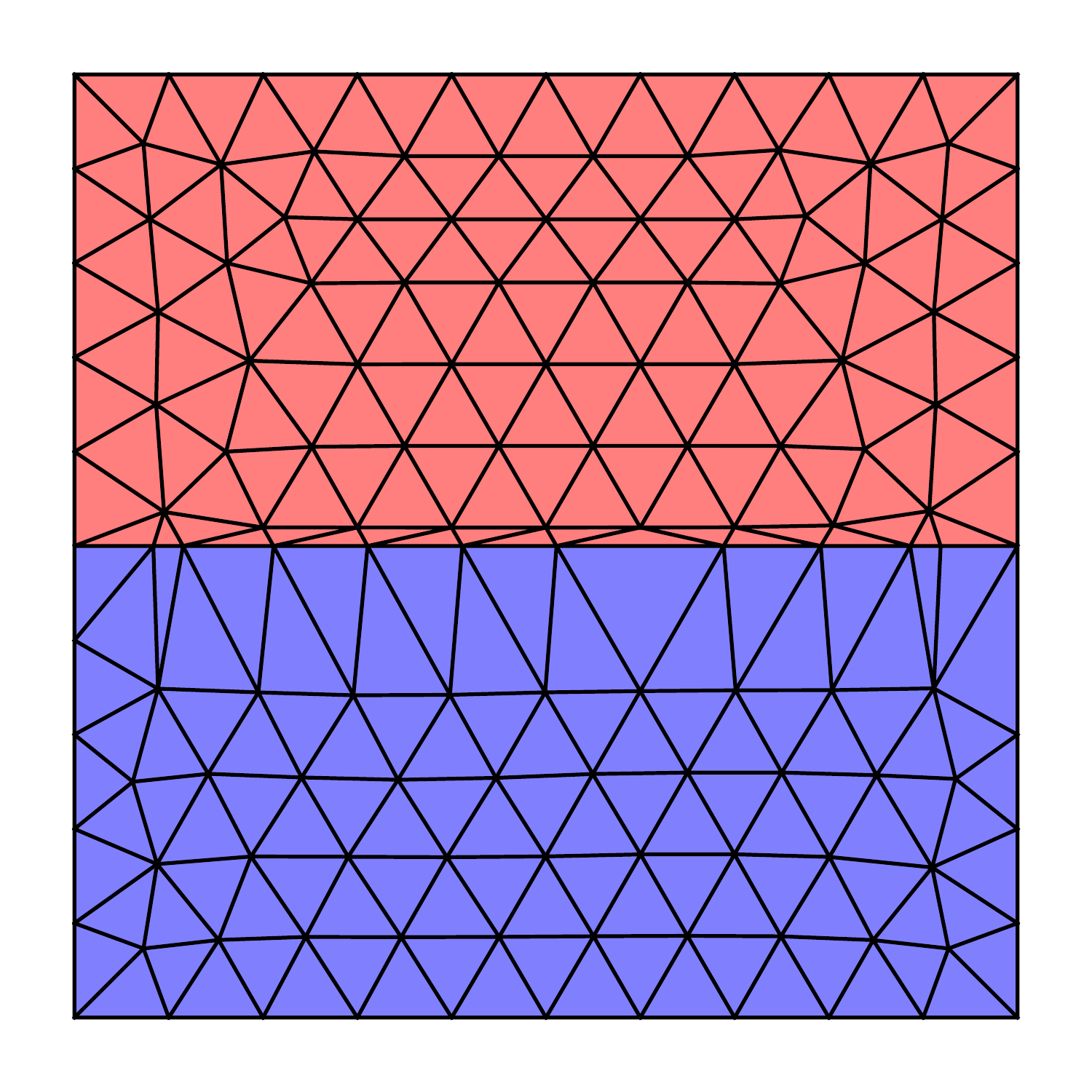}&
    \includegraphics[width = 0.38 \textwidth]{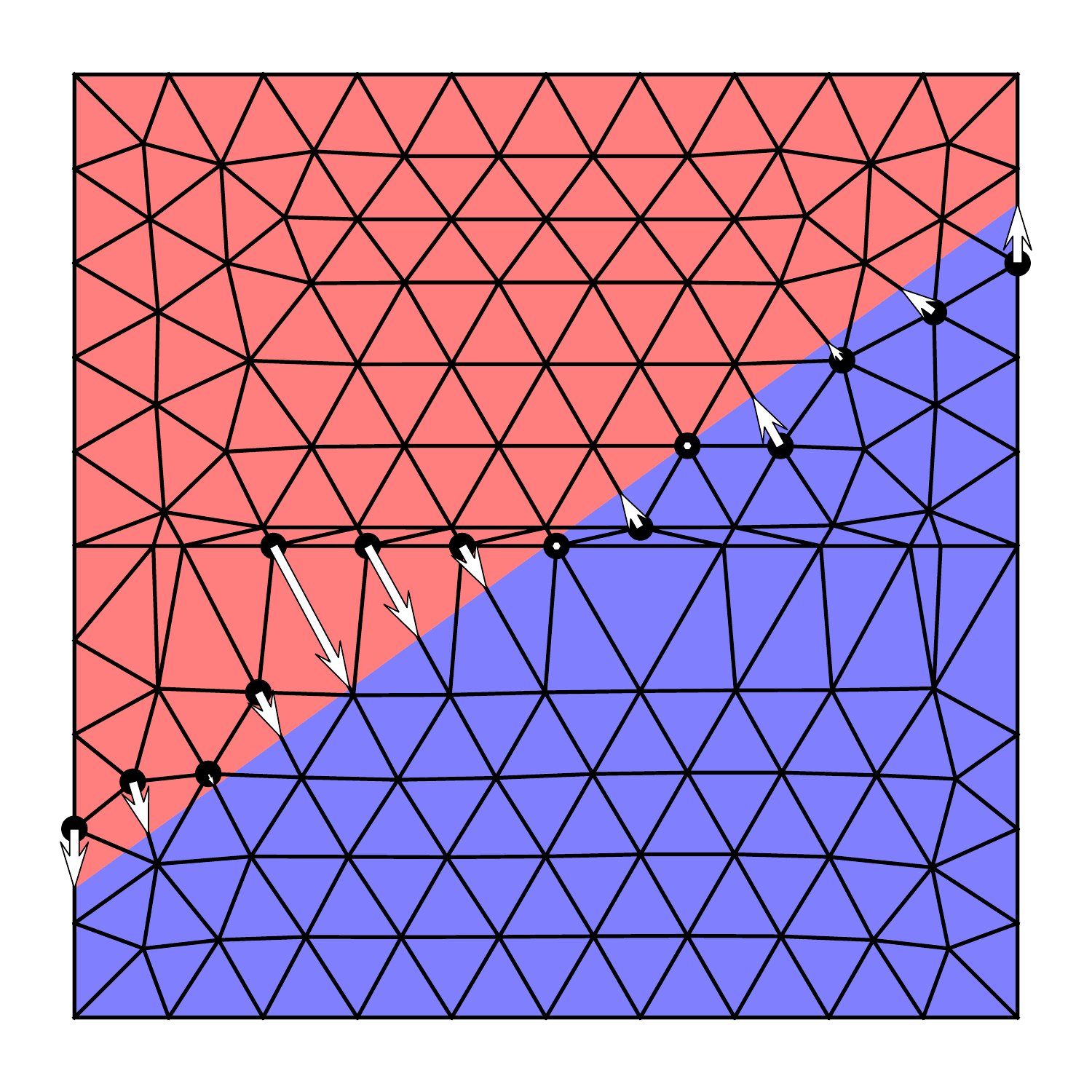}\\
     (c) Mesh and function at $t=\pi/5$ &  
     (d) Relaxation  \\  
    \includegraphics[width = 0.38 \textwidth]{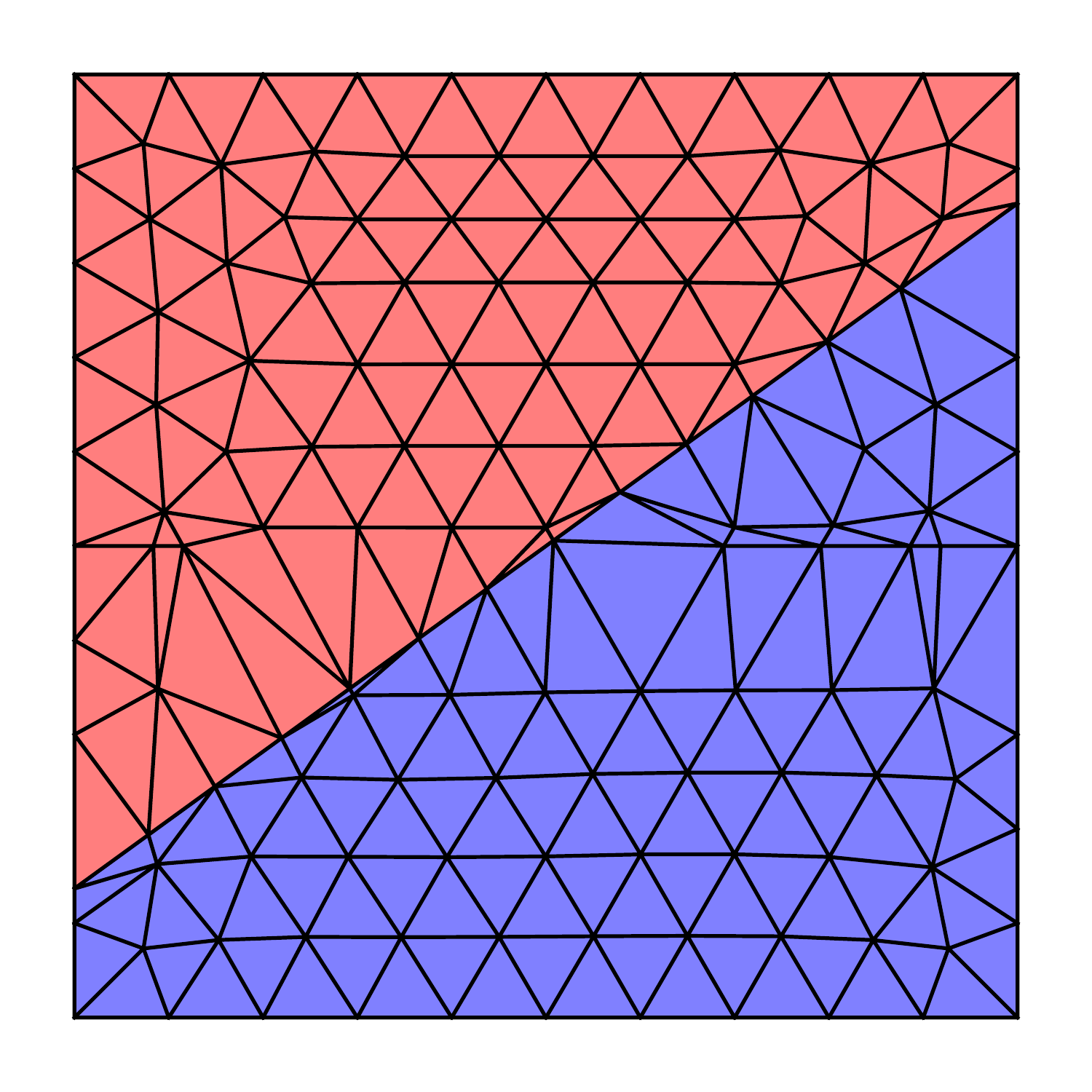}&
    \includegraphics[width = 0.38 \textwidth]{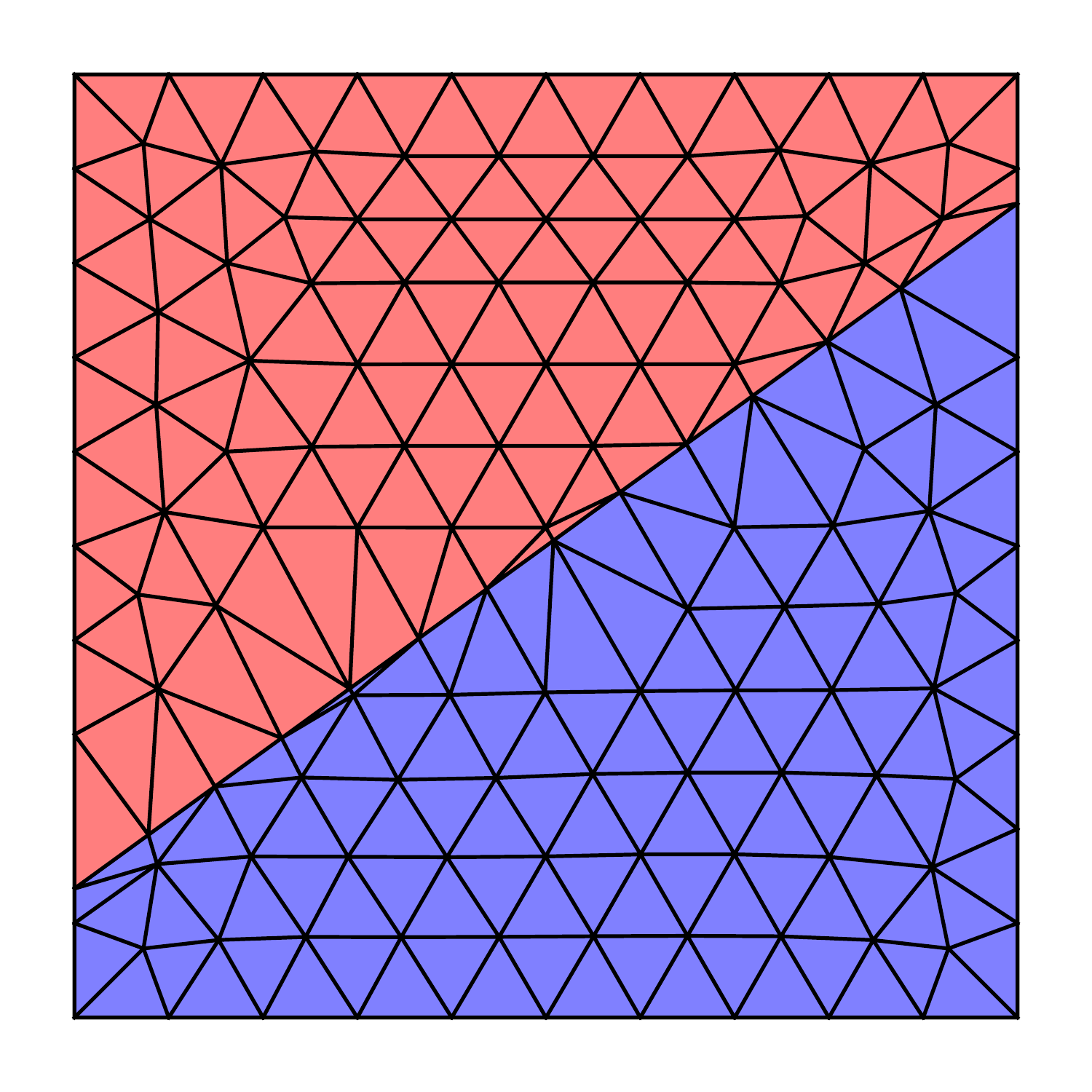}\\
  \end{tabular}    
\caption{Illustration of the relaying algorithm \label{fig:rel}}
\end{figure}
 A very important ingredient of X-MESH concerns the deformation of the mesh.
We want to develop a front tracking method where the connectivity of the
mesh remains unchanged but where topological changes of the interfaces
are possible. For this, we propose an original method that we call relaying,
by analogy with the passage of the baton in a relay race.

To explain relaying, we will use a manufactured example: we consider the function
$$f(x,y,t) = y \cos(t) + x \sin(t)$$
and we ask a mesh to track the iso zero
$$y = \tan(t) ~x$$
of this function at any time $t$.
The considered domain is a unit square centered at the origin.

At any time $t$, the mesh is supposed to represent exactly the iso-zero:
Figure \ref{fig:rel}(a) shows the mesh at $t=0$. This adaptation has been
realized with the current algorithm, starting from a standard triangular
mesh generated by gmsh \cite{gmsh}. The red color represents the
zone where $f(x,y,t) >0$ while blue color represents the zone
where $f <0$. Figure \ref{fig:rel}(b) shows function $f$ at
time $t=\pi/5$.  Nodes colored in black are called active nodes.
We define two subsets of the set of nodes of the mesh:
\begin{enumerate}
\item Subset $S_1$ contains both the nodes that were on the front at the previous
timestep (here at $t=0$) and the nodes that have changed of color at
current timestep (here $t =\pi/5$) -- 
going from red to blue or vice and versa.  
\item Subset $S_2$ contain all the nodes that belong to the edges that
  are crossed by the interface at current timestep (here the iso-zero of $f(x,y,\pi/5)$).
\end{enumerate}
The set $S$ of active nodes (in black dots on  Figure \ref{fig:rel}(b)) belong to
the intersection $S = S_1 \cap S_2$ of those two sets.
Active nodes
will move towards the front at current timestep.
Choosing active nodes in $S_1$ ensures that only the nodes upstream of the front move towards it,
thus ensuring the relaying. Choosing active nodes in $S_2$ ensures
that every edge crossed by the interface will have at least one of its
nodes going towards the front.

Another design choice related to the relaying algorithm is
related to moving the active nodes towards the front. We decided here to move the
nodes \emph{along the existing edges of the mesh}. Looking at Figure \ref{fig:rel}(b), we can see that,
most of the time, several edges are possible: we choose to use the shortest one, except on
the boundaries of the domain where the boundary edge is privileged.

The relaying process is extremely simple and powerful:
it handles automatically front nucleation, annihilation and
coalescence. In the nucleation case, the active nodes correspond only
to  nodes for which the sign of $f$ changes since there was no
  previous nodes on the front.
These active nodes will then move to target points corresponding to
the nucleation boundary. Annihilation is the inverse process to
nucleation in which the front becomes a small closed contour and
eventually vanishes. In this case, there are no targets since $f \neq
0$ and thus the front becomes also empty since all active nodes are
removed from the front.

Finally, coalescence is also automatic for about the same reasons as annihilation. In the coalescence
zone, there are no longer any targets and active nodes are removed from the front.

We preferred to move the nodes along the edges compared to, for example, projecting them on the front.
We have one very good reason for doing that. Relaying is very powerful, but it has a price:
the necessity of creating arbitrarily small elements, or even \emph{elements of zero measure.}

In \cite{Babuska1976}, authors prove that a sufficient condition for having convergent finite
element solution when the mesh is refined is that angles should be
bounded from above and thus forbid zero- or obtuse nearly zero-measure
elements.
Mesh algorithms are thus supposed to try their best to
generate \emph{quality meshes} in that sense \cite{shewchuk2002good}.

Babu{\v{s}}ka's conclusion has been mitigated since. In \cite{Hannukainen2012}, it is shown that there exist meshes that do not respect the angle
condition and that converge anyway.
Some authors have also wisely noted that the angle condition is only a sufficient
condition \cite{hannukainen2012maximum} and that this condition can
be largely weakened \cite{Kucera2016}.
In reality, it has been proven
that optimally convergent finite element solutions can be obtained with
meshes that are \emph{visually unpleasant}.
The family of computationally acceptable meshes can thus
include patterns of badly shaped elements \cite{Duprez2019} like
isolated caps (a cap is a triangle with an angle close to 180 degrees)
but also can consist in more complex
structures as bands or clusters. 
 In his paper \cite{Kucera2016}, Pr. Ku{\v{c}}era
claims that \emph{``one can fabricate very strange triangulations
  satisfying the assumptions of the theory presented.
  Such meshes perhaps do not have any value from the practical point
  of view''}.
In our perspective, these ``wild'' meshes are indeed very useful from
the practical point of view and are at the core of what is proposed here.

The presented algorithm does not preclude very small or even inverted element in the course of the iterations. This is quite natural since the front may 
transition from some nodes to other nodes.
It is the matrix coming from the diffusive term which will have the worst conditioning as 
the element becomes very small because it involves two deformation gradients.
To avoid issues with matrix conditioning in the solver, we set a lower threshold to the absolute
value of the Jacobian. This value is $A_\mathrm{tol}/A^0$ where $A_\mathrm{tol}$ is an area threshold
and $A^0$ is the area of the element in the initial mesh at $t=0$.
We transform the element Jacobian \answ{(see equation \eqref{eq:kinematicDef})} with the following formula
\begin{equation}
    J = \max( \mid J \mid, A_\mathrm{tol}/A^0) \; \mathrm{sign}(J).
\label{eq:jmax}
\end{equation}
which may be viewed as an added volume approach.
We found that the $\mathrm{sign}$ factor in the above relation 
was important in practice. In other words, if the element is inverted, 
the negative Jacobian sign is kept.  We have observed that, keeping
the jacobian negative allows Newton's algorithm to force the mesh to
return to a situation where the Jacobian is positive.
On the other hand, using $ J = \max( \mid J \mid, A_\mathrm{tol}/A^0)$
makes tangled elements acceptable to the solver and leads to meshes 
that contain inverted elements at convergence.

Visually unpleasant, yet acceptable patterns in a mesh only harm the
conditioning of the finite element matrices \cite{Duprez2019}. 
The solution proposed in this paper -- to bound the value of the
determinant of the Jacobian of the elements from below -- is not the
most elegant way of controlling this conditioning. 
We have developed two more rigorous approaches that enable to
handle zero measure elements without modifying the Jacobian.
This work will be submitted very soon.

Nevertheless, there are some unacceptable mesh patterns: the only really problematic case is the one
related to long bands of caps. The relaying algorithm we propose allows to never generate this kind of pattern:
moving the nodes along the edges allows us to create only isolated caps, which allows us to ensure
the convergence of our computations.  On Figure \ref{fig:rel}(a), we clearly see that most of the small size elements
on the front are needles (a needle is a triangular elements having an angle close to 0 degrees) and that only
one isolated cap is present.

Figure \ref{fig:rel}(c) shows the mesh after relocalization of the
nodes. The elements downstream of the front have a size that can be up
to twice their initial size, which is not ideal. We therefore end our
relaying algorithm with a so-called relaxation phase where the nodes
that have left the front return to their original position in the
initial mesh (see  Figure \ref{fig:rel}(d)).

\begin{figure}
  \begin{tabular}{ccccc}
    \includegraphics[width = 0.22 \textwidth]{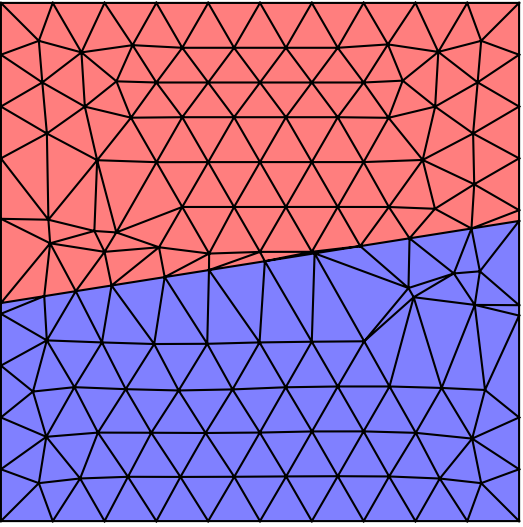}&
    \includegraphics[width = 0.22 \textwidth]{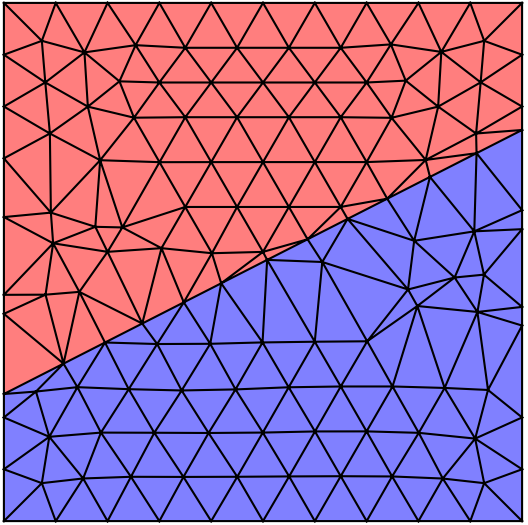}&
    \includegraphics[width = 0.22 \textwidth]{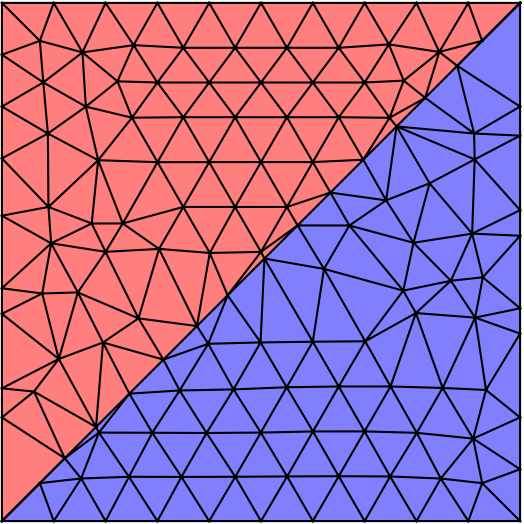}&
    \includegraphics[width = 0.22 \textwidth]{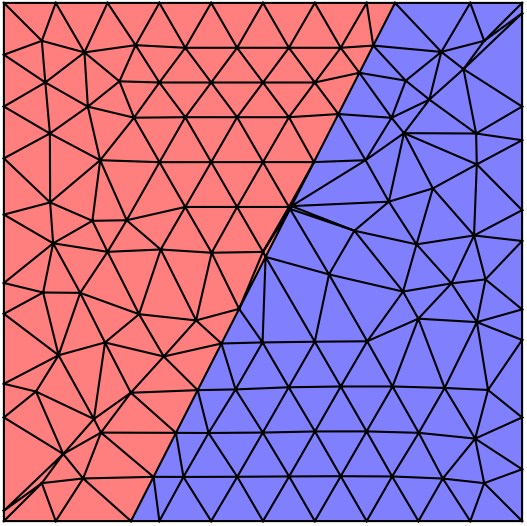}\\
    \includegraphics[width = 0.22 \textwidth]{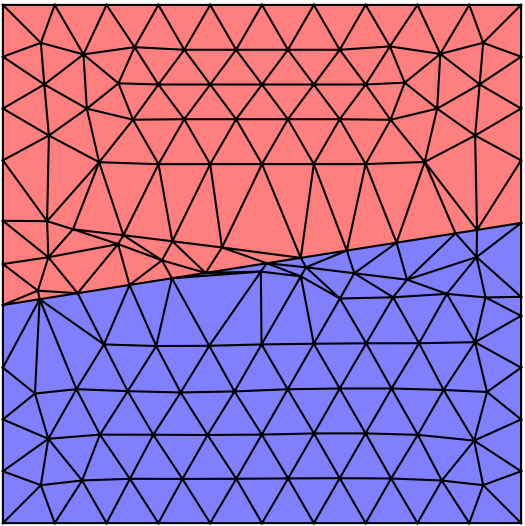}&
    \includegraphics[width = 0.22 \textwidth]{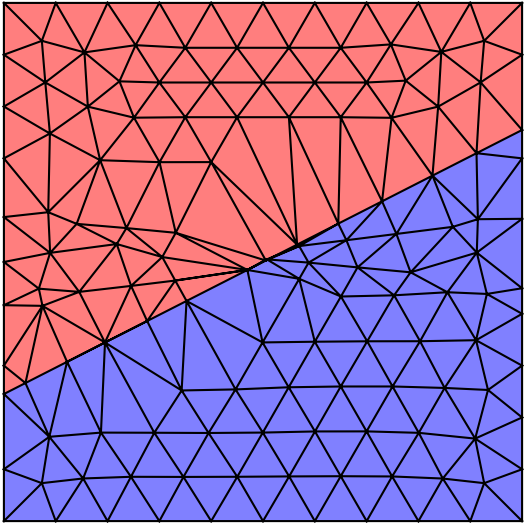}&
    \includegraphics[width = 0.22 \textwidth]{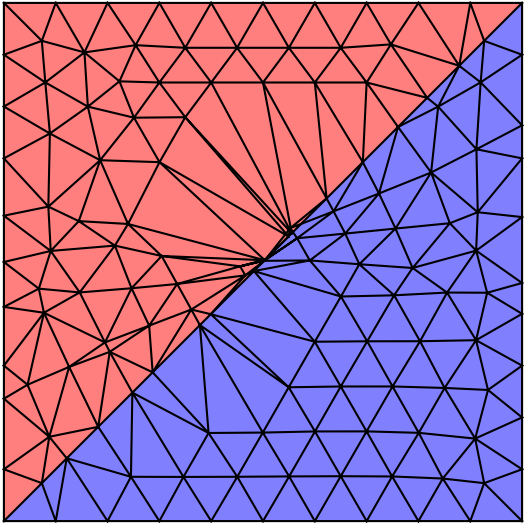}&
    \includegraphics[width = 0.22 \textwidth]{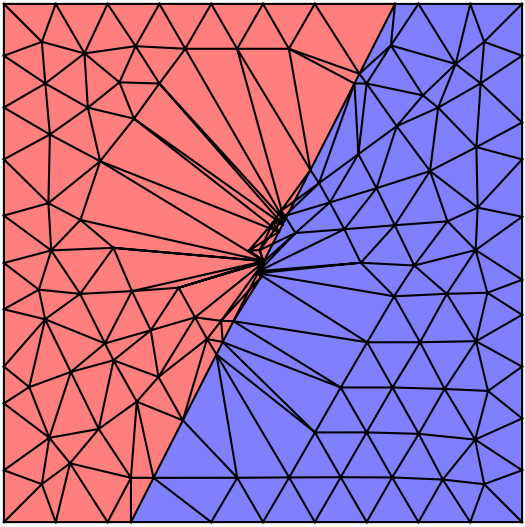}\\
  \end{tabular}    
\caption{Top figures show evolution of the mesh when active nodes are chosen upstream of the front (relaying) and bottom figure show the same evolution, but when active nodes are chosen dowstream of the front.\label{fig:antirelay}}
\end{figure}

One would think that any other strategy other than relaying would be the same.
Figure \ref{fig:antirelay} shows the evolution of the mesh if, instead of taking
the active nodes upstream of the front, we took them downstream. In this case,
nodes tend to accumulate downstream of the front, quickly leading to an invalid mesh.


%% file: stefan_jcp_third.bbl
\begin{thebibliography}{10}
\expandafter\ifx\csname url\endcsname\relax
  \def\url#1{\texttt{#1}}\fi
\expandafter\ifx\csname urlprefix\endcsname\relax\def\urlprefix{URL }\fi
\expandafter\ifx\csname href\endcsname\relax
  \def\href#1#2{#2} \def\path#1{#1}\fi

\bibitem{gupta2003classical}
S.~Gupta, The classical Stefan problem: basic concepts, Elsevier Amsterdam,
  2003.

\bibitem{Koga2020}
M.~Koga, S.and~Krstic, {Phase Change Model: Stefan Problem}, Springer
  International Publishing, Cham, 2020, pp. 1--13.
\newblock \href {https://doi.org/10.1007/978-3-030-58490-0_1}
  {\path{doi:10.1007/978-3-030-58490-0_1}}.

\bibitem{friedman1959free}
A.~Friedman, {Free boundary problems for parabolic equations I. Melting of
  solids}, Journal of Mathematics and Mechanics (1959) 499--517.

\bibitem{Jaafar2017}
M.~Jaafar, D.~Rousse, S.~Gibout, J.-P. B{\'{e}}d{\'{e}}carrats,
  \href{http://dx.doi.org/10.1016/j.rser.2017.02.050}{{A review of dendritic
  growth during solidification: Mathematical modeling and numerical
  simulations}}, Renewable and Sustainable Energy Reviews (2017).
\newblock \href {https://doi.org/10.1016/j.rser.2017.02.050}
  {\path{doi:10.1016/j.rser.2017.02.050}}.
\newline\urlprefix\url{http://dx.doi.org/10.1016/j.rser.2017.02.050}

\bibitem{Back2014}
J.~Back, S.~W. McCue, M.~H.-N. Hsieh, T.~Moroney, The effect of surface tension
  and kinetic undercooling on a radially-symmetric melting problem, Applied
  Mathematics and Computation 229 (2014) 41--52.

\bibitem{King2006}
J.~R. King, J.~D. Evans,
  \href{http://www.siam.org/journals/siap/65-5/60528.html}{Regularization by
  kinetic undercooling of blow-up in the ill-posed stefan problem},
  http://dx.doi.org/10.1137/04060528X 65 (2006) 1677--1707.
\newblock \href {https://doi.org/10.1137/04060528X}
  {\path{doi:10.1137/04060528X}}.
\newline\urlprefix\url{http://www.siam.org/journals/siap/65-5/60528.html}

\bibitem{Huerta2004}
J.~Donea, A.~Huerta, J.-P. Ponthot, A.~Rodríguez-Ferran, {Arbitrary
  Lagrangian–Eulerian Methods}, John Wiley \& Sons, Ltd, 2004, Ch.~14, pp.
  1--38.
\newblock \href {https://doi.org/https://doi.org/10.1002/0470091355.ecm009}
  {\path{doi:https://doi.org/10.1002/0470091355.ecm009}}.

\bibitem{LOUBERE20104724}
R.~Loubère, P.-H. Maire, M.~Shashkov, J.~Breil, S.~Galera,
  \href{https://www.sciencedirect.com/science/article/pii/S002199911000121X}{{ReALE:
  A reconnection-based arbitrary-Lagrangian–Eulerian method}}, Journal of
  Computational Physics 229~(12) (2010) 4724--4761.
\newblock \href {https://doi.org/https://doi.org/10.1016/j.jcp.2010.03.011}
  {\path{doi:https://doi.org/10.1016/j.jcp.2010.03.011}}.
\newline\urlprefix\url{https://www.sciencedirect.com/science/article/pii/S002199911000121X}

\bibitem{boscheri2014high}
W.~Boscheri, M.~Dumbser, D.~Balsara, High-order {ADER-WENO ALE} schemes on
  unstructured triangular meshes—application of several node solvers to
  hydrodynamics and magnetohydrodynamics, International Journal for Numerical
  Methods in Fluids 76~(10) (2014) 737--778.

\bibitem{BARLOW2016603}
A.~Barlow, P.~Maire, W.~J. Rider, R.~N. Rieben, M.~J. Shashkov,
  \href{https://www.sciencedirect.com/science/article/pii/S0021999116302807}{{Arbitrary
  Lagrangian–Eulerian methods for modeling high-speed compressible
  multimaterial flows}}, Journal of Computational Physics 322 (2016) 603--665.
\newblock \href {https://doi.org/https://doi.org/10.1016/j.jcp.2016.07.001}
  {\path{doi:https://doi.org/10.1016/j.jcp.2016.07.001}}.
\newline\urlprefix\url{https://www.sciencedirect.com/science/article/pii/S0021999116302807}

\bibitem{BURTON2018492}
D.~Burton, N.~Morgan, M.~Charest, M.~Kenamond, J.~Fung,
  \href{https://www.sciencedirect.com/science/article/pii/S0021999117308550}{{Compatible,
  energy conserving, bounds preserving remap of hydrodynamic fields for an
  extended ALE scheme}}, Journal of Computational Physics 355 (2018) 492--533.
\newblock \href {https://doi.org/https://doi.org/10.1016/j.jcp.2017.11.017}
  {\path{doi:https://doi.org/10.1016/j.jcp.2017.11.017}}.
\newline\urlprefix\url{https://www.sciencedirect.com/science/article/pii/S0021999117308550}

\bibitem{Baines2009}
M.~J. Baines, M.~E. Hubbard, P.~K. Jimack, R.~Mahmood, {A moving-mesh finite
  element method and its application to the numerical solution of phase-change
  problems}, Communications in Computational Physics 6~(3) (2009) 595--624.
\newblock \href {https://doi.org/10.4208/cicp.2009.v6.595}
  {\path{doi:10.4208/cicp.2009.v6.595}}.

\bibitem{Gros2018}
E.~Gros, G.~Anjos, J.~Thome,
  \href{https://www.sciencedirect.com/science/article/pii/S0096300318306180}{Moving
  mesh method for direct numerical simulation of two-phase flow with phase
  change}, Applied Mathematics and Computation 339 (2018) 636--650.
\newblock \href {https://doi.org/https://doi.org/10.1016/j.amc.2018.07.052}
  {\path{doi:https://doi.org/10.1016/j.amc.2018.07.052}}.
\newline\urlprefix\url{https://www.sciencedirect.com/science/article/pii/S0096300318306180}

\bibitem{Zhang2019}
Y.~Zhang, A.~Chandra, F.~Yang, E.~Shams, O.~Sahni, M.~Shephard, A.~A. Oberai,
  \href{https://doi.org/10.1016/j.jcp.2019.04.039}{{A locally discontinuous ALE
  finite element formulation for compressible phase change problems}}, Journal
  of Computational Physics 393 (2019) 438--464.
\newblock \href {https://doi.org/10.1016/j.jcp.2019.04.039}
  {\path{doi:10.1016/j.jcp.2019.04.039}}.
\newline\urlprefix\url{https://doi.org/10.1016/j.jcp.2019.04.039}

\bibitem{Chen1997a}
S.~Chen, B.~Merriman, S.~Osher, P.~Smereka, {A simple level set method for
  solving Stefan problems}, Journal of Computational Physics 135~(1) (1997)
  8--29.
\newblock \href {https://doi.org/10.1006/jcph.1997.5721}
  {\path{doi:10.1006/jcph.1997.5721}}.

\bibitem{Shaikh2016}
J.~Shaikh, A.~Sharma, R.~Bhardwaj,
  \href{http://dx.doi.org/10.1016/j.ijheatmasstransfer.2015.12.074}{{On
  sharp-interface level-set method for heat and/or mass transfer induced Stefan
  problem}}, International Journal of Heat and Mass Transfer 96 (2016)
  458--473.
\newblock \href {https://doi.org/10.1016/j.ijheatmasstransfer.2015.12.074}
  {\path{doi:10.1016/j.ijheatmasstransfer.2015.12.074}}.
\newline\urlprefix\url{http://dx.doi.org/10.1016/j.ijheatmasstransfer.2015.12.074}

\bibitem{Vasilev2020}
V.~Vasil'ev, M.~Vasilyeva, {An accurate approximation of the two-phase stefan
  problem with coefficient smoothing}, Mathematics 8~(11) (2020) 1--25.
\newblock \href {https://doi.org/10.3390/math8111924}
  {\path{doi:10.3390/math8111924}}.

\bibitem{Merle2002}
R.~Merle, J.~Dolbow,
  \href{http://www.springerlink.com/index/RNQ9AG3XN8E8QHK5.pdf}{{Solving
  thermal and phase change problems with the eXtended finite element method}},
  Computational Mechanics 28~(5) (2002) 339--350.
\newblock \href {https://doi.org/10.1007/s00466-002-0298-y}
  {\path{doi:10.1007/s00466-002-0298-y}}.
\newline\urlprefix\url{http://www.springerlink.com/index/RNQ9AG3XN8E8QHK5.pdf}

\bibitem{Ji2002}
H.~Ji, D.~Chopp, J.~Dolbow,
  \href{http://www3.interscience.wiley.com/journal/93518898/abstract}{{A hybrid
  extended finite element/level set method for modeling phase
  transformations}}, International Journal For Numerical Methods in Engineering
  54 (2002) 1209--1233.
\newline\urlprefix\url{http://www3.interscience.wiley.com/journal/93518898/abstract}

\bibitem{He2021}
M.~He, Q.~Yang, N.~Li, X.~Feng, N.~Liu, {An Extended Finite Element Method for
  heat transfer with phase change in frozen soil}, Soil Mechanics and
  Foundation Engineering 57~(6) (2021) 497--505.
\newblock \href {https://doi.org/10.1007/s11204-021-09698-z}
  {\path{doi:10.1007/s11204-021-09698-z}}.

\bibitem{Wang2015}
L.~Wang, P.-O. Persson, {A high-order discontinuous Galerkin method with
  unstructured space–time meshes for two-dimensional compressible flows on
  domains with large deformations}, Computers \& Fluids 118 (2015) 53--68.
\newblock \href {https://doi.org/10.1016/J.COMPFLUID.2015.05.026}
  {\path{doi:10.1016/J.COMPFLUID.2015.05.026}}.

\bibitem{GABURRO2020109167}
E.~Gaburro, W.~Boscheri, S.~Chiocchetti, C.~Klingenberg, V.~Springel,
  M.~Dumbser,
  \href{https://www.sciencedirect.com/science/article/pii/S0021999119308721}{{High
  order direct Arbitrary-Lagrangian-Eulerian schemes on moving Voronoi meshes
  with topology changes}}, {Journal of Computational Physics} 407 (2020)
  109167.
\newblock \href {https://doi.org/https://doi.org/10.1016/j.jcp.2019.109167}
  {\path{doi:https://doi.org/10.1016/j.jcp.2019.109167}}.
\newline\urlprefix\url{https://www.sciencedirect.com/science/article/pii/S0021999119308721}

\bibitem{gaburro2021unified}
E.~Gaburro, {A unified framework for the solution of hyperbolic PDE systems
  using high order direct Arbitrary-Lagrangian--Eulerian schemes on moving
  unstructured meshes with topology change}, Archives of Computational Methods
  in Engineering 28~(3) (2021) 1249--1321.

\bibitem{re2017interpolation}
B.~Re, C.~Dobrzynski, A.~Guardone, {An interpolation-free ALE scheme for
  unsteady inviscid flows computations with large boundary displacements over
  three-dimensional adaptive grids}, Journal of Computational Physics 340
  (2017) 26--54.

\bibitem{springel2010pur}
V.~Springel, {E pur si muove: Galilean-invariant cosmological hydrodynamical
  simulations on a moving mesh}, Monthly Notices of the Royal Astronomical
  Society 401~(2) (2010) 791--851.

\bibitem{Boffi2004}
D.~Boffi, L.~Gastaldi, {Stability and geometric conservation laws for ALE
  formulations}, Computer Methods in Applied Mechanics and Engineering
  193~(42-44) (2004) 4717--4739.
\newblock \href {https://doi.org/10.1016/j.cma.2004.02.020}
  {\path{doi:10.1016/j.cma.2004.02.020}}.

\bibitem{azaiez2016two}
M.~Azaiez, F.~Jelassi, M.~Mint~Brahim, J.~Shen, {Two-phase Stefan problem with
  smoothed enthalpy}, Communications in Mathematical Sciences 14~(6) (2016)
  1625--1641.

\bibitem{gmsh}
C.~Geuzaine, J.-F. Remacle, {Gmsh: A 3-D finite element mesh generator with
  built-in pre-and post-processing facilities}, International journal for
  numerical methods in engineering 79~(11) (2009) 1309--1331.

\bibitem{Babuska1976}
I.~Babuska, A.~Aziz, On the angle condition in the finite element method, SIAM
  Journal on Numerical Analysis 13 (1976) 214--226.
\newblock \href {https://doi.org/https://doi.org/10.1137/0713021}
  {\path{doi:https://doi.org/10.1137/0713021}}.

\bibitem{shewchuk2002good}
J.~Shewchuk, What is a good linear finite element? interpolation, conditioning,
  anisotropy, and quality measures (preprint), University of California at
  Berkeley 73 (2002) 137.

\bibitem{Hannukainen2012}
A.~Hannukainen, S.~Korotov, M.~Kř{\'{i}}{\v{z}}ek, {The maximum angle
  condition is not necessary for convergence of the finite element method},
  Numerische Mathematik 120~(1-6) (2012) 79--88.
\newblock \href {https://doi.org/10.1007/s00211-011-0403-2}
  {\path{doi:10.1007/s00211-011-0403-2}}.

\bibitem{hannukainen2012maximum}
A.~Hannukainen, S.~Korotov, M.~K{\v{r}}{\'\i}{\v{z}}ek, The maximum angle
  condition is not necessary for convergence of the finite element method,
  Numerische mathematik 120~(1) (2012) 79--88.

\bibitem{Kucera2016}
V.~Ku{\v{c}}era, \href{http://arxiv.org/abs/1601.02942}{{On necessary and
  sufficient conditions for finite element convergence}} (2016).
\newblock \href {http://arxiv.org/abs/1601.02942} {\path{arXiv:1601.02942}}.
\newline\urlprefix\url{http://arxiv.org/abs/1601.02942}

\bibitem{Duprez2019}
M.~Duprez, V.~Lleras, A.~Lozinski, {Finite element method with local damage of
  the mesh}, ESAIM: Mathematical modelling and numerical analysis 53 (2019)
  1--10.
\newblock \href {http://arxiv.org/abs/arXiv:1401.0559v1}
  {\path{arXiv:arXiv:1401.0559v1}}.

\bibitem{Carslaw1959}
H.~S. Carslaw, J.~Jaeger, {Conduction of heat in solids}, Clarendon Press,
  1959.

\end{thebibliography}
